\documentclass[10pt,aps,prd,twocolumn,showpacs,superscriptaddress,longbibliography,nofootinbib]{revtex4-2}
\usepackage{mathrsfs,amsmath,amsthm,latexsym,amssymb,amsfonts,epsfig,cancel,enumerate,graphicx,txfonts,diagbox}

\usepackage{multirow}
\usepackage[T1]{fontenc}
\usepackage[utf8]{inputenc}
\usepackage[colorlinks=true,linkcolor=blue,citecolor=blue,urlcolor=blue]{hyperref}
\usepackage{orcidlink}

\usepackage{xcolor}
\usepackage{booktabs}
\usepackage{graphicx}
\usepackage{subcaption}

\definecolor{navy}{RGB}{0,0,150}
\usepackage{appendix}
\allowdisplaybreaks
\newcommand{\RGU}{Department of Physics, The Assam Royal Global University, Guwahati-781035, Assam, India}
\newcommand{\ABTU}{Department of Physics, Al-Hussein Bin Talal University, 71111, Ma’an, Jordan}
\newcommand{\UCC}{Centro de Investigaci\'{o}n en Ciencias del Espacio y F\'{i}sica Te\'{o}rica (CICEF), Universidad Central de Chile, La Serena 1710164, Chile}

\begin{document}

\title{Some phenomenological aspects of a quantum-corrected Reissner–Nordstr\"{o}m black hole: quasi-periodic oscillations, scalar perturbations and thermal fluctuations
}

\author{Faizuddin Ahmed\orcidlink{0000-0003-2196-9622}}
\email{faizuddinahmed15@gmail.com}
\affiliation{\RGU}

\author{Ahmad Al-Badawi\orcidlink{0000-0002-3127-3453}}
\email{ahmadbadawi@ahu.edu.jo}
\affiliation{\ABTU}

\author{Mohsen Fathi\orcidlink{0000-0002-1602-0722}}
\email{mohsen.fathi@ucentral.cl (Corresp. author)}
\affiliation{\UCC}

\begin{abstract}
In this work, we investigate several phenomenological aspects of a covariant quantum-corrected Reissner-Nordstr\"{o}m  black hole characterized by the mass $M$, electric charge $Q$, and the quantum correction parameter $\zeta$. We first study the motion of neutral test particles and derive the fundamental orbital and epicyclic frequencies, which are then employed to analyze different quasi–periodic oscillation (QPO) models. Using observational QPO data from stellar–mass, intermediate–mass, and supermassive black hole candidates, we perform a Bayesian parameter estimation through a Markov Chain Monte Carlo (MCMC) analysis and obtain constraints on the black hole parameters. The results show that the presence of the quantum correction significantly affects the location of the QPO radii and the separation between the QPO orbit and the ISCO. We then examine the scalar perturbations by deriving the Schr\"{o}dinger-like radial equation and the corresponding effective potential. The influence of the parameters $Q$ and $\zeta$ on the perturbation potential and stability of the spacetime is discussed. Furthermore, we compute the greybody factor and the energy emission rate in the high-frequency (geometric–optics) regime, showing how the quantum correction modifies the absorption probability and radiation spectrum. 
Finally, we study the effect of thermal fluctuations on the black hole entropy and obtain the logarithmic corrections to the Bekenstein-Hawking area law. We show that these corrections become important for small black holes, while for large horizon radius the standard thermodynamic behavior is recovered. Our analysis demonstrates that the quantum correction parameter leaves observable imprints on both dynamical and thermodynamical properties of the spacetime and can be constrained through QPO observations.\\

\href{https://doi.org/10.48550/arXiv.2602.15551}{arXiv:2602.15551 [gr-qc]}
\end{abstract}

\maketitle

\tableofcontents

\section{Introduction}

Black holes (BHs), predicted by General Relativity (GR), have now been confirmed through various observational means~\cite{LIGO2016,EHTL1,EHTL12}. Their intense gravitational fields make them ideal laboratories for testing diverse gravitational theories. Simultaneously, BHs possess enormous energy. Phenomena such as high-energy radiation and relativistic jets produced by the accretion disks surrounding BHs~\cite{Falcke1993} constitute some of the most active and energetically efficient astrophysical processes in the Universe. It was generally believed that this immense energy was inaccessible to external extraction. This view changed when Penrose, based on the existence of negative-energy orbits within the ergoregion, proposed a mechanism for extracting energy from a rotating BH, later known as the Penrose process~\cite{Penrose:1971uk}. This proposal fundamentally transformed our understanding of BHs, redefining them from passive gravitational endpoints into active media capable of exchanging energy with their surroundings.

The study of particle motion in the vicinity of a BH has become a topic of significant interest in astrophysics. This interest arises from the extremely strong gravitational field of BHs, which profoundly influences the surrounding spacetime and the motion of nearby particles. Such environments provide a natural laboratory for testing the theoretical predictions of Einstein’s general theory of relativity under extreme conditions.

In the strong gravitational field near the event horizon, both particles and light deviate from their original trajectories and follow curved paths determined by the spacetime geometry. It has been suggested that the motion of particles is not identical around every BH; rather, it depends on the specific properties of the BH, such as its mass, charge, and angular momentum. Studying these differences in particle motion allows us to better understand the corresponding physical characteristics of BHs, particularly in extreme regimes. Particles moving around a BH may follow a variety of trajectories, including stable and unstable circular orbits, spiral paths, or even chaotic motion. Analyzing these different types of trajectories is essential for exploring the structure of the surrounding spacetime. Due to the intense gravitational field near a BH, particles may also experience significant relativistic effects such as gravitational redshift and time dilation. These effects become especially pronounced near the innermost stable circular orbit (ISCO), which represents the smallest radius at which a particle can maintain a stable circular orbit \cite{RB1996}.

The motion of test particles in the vicinity of BHs reveals intriguing connections to the QPOs observed in the X-ray emission of certain microquasars \cite{JH2003}. These motions are astrophysically significant, as they provide strong evidence supporting the predictions of GR in strong gravitational fields. Moreover, QPOs serve as valuable probes for determining fundamental BH parameters, such as mass and spin \cite{MQ2023,JR2022}. Since their discovery around thirty years ago, QPOs have been widely used as diagnostic tools in astrophysics, enhancing our understanding of BH behavior and the physics of accretion processes. Several theoretical models have been proposed to explain the origin of QPOs in astrophysical systems, including (i) hot spot models, (ii) disk seismic models, (iii) resonance models, and (iv) warped disk models \cite{LR2013}. Despite these efforts, the exact physical mechanisms responsible for QPOs remain unclear, making them a central topic of research in both GR and alternative gravity theories \cite{Rezolla2003,Jiang2021,Torok2011,Stuchlik2011,Stuchlik2013,Ashraf2025}. 

In addition, thin Keplerian accretion disks around compact objects are characterized by stable circular orbits, which play a crucial role in shaping the flow dynamics \cite{Abramowicz2010,Zanotti2003}. The oscillatory behavior observed in such systems is influenced by epicyclic frequencies, which can be investigated using epicyclic or geodesic models that describe QPOs near supermassive BHs at galactic centers \cite{Alston2016}. The dynamics become more complex when electromagnetic interactions between charged regions and the strong magnetic fields surrounding BHs are considered. Including these effects provides deeper insight into the properties of epicyclic oscillations \cite{Kolos2015,Tursunov2016,Martin2017,Zdenek2016}. Observed QPOs in X-ray emission are thought to be closely linked to oscillatory motion in regions near the innermost stable circular orbit (ISCO) \cite{Bambi2017}. In low-mass X-ray binaries, which may contain a neutron star as one of the components, characteristic QPOs often display multiple peaks \cite{Ingram2010,Rayimbaev2023,Torok2005}. Various physical processes are proposed to generate twin-peaked or multi-peaked QPOs, reflecting differences in the underlying motion \cite{Germana2017,Luniano2003}. Developing specific theoretical models for each type of QPO is essential to explain their origin under varying astrophysical conditions. By analyzing these oscillatory features, astronomers can improve their understanding of extreme accretion environments and the remarkable physical phenomena occurring near compact objects.

Scalar perturbations are small disturbances in a scalar field $\phi$ propagating in curved spacetime, typically around BHs. They represent the simplest type of perturbation and are widely used to study BH stability, wave propagation, and quasinormal modes (QNMs) \cite{Berti2009,Konoplya2011}. Scalar fields can be fundamental, such as the inflaton or dilaton, or effective fields describing physical quantities like density fluctuations \cite{Chandrasekhar1983}. The dynamics of scalar perturbations are characterized by QNMs, which are damped oscillations with complex frequencies $\omega = \omega_R + i \omega_I$, where $\omega_I < 0$ indicates stability. QNMs describe the characteristic “ringing” of BHs following perturbations and are independent of the initial disturbance \cite{Kokkotas1999}. Measuring QNMs allows for the determination of BH parameters including mass, spin, and charge, and provides tests of GR in strong-field regimes \cite{Berti2009}. Scalar perturbations are also used to probe BH stability. Schwarzschild and Kerr BHs are stable under such perturbations. Stability is often linked to the positivity of the effective potential $V_\text{eff}(r)$ outside the event horizon \cite{Chandrasekhar1983}.

Recent observations of gravitational waves (GWs) arising from the relativistic mergers of compact objects have opened a new window for probing the nature and properties of these objects. In light of the recent LIGO and Virgo detections \cite{LIGO2016, VGW151226, VIRGO2017, Virgo2017GW170814, Virgo2017GW170817}, GW astronomy is poised to provide unprecedented insights into the gravitational dynamics and astrophysical characteristics of compact binaries. Although current observations do not yet resolve the detailed structure of spacetime near compact objects beyond the photon sphere, future detections are expected to reveal more about their near-horizon geometry. In particular, precise measurements of the ringdown phase, dominated by a sequence of QNMs, will offer direct information about the BH parameters and spacetime geometry \cite{Vishveshwara1970}. Moreover, any deviations from the classical BH paradigm at near-horizon scales-such as quantum corrections or exotic compact structures-could manifest as secondary signals or “echoes” in the GWs' waveform following the primary ringdown \cite{Cardoso2016, Cardoso2016b}. These observations have the potential to probe the physics of gravity in the strong-field regime and to test fundamental aspects of BH structure. Several analytical and numerical techniques are employed to study scalar perturbations:  
\begin{itemize}  
    \item \textbf{WKB approximation:} Provides semi-analytical estimates of QNMs \cite{Schutz1985,Iyer1987}.  
    \item \textbf{Time-domain integration:} Simulates the evolution of perturbations, capturing ringdown behavior and late-time tails.  
    \item \textbf{Continued fraction method:} Offers precise computation of QNMs for Kerr and charged BHs \cite{Leaver1985}.  
\end{itemize}
Understanding scalar perturbations is crucial for analyzing wave propagation, energy emission, and stability around BHs, as well as for testing alternative theories of gravity through astrophysical observations \cite{Zanotti2003,Bambi2017}.

This study investigates the dynamics of particles and the resulting QPOs in the spacetime of a quantum-corrected Reissner-Nordstr\"{o}m (RN) BH. The paper is organized as follows. In Sec.~\ref{sec:2}, we first provide a brief review on the quantum-corrected RN BH and analyze its horizon structure. We then investigate the motion of test particles around this BH in Sec.~\ref{sec:3}. Fundamental frequencies were discussed in Sec.~\ref{sec:4}. The scalar perturbations of the BH and greybody factors in Sec.~\ref{sec:5}. The energy emission rate in Sec. \ref{sec:6} and thermal fluctuations in Sec.\ref{sec:7}. Finally, the conclusions in Sec.~\ref{sec:8}.

Throughout this paper, we adopt geometric units with $G=c=1$, and for numerical calculations we set the BH mass $M=1$.

\section{Quantum-corrected RN BH}\label{sec:2}

Recently, a covariant quantum-corrected RN BH solution has been obtained by solving the equations of motion derived from the effective Hamiltonian constraint~\cite{Yang:2025ufs}. In the present work, we restrict our analysis to the case of vanishing cosmological constant, i.e., $\Lambda = 0$. Unless otherwise specified in the following text, the term ``quantum-corrected BH" refers specifically to the covariant quantum-corrected RN BH introduced above. In Schwarzschild coordinates, its line element is given as follows \cite{Chen2026,Qian2026}:
\begin{equation}
	{\rm d}s^2=-f(r){\rm d}t^2+\frac{1}{f(r)}{\rm d}r^2+ r^2 {\rm d}\theta^2+ r^2 \sin^2{\theta} {\rm d}\phi^2,\label{metric}
\end{equation}
with
\begin{equation}
	f(r)= \left(1-\frac{2 M}{r}+\frac{Q^2}{r^2}\right)\left[1+\frac{\zeta ^2}{r^2}\left(1-\frac{2 M}{r}+\frac{Q^2}{r^2}\right)\right]. \label{metric1}
\end{equation}
The corresponding electromagnetic 4-potential $A_a$ and electromagnetic field tensor $F_{ab}$ have the following form
\begin{eqnarray}
	A_a &=&-\frac{Q}{r}({\rm d}t)_a, \label{Aa} \\
	F_{ab}&=&-\frac{Q}{r^2} ({\rm d}t)_a \wedge ({\rm d}r)_b.
	\label{AF}
\end{eqnarray}
Here, $M$, $Q$, and $\zeta$ stand for the BH mass, charge, and quantum parameter, respectively. Clearly, when $\zeta = 0$, the solution reduces to the classical RN BH solution.

The BH horizon is defined by the roots of $g^{rr} = 0$, which in this quantum-corrected BH spacetime is expressed as
\begin{equation}
	\left(1-\frac{2 M}{r}+\frac{Q^2}{r^2}\right)\left[1+\frac{\zeta ^2}{r^2}\left(1-\frac{2 M}{r}+\frac{Q^2}{r^2}\right)\right]=0. \label{horizon}
\end{equation}
We find that when $Q < M$, Eq.~\eqref{horizon} admits at least two roots, denoting the inner ($r_-$) and outer horizons ($r_+=r_h$) as in the RN BH, namely
\begin{equation}
	r_{\pm}=M \pm \sqrt{M^2-Q^2}.
\end{equation}
Moreover, the term containing the quantum parameter in Eq.~\eqref{horizon} may induce additional horizons in this quantum-corrected BH. This yields the following equation from Eq.~\eqref{horizon}:
\begin{equation}
	F(r) \equiv 1+\frac{\zeta ^2}{r^2}\left(1-\frac{2 M}{r}+\frac{Q^2}{r^2}\right)=0.\label{Fr}
\end{equation}
It is straightforward to show that the root obtained from Eq.~\eqref{Fr} is less than $r_+$.

\section{Dynamics of Neutral particles}\label{sec:3}

In this section, we investigate the dynamics of test particles around the quantum-corrected RN BH. Using the Hamiltonian formalism, we derive the effective potential, from which the specific energy and angular momentum of particles in circular orbits can be obtained by imposing the circular orbit conditions. Furthermore, we analyze the marginally stable circular orbits. The dynamics of neutral particles around BHs in various configurations have been extensively studied in the literature (see \cite{Battista:2023iyu,Wang:2025fmz,FJ1,FA1,FA2,FA4,FA6} and related references therein).

The Hamiltonian function describing dynamics of test particles is given by
\begin{equation}
H = \frac{1}{2} g^{\mu\nu} p_\mu p_\nu + \frac{1}{2}\mu^2 ,\label{aa1}
\end{equation}
where $\mu$ represents the mass of a test particle with its four-momentum expressed as $p^\alpha = \mu u^\alpha$, where $u^\alpha = \dfrac{dx^\alpha}{d\tau}$ denotes the four-velocity, and $\tau$ signifies the particle's proper time. The equations of motion can be articulated by Hamiltonian dynamics as follows:
\begin{equation}
\frac{dx^\alpha}{d\zeta} = \mu u^\alpha = \frac{\partial H}{\partial p_\alpha},
\qquad
\frac{dp_\alpha}{d\zeta} = -\frac{\partial H}{\partial x^\alpha},
\label{aa2}
\end{equation}
in this context, $\zeta = \tau/\mu$ functions as the affine parameter. Due to the symmetries present in BH spacetime, two conserved quantities emerge: specific energy ${\rm E}$ and specific angular momentum ${\rm L}$, defined as follows:
\begin{equation}
\frac{p_t}{\mu}
= -f(r) \frac{dt}{d\tau}= -\mathbb{E},\label{aa3}
\end{equation}
\begin{equation}
\frac{p_\phi}{\mu}= r^2 \sin^2\theta \, \frac{d\phi}{d\tau}= \mathbb{L}.\label{aa4}
\end{equation}
And the conjugate momentum associated with the $\theta$ coordinate is given by
\begin{equation}
    \frac{p_{\theta}}{\mu}=r^2 \frac{d\theta}{d\tau}.\label{aa5}
\end{equation}
Here the values $\mathbb{E} = {\rm E}/m$ and $\mathbb{L} = {\rm L}/m$ denote the particular energy and the specific angular momentum, respectively. The elements of four-velocity $u^\alpha$, specifically the temporal component $u^t$, the azimuthal component $u^\phi$, and the radial component $u^r$, satisfy the subsequent equations of motion:
\begin{equation}
\frac{dt}{d\tau}= \frac{\mathbb{E}}{f(r)},\label{aa6}
\end{equation}
\begin{equation}
\frac{d\theta}{d\tau}=\frac{p_{\theta}}{\mu r^2},\label{aa7}  
\end{equation}
\begin{equation}
\frac{d\phi}{d\tau}= \frac{\mathbb{L}}{r^2 \sin^2\theta},\label{aa8}
\end{equation}
And
\begin{equation}
\left(\frac{dr}{d\tau}\right)^2+\left(- \epsilon + \frac{\mathbb{L}^2}{r^2 \sin^2\theta}+\frac{p^2_{\theta}}{\mu^2 r^2}\right) f(r)= \mathbb{E}^2 ,
\label{aa9}
\end{equation}
in this context, $\epsilon=-1$ pertains to timelike particles, whereas $\epsilon=0$ is relevant for null (light-like) test particles. A dot on a quantity signifies differentiation concerning the proper time $\tau$.

The Hamiltonian $H$ Eq. (\ref{aa1}) using the above reduces as
\begin{equation}
\begin{split}
H=\frac{f(r)}{2} p^2_{r}+\frac{p^2_{\theta}}{2 r^2}+\frac{\mu^2}{2 f(r)}\left(U_{\rm eff}(r, \theta)-\mathbb{E}^2\right),\label{aa10}
\end{split}
\end{equation}
where
\begin{align}
U_{\rm eff}(r, \theta)&=\left(1+ \frac{\mathbb{L}^2}{r^2 \sin^2 \theta}\right) f(r)=\left(1+ \frac{\mathbb{L}^2}{r^2 \sin^2 \theta}\right)\nonumber\\
&\left[\left(1-\frac{2 M}{r}+\frac{Q^2}{r^2}\right)+\frac{\zeta ^2}{r^2}\left(1-\frac{2 M}{r}+\frac{Q^2}{r^2}\right)^2\right].\label{aa11}
\end{align}
From the above expression, it is clear that the presence of quantum correction parameter $\zeta$ modifies the effective potential governing the particle dynamics in comparison to the result for RN BH case. 

\begin{figure*}[ht!]
    \centering
    \includegraphics[width=0.5\linewidth]{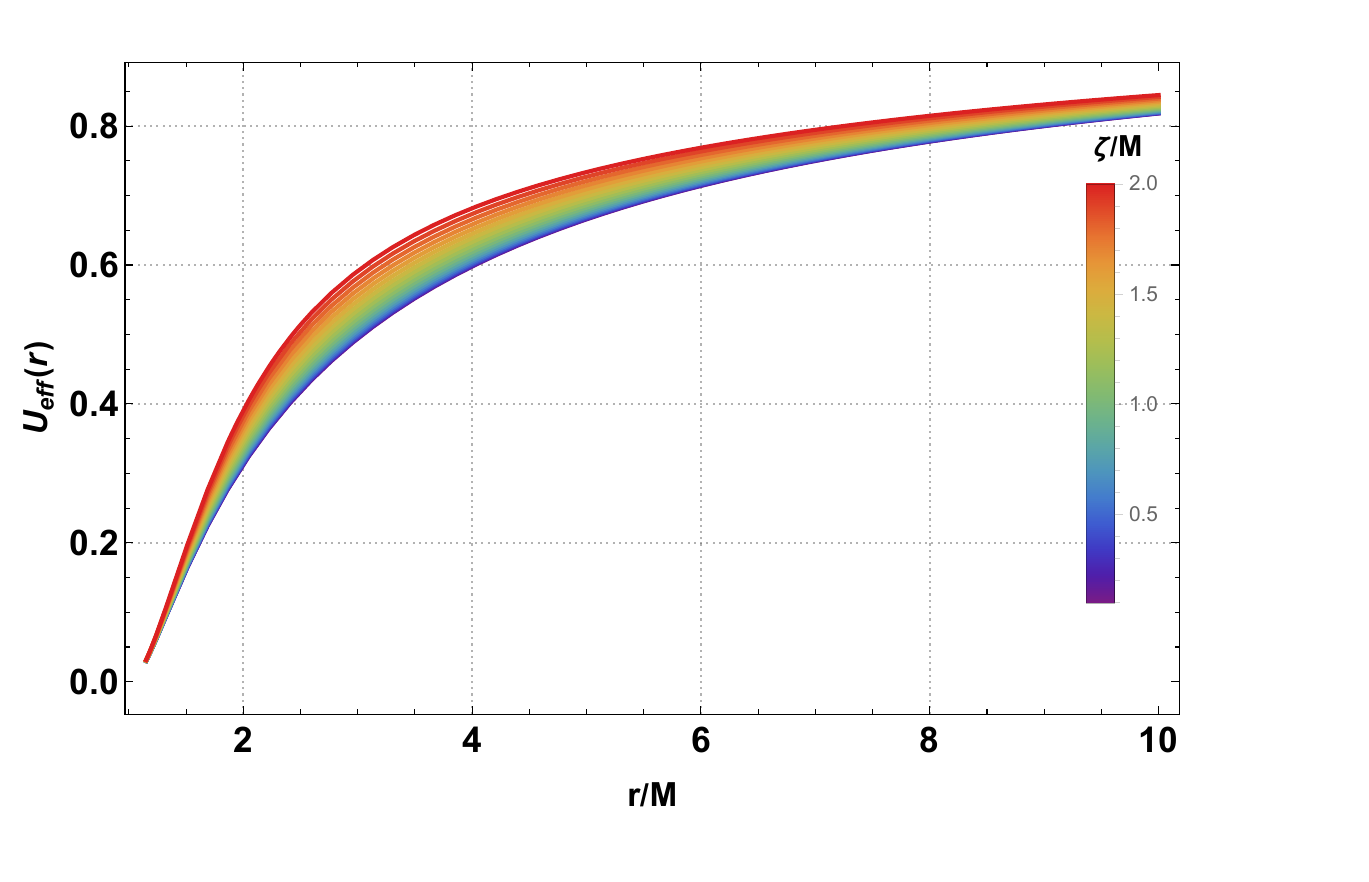}
    \caption{Behavior of the effective potential $U_{\rm eff}$ as a function of dimensionless radial distance $r/M$ for different values of quantum correction parameter $\zeta$. Here $Q/M=1$.}
    \label{fig:potential}
\end{figure*}

\begin{itemize}
    \item When $\zeta=0$, corresponding to the absence of quantum correction, the effective potential simplifies as,
    \begin{align}
U_{\rm eff}(r, \theta)=\left(1+ \frac{\mathbb{L}^2}{r^2 \sin^2 \theta}\right)\left(1-\frac{2 M}{r}+\frac{Q^2}{r^2}\right).\label{aa11a}
\end{align}
\end{itemize}

\begin{figure*}[ht!]
    \centering
    \includegraphics[width=0.48\linewidth]{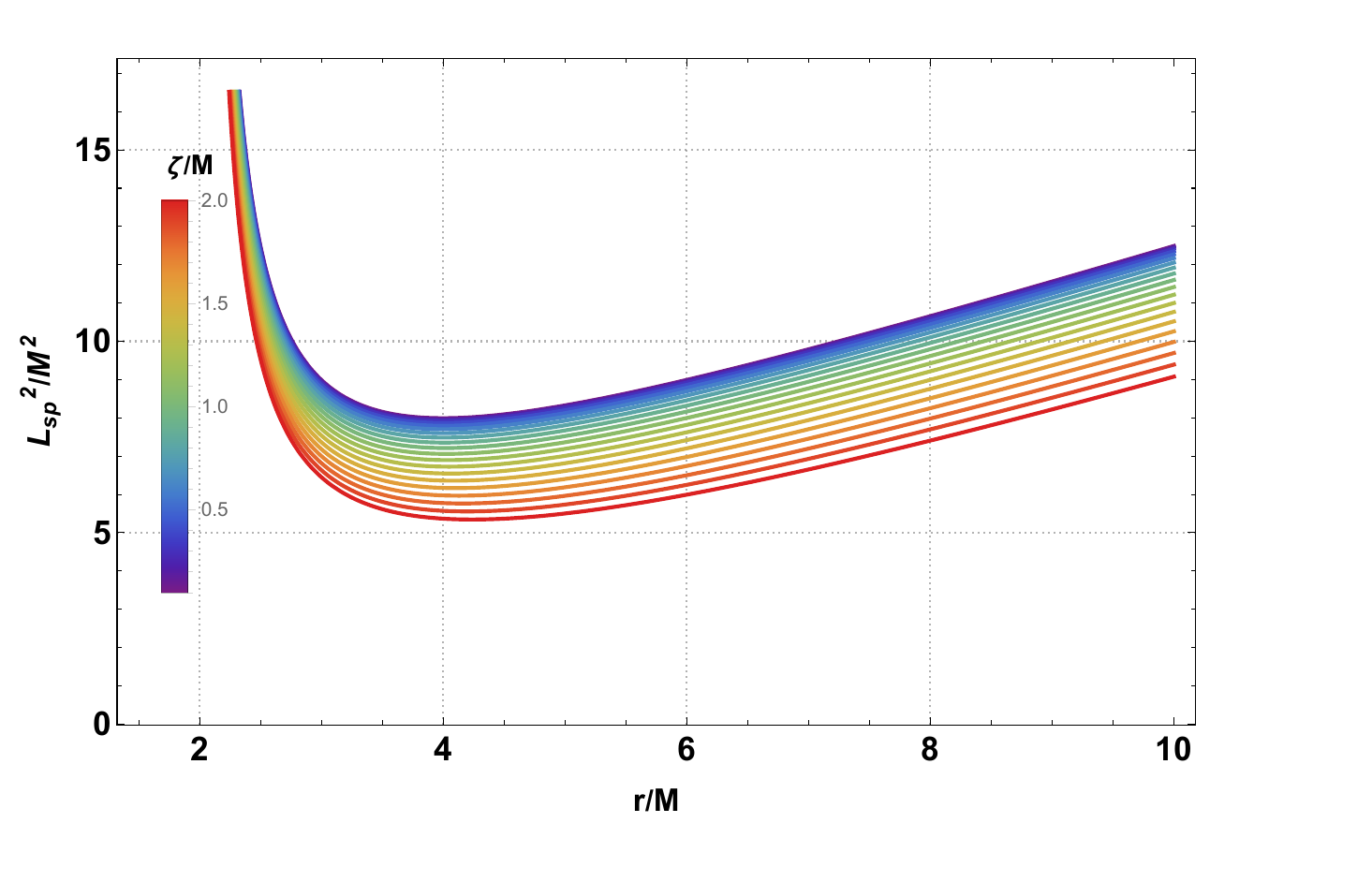}\quad
    \includegraphics[width=0.5\linewidth]{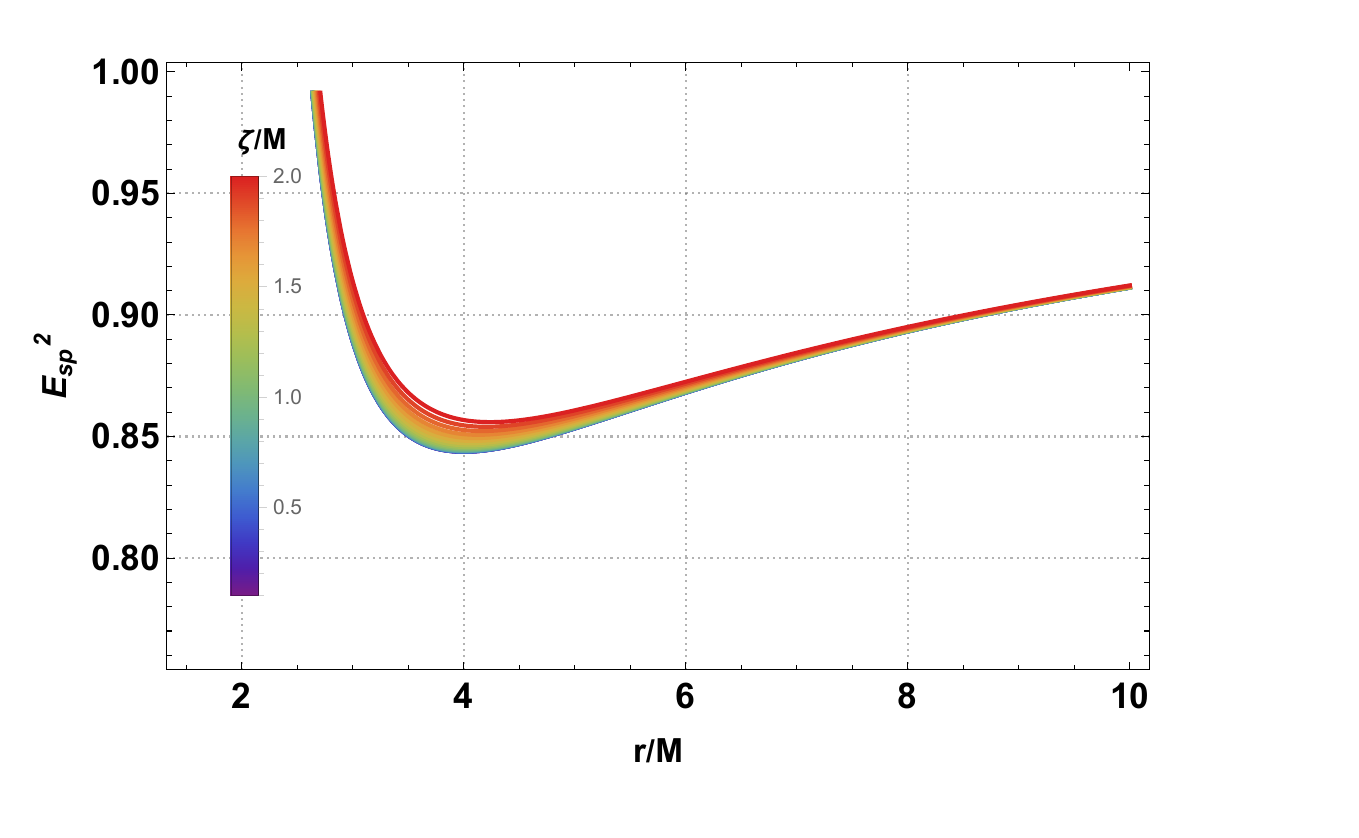}
    \caption{Behavior of the squared specific angular momentum per unit mass and squared specific energy of test particles as functions of the dimensionless radial coordinate \( r/M \) for various values of the quantum correction parameter \(\zeta\). Here $Q/M=1$.}
    \label{fig:specific}
\end{figure*}

For circular motion, the conditions $\frac{dr}{d\tau}=0$ and $\frac{d^2r}{d\tau^2}=0$ must be satisfied. These conditions imply the following two relations:
\begin{equation}
\mathbb{E}^2=U_\text{eff}(r, \theta),\qquad \partial_r U_\text{eff}(r,\theta)=0.\label{aa12}
\end{equation}
Using potential (\ref{aa11}) and after simplification of Eq. (\ref{aa12}) results

\begin{widetext}
\begin{align}
\mathbb{L}^2_{\mathrm{sp}} &= 
\frac{r^3 f'(r)}{2 f(r) - r f'(r)} 
= r^2\,\frac{\frac{M}{r} - \frac{Q^2}{r^2} 
+ \frac{2 \zeta^2}{r^2} \left(1 - \frac{2M}{r} + \frac{Q^2}{r^2}\right) 
\left(\frac{M}{r} - \frac{Q^2}{r^2}\right)
- \frac{\zeta^2}{r^2} \left(1 - \frac{2M}{r} + \frac{Q^2}{r^2}\right)^2}
{1 - \frac{3M}{r} + \frac{2 Q^2}{r^2}
+ \frac{2\zeta^2}{r^2} \left(1-\frac{2M}{r}+\frac{Q^2}{r^2}\right)^2
- \frac{\zeta^2}{r} \left(1-\frac{2M}{r}+\frac{Q^2}{r^2}\right)
\left(\frac{2M}{r^2}-\frac{2Q^2}{r^3}\right)}, 
\label{aa13} \\
\mathbb{E}^2_{\mathrm{sp}} &= 
\frac{2 f^2(r)}{2 f(r) - r f'(r)}
= \frac{\left[\left(1-\frac{2 M}{r}+\frac{Q^2}{r^2}\right)
\left\{1+\frac{\zeta ^2}{r^2}\left(1-\frac{2 M}{r}+\frac{Q^2}{r^2}\right)\right\}\right]^2}
{1 - \frac{3M}{r} + \frac{2 Q^2}{r^2}
+ \frac{2\zeta^2}{r^2} \left(1-\frac{2M}{r}+\frac{Q^2}{r^2}\right)^2
- \frac{\zeta^2}{r} \left(1-\frac{2M}{r}+\frac{Q^2}{r^2}\right)
\left(\frac{2M}{r^2}-\frac{2Q^2}{r^3}\right)}.
\label{aa14}
\end{align}
\end{widetext}

\begin{itemize}
    \item When $\zeta=0$, corresponding to the absence of quantum-correction, the particular angular momentum and particular energy of test particles revolving in circular orbits simplifies as,
    \begin{align}
        \mathbb{L}^2_{\mathrm{sp}}=r^2\,\frac{\left(\frac{M}{r} - \frac{Q^2}{r^2}\right)}{1 - \frac{3M}{r} + \frac{2 Q^2}{r^2}},\qquad 
      \mathbb{E}^2_{\mathrm{sp}}=\frac{\left(1-\frac{2 M}{r}+\frac{Q^2}{r^2}\right)^2}{1 - \frac{3M}{r} + \frac{2 Q^2}{r^2}}
    \end{align}

    \item When $Q=0$, corresponding to the absence of electric charge, the selected space-time simplifies to quantum-corrected Schwrazschild BH \cite{Zhang2025}. In that limit, we find
    \begin{align}
    \mathbb{L}^2_{\mathrm{sp}}&=r^2\,\frac{\frac{M}{r} 
+ \frac{2 \zeta^2}{r} \left(1 - \frac{2M}{r} \right) \left(\frac{M}{r^2}\right)
- \frac{\zeta^2}{r^2} \left(1 - \frac{2M}{r}\right)^2}{1 - \frac{3M}{r}
+ \frac{2\zeta^2}{r^2} \left(1-\frac{2M}{r}\right)^2
- \frac{\zeta^2}{r} \left(1-\frac{2M}{r}\right)
\left(\frac{2M}{r^2}\right)},\nonumber\\
\mathbb{E}^2_{\mathrm{sp}}&=\frac{\left[\left(1-\frac{2 M}{r}\right)\left\{1+\frac{\zeta ^2}{r^2}\left(1-\frac{2 M}{r}\right)\right\}\right]^2}{1 - \frac{3M}{r}
+ \frac{2\zeta^2}{r^2} \left(1-\frac{2M}{r}\right)^2
- \frac{\zeta^2}{r} \left(1-\frac{2M}{r}\right)
\left(\frac{2M}{r^2}\right)}.
    \end{align}
\end{itemize}

From the above analysis, it is clear that both the electric charge of BH and quantum correction modifies these physical quantities associated with test particles motion in circular orbits in comparison to the standard Schwarzschild black hole case.

The minimum and maximum values of the effective potential correspond to stable and unstable circular orbits, respectively. In Newtonian gravity, the ISCO does not have a minimum bound on the radius; the effective potential always exhibits a minimum for any given angular momentum. However, this behavior changes when the effective potential depends not only on the angular momentum of the particle but also on additional factors, such as spacetime curvature or external fields. 

In GR, the effective potential for particles orbiting near a Schwarzschild BH features two extrema-one minimum and one maximum-for a given angular momentum. At $r = 3\,r_s$, where $r_s$ is the Schwarzschild radius, these two extrema coincide at a critical value of angular momentum, marking the location of the ISCO. This point defines the transition from stable to unstable circular orbits. The ISCO can be determined by applying the following conditions to the effective potential $U_\text{eff}(r)$:
\begin{align}
U_\mathrm{eff} &= \mathbb{E}^2, \label{aa15} \\[1em]
\frac{d U_\mathrm{eff}}{d r} &= 0, \quad \text{(circular orbit condition)}, \label{aa16} \\[1em]
\frac{d^2 U_\mathrm{eff}}{d r^2} &= 0, \quad \text{(marginal stability condition)}. \label{aa17}
\end{align}
While Eq.~\eqref{aa16} determines the radii of stable circular orbits $r_C$, Eqs.~\eqref{aa16}--\eqref{aa17} together provide the criterion for determining the radius of the innermost stable circular orbit, namely $r_{\mathrm{ISCO}}$. Taking into account Eq.~\eqref{aa13}, $r_{\mathrm{ISCO}}$ is obtained from the following equation:

\begin{widetext}
\begin{equation}
\Biggl( \frac{9 M Q^2}{r} + M (r - 6 M) - \frac{4 Q^4}{r^2} \Biggr)
\Biggl[ r^2 + 3 \zeta^2 - \frac{6 M \zeta^2}{r} + \frac{3 \zeta^2 Q^2}{r^2} \Biggr]
=
2 \zeta^4 \Biggl( 1 - \frac{2M}{r} + \frac{Q^2}{r^2} \Biggr)^2
\Biggl[ 2 - \frac{13 M}{r} + \frac{24 M^2 + 8 Q^2}{r^2} - \frac{33 M Q^2}{r^3} + \frac{12 Q^4}{r^4} \Biggr].
\label{rISCO_final}
\end{equation}
\end{widetext}

The exact analytical solution of the above polynomial equation in $r$ will give us the ISCO radius. One can see that this polynomial is higher order more than 4, so its analytical solution is quite a challenging task. However, one can find numerical values of ISCO radius by selecting suitable values of the geometric parameters $(Q, \eta)$.

In the limit $Q=0$, corresponding to the absence of electric charge, the selected space-time reduces to the quantum-corrected Schwraschild BH \cite{Zhang2025}. In that limit, the above polynomial simplifies as,
\begin{widetext}
\begin{equation}
M\, r^6 (r - 6 M) \Bigl[r^4 + 3 \zeta^2 r (r - 2 M)\Bigr]
= 2 \zeta^4 r^4 (r - 2 M)^2 \Bigl( 24 M^2 - 13 M r + 2 r^2 \Bigr).
\label{eq:wide_example}
\end{equation}
\end{widetext}

\begin{figure}[ht!]
    \centering
    \includegraphics[width=1\linewidth]{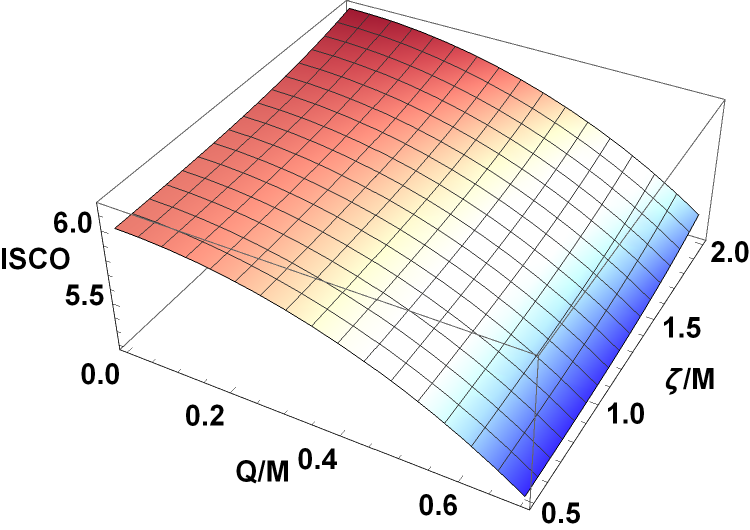}
    \caption{The ISCO radius $r_{\rm ISCO}/M$ given in Eq. (\ref{rISCO_final}) of quantum-corrected RN BH for various values of the quantum parameter $\zeta/M$ and charge $Q/M$.}
    \label{figA015}
\end{figure}

Moreover, in the limit $\zeta=0$ corresponding to the absence of quantum correction, the above polynomial relation reduces to
\begin{equation}
    r^3 - 6 M r^2 + 9 Q^2 r - \frac{4 Q^4}{M} = 0
\end{equation}
whose relation valued solution is given by
\begin{widetext}
\begin{align}
    r_{\rm ISCO} &= 2M + 
\sqrt[3]{-\left(-8 M^3 + 9 M Q^2 - \frac{2 Q^4}{M}\right)+ \sqrt{\left(-8 M^3 + 9 M Q^2 - \frac{2 Q^4}{M}\right)^2 + \left(3 Q^2-4 M^2\right)^3}} \nonumber\\
&\quad+ \sqrt[3]{-\left(-8 M^3 + 9 M Q^2 - \frac{2 Q^4}{M}\right)- \sqrt{\left(-8 M^3 + 9 M Q^2 - \frac{2 Q^4}{M}\right)^2 + \left(3 Q^2-4 M^2\right)^3}}
\end{align}
\end{widetext}
which is similar to that for RN BH without any correction.\\ The dependence of the ISCO radius on the electric charge $Q$ and the quantum-corrected parameter $\zeta$ is illustrated in Fig.~\ref{figA015}. The figure indicates that increasing $Q$ results in a smaller ISCO radius. In contrast, larger values of $\zeta$ lead to an expansion of the ISCO radius. 

\begin{figure}[ht!]
    \centering
    \includegraphics[width=1\linewidth]{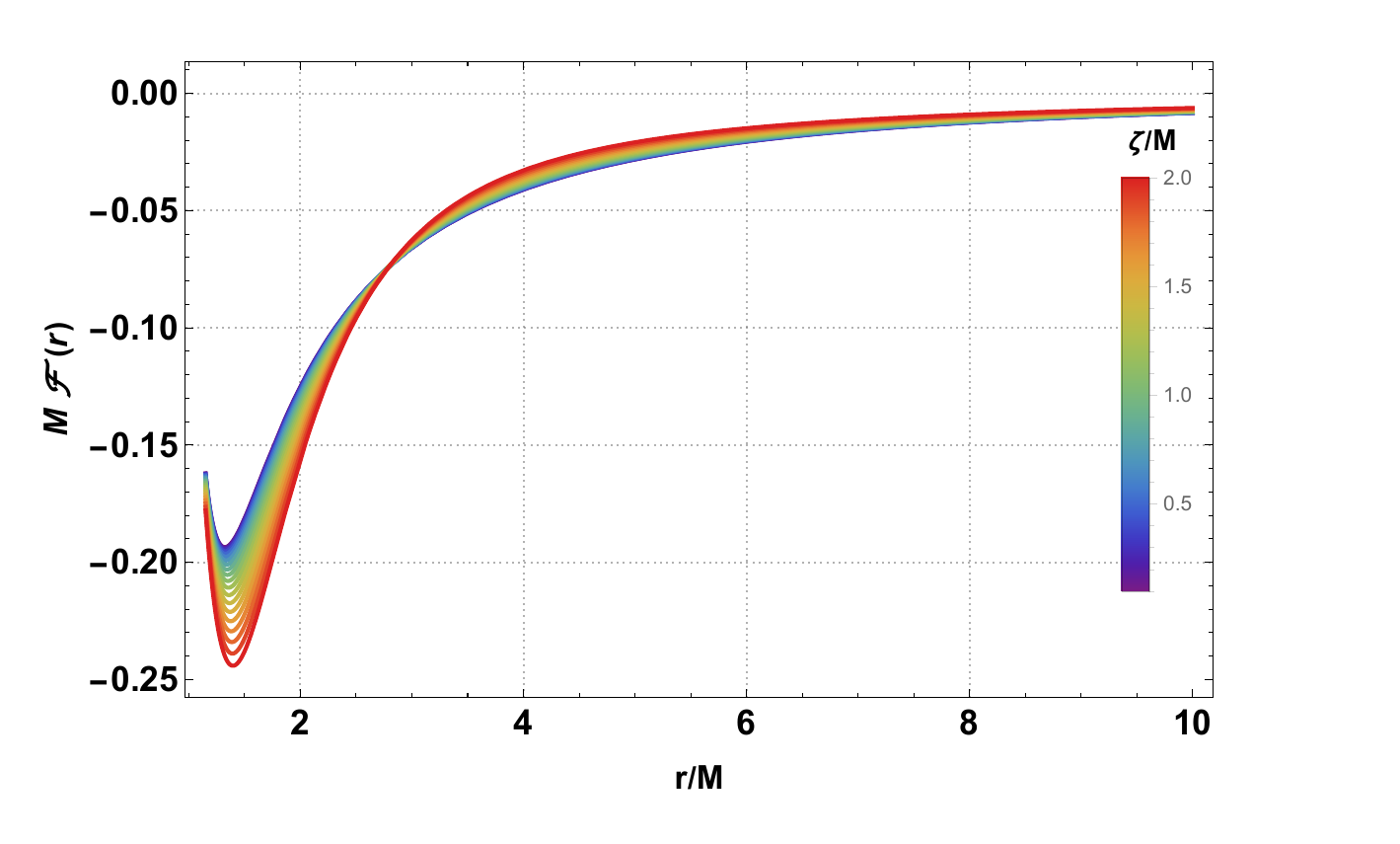}
    \caption{Behavior of the mass times the effective force $M\,\mathcal{F}$ as a function of dimensionless radial distance $r/M$ by varying quantum correction parameter.}
    \label{fig:placeholder}
\end{figure}

The effective force experienced by neutral test particles can be determined using the effective potential $U_\text{eff}$ which governs the particle dynamics in the gravitational field. This force indicates whether the test particles is attracted towards or move away from BH. This force is the negative gradient of the effective potential and mathematically it is defined by  
\begin{equation}
    \mathcal{F}=-\frac{1}{2}\,\frac{dU_\text{eff}}{dr}.\label{aa19}
\end{equation}
Using potential (\ref{aa11}), we find the effective radial force on the equatorial plane ($\theta=\pi/2$)

\begin{align}
\mathcal{F}&=-\frac{f'(r)}{2}+\frac{\mathbb{L}^2}{2 r^3} \left(2 f(r)-r f'(r)\right)\nonumber\\
&=\left(-\frac{M}{r^2}+\frac{Q^2}{r^3}\right)
\left[1+\frac{2\zeta^2}{r^2}
\left(1-\frac{2M}{r}+\frac{Q^2}{r^2}\right)\right]\nonumber\\
&+\frac{\zeta^2}{r^3}
\left(1-\frac{2M}{r}+\frac{Q^2}{r^2}\right)^2+\frac{\mathbb{L}^2}{r^3}\Bigg[1-\frac{3M}{r}+\frac{2Q^2}{r^2}\nonumber\\
&+\frac{2\zeta^2}{r^2}
\left(1-\frac{2M}{r}+\frac{Q^2}{r^2}\right)^2\nonumber\\
&-\frac{\zeta^2}{r}
\left(1-\frac{2M}{r}+\frac{Q^2}{r^2}\right)
\left(\frac{2M}{r^2}-\frac{2Q^2}{r^3}\right)\Bigg].\label{aa20}   
\end{align}

\begin{itemize}
    \item When $\zeta=0$ corresponding to the absence of quantum-correction, the effective force simplifies as,
    \begin{align}
\mathcal{F}=\left(-\frac{M}{r^2}+\frac{Q^2}{r^3}\right)+\frac{\mathbb{L}^2}{r^3}\left[1-\frac{3M}{r}+\frac{2Q^2}{r^2}\right].\label{aa21} 
\end{align}

When $Q=0$, corresponding to the absence of electric charge of the BH, the selected space-time simplifies to the quantum-corrected Schwarzschild BH \cite{Zhang2025}. In that case, the effective force simplifies as,
\begin{align}
\mathcal{F}&=\left(-\frac{M}{r^2}\right)\left(1+\frac{2\zeta^2}{r^2}\left(1-\frac{2M}{r}\right)\right)+\frac{\zeta^2}{r^3}
\left(1-\frac{2M}{r}\right)^2\nonumber\\
&+\frac{\mathbb{L}^2}{r^3}\left[1-\frac{3M}{r}+\frac{2\zeta^2}{r^2}\left(1-\frac{2M}{r}\right)^2-\frac{\zeta^2}{r}\left(1-\frac{2M}{r}\right)\left(\frac{2M}{r^2}\right)\right].\label{aa22}   
\end{align}
\end{itemize}

Form the above analysis, it is clear that the presence of quantum-correction in the space-time geometry modifies the effective force experienced by test particles in comparison to that for the standard BH solution.

\section{Fundamental Frequencies}\label{sec:4}

To study the quasi-periodic oscillations (QPOs) properties of the quantum-corrected RN BH, we calculate the fundamental frequencies of massive neutral test particles orbiting the BH. In this work, we adopt one of the simplest and most widely used approaches, where the accretion dynamics is modeled by test particles moving along timelike geodesics in the given spacetime geometry. Despite its simplicity, this model captures the essential physical mechanisms responsible for QPOs observables around compact astrophysical objects.

We consider small perturbations around stable circular equatorial orbits. When a particle is slightly displaced from its equilibrium position, its motion can be described by radial and vertical epicyclic oscillations, which to leading order behave as linear harmonic oscillators. This perturbative framework allows us to define the Keplerian, radial, and vertical epicyclic frequencies measured by a distant observer, and to analyze the stability properties of circular orbits in the quantum-corrected RN spacetime.  It is worth noting that QPOs for quantum-corrected uncharged BHs have recently been studied in Ref.~\cite{Zhang2025}. Subsequent analyses using Event Horizon Telescope (EHT) and X-ray binary observations, as well as astrophysical constraints from the orbit of the S2 star, were performed in Refs.~\cite{PDU2026,JCAP2026}. In the present work, we extend these investigations to the charged sector by considering a quantum-corrected RN black hole and analyzing its associated QPO properties.

\subsection{Keplerian frequency}\label{subsec:Keplerian}

It is well known that the angular velocity measured by a distant observer is given by $\Omega_K = d\phi/dt$. For a spacetime of the form \eqref{metric}, this quantity can be written as
\begin{equation}
\Omega_K^2 = \frac{f'(r)}{2r}.
\label{eq:Omega_K}
\end{equation}
For fixed values of $(r,\theta)$, the angular frequency can be converted into a physical frequency measured in Hertz (Hz) according to \cite{shahzadi_epicyclic_2021,jumaniyozov_circular_2024,rahmatov_qpos_2024,mustafa_particle_2025,jumaniyozov_radiative_2025,shermatov_circular_2026}
\begin{equation}
\nu_{(K,r,\theta)} = \frac{1}{2\pi}\frac{c^3}{G M}\,\Omega_{(K,r,\theta)} \; \mathrm{[Hz]},
\label{eq:nu-K}
\end{equation}
which allows for direct comparison with astrophysical observations. In the above expression, we use $c = 3\times 10^8\,\mathrm{m\,s^{-1}}$ and $G = 6.67\times 10^{-11}\,\mathrm{m^3\,kg^{-1}\,s^{-2}}$.

\subsection{{Harmonic oscillations}}\label{subsec:harmonic}

If the circular orbits of particles around the BH are perturbed as $r \to r_0 + \delta r$ and $\theta \to \theta_0 + \delta \theta$, respectively, along the radial and vertical directions, the effective potential can be expanded around the equilibrium position as
\begin{eqnarray}
U_{\mathrm{eff}}(r,\theta) &\approx& U_{\mathrm{eff}}(r_0,\theta_0)
+ \delta r\, \partial_r U_{\mathrm{eff}}(r,\theta)\Big{|}_{r_0,\theta_0}
+ \delta \theta\, \partial_\theta U_{\mathrm{eff}}(r,\theta)\Big{|}_{r_0,\theta_0} \nonumber\\
&& + \frac{1}{2}\delta r^2\, \partial_r^2 U_{\mathrm{eff}}(r,\theta)\Big{|}_{r_0,\theta_0}
+ \frac{1}{2}\delta \theta^2\, \partial_\theta^2 U_{\mathrm{eff}}(r,\theta)\Big{|}_{r_0,\theta_0} \nonumber\\
&& + \delta r\,\delta \theta\, \partial_r \partial_\theta U_{\mathrm{eff}}(r,\theta)\Big{|}_{r_0,\theta_0}
+ \mathcal{O}(\delta r^3, \delta \theta^3).
\label{eq:Ueff_expand}
\end{eqnarray}
For circular orbits, the first-order derivatives vanish due to the circular-orbit condition given by Eq.~\eqref{aa16}. Moreover, the mixed and higher-order terms do not contribute under the assumption of small perturbations and orbital stability. As a result, only the second-derivative terms in the expansion remain relevant.

To obtain physical quantities measured by a distant observer, we transform from the affine parameter $\lambda$ to the coordinate time $t$ by using $dt/d\lambda = u^t = \mathbb{E}/f(r)$.

Substituting Eq.~\eqref{aa11} into Eq.~\eqref{eq:Ueff_expand}, one obtains the equations governing harmonic oscillations,
\begin{equation}
\frac{d^2(\delta_x)}{dt^2} + \Omega_x^2\, \delta_x = 0, \qquad x \in (r,\theta),
\label{eq:harmonic_0}
\end{equation}
where $\delta_x$ denote the radial and vertical displacements, and $\Omega_x$ are the corresponding epicyclic frequencies measured by a distant observer.

By considering test particles initially located on the equatorial plane, the radial and vertical epicyclic frequencies take the form
\begin{eqnarray}
\Omega_r^2 &=& -\frac{1}{2 g_{rr} (u^t)^2}\,
\partial_r^2 U_{\mathrm{eff}}(r,\theta) \bigg{|}_{r=r_c,\theta = \pi/2} \nonumber\\
\nonumber\\
&=&\Omega_K^2\left(\frac{r f f''}{f'}+3f-2r f'\right)\bigg{|}_{r=r_C},
\label{aa28} \\
\Omega_\theta^2 &=& -\frac{1}{2 g_{\theta\theta} (u^t)^2}\Big{|}\,
\partial_\theta^2 U_{\mathrm{eff}}(r,\theta) \bigg{|}_{r=r_c,\theta = \pi/2} \nonumber\\
\nonumber\\
&=&\Omega_K^2,
\label{aa29}
\end{eqnarray}
to obtain which, we have considered the relation
\begin{equation}
u^t=\sqrt{\frac{2}{2f-rf'}}.\label{aa25}    
\end{equation}
Also, as it can be observed, $\Omega_\theta = \Omega_\phi = \Omega_K$. Considering the lapse function of the quantum-corrected RN spacetime in Eq. \eqref{metric1}, one obtains 
\begin{equation}
    \Omega_K^2 = \frac{Mr-Q^2}{r^4}-\frac{X^2\zeta^2}{r^8},
    \label{eq:Omega_K2}
\end{equation}
where $X^2=[3Q^2+r(r-4M)][Q^2+r(r-2M)]$. Furthermore, we get
\begin{eqnarray}
    \Omega_r^2 &=& \Omega_K^2\Biggl\{
    \frac{1}{r^6}\Bigl[9MQ^2r+Mr^2(r-6M)-4Q^4\Bigr] \nonumber\\
    && + \frac{3\xi^2}{r^{10}}\Bigl[9 M Q^2 r+M r^2 (r-6 M)-4 Q^4\Bigr] \Bigl[r (r-2 M)+Q^2\Bigr]\nonumber\\
    &&-\frac{\zeta^4}{r^{14}}\Bigl[r (r-2 M)+Q^2\Bigr]^2\Bigl[
     Q^2 r (8 r-33 M)+12 Q^4\nonumber\\
     && + r^2 (24 M^2-13 M r+2 r^2)
    \Bigr]
    \Biggr\}.
    \label{eq:Omega_r}
\end{eqnarray}
Again, in order to obtain the observationally relevant counterparts of the above frequencies, we make use of the relation \eqref{eq:nu-K}.

\begin{figure*}[t]
    \centering
    \includegraphics[width=5.4cm]{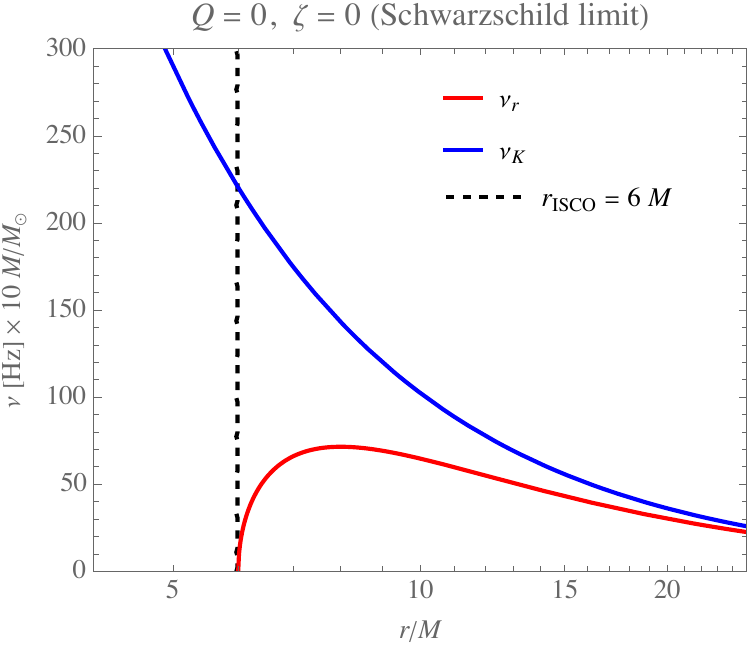}\quad
    \includegraphics[width=5.4cm]{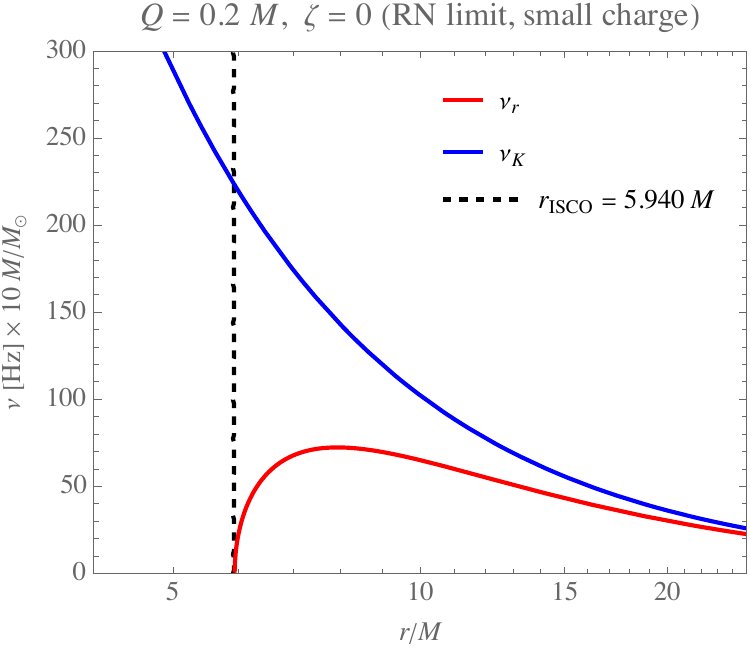}\quad
    \includegraphics[width=5.4cm]{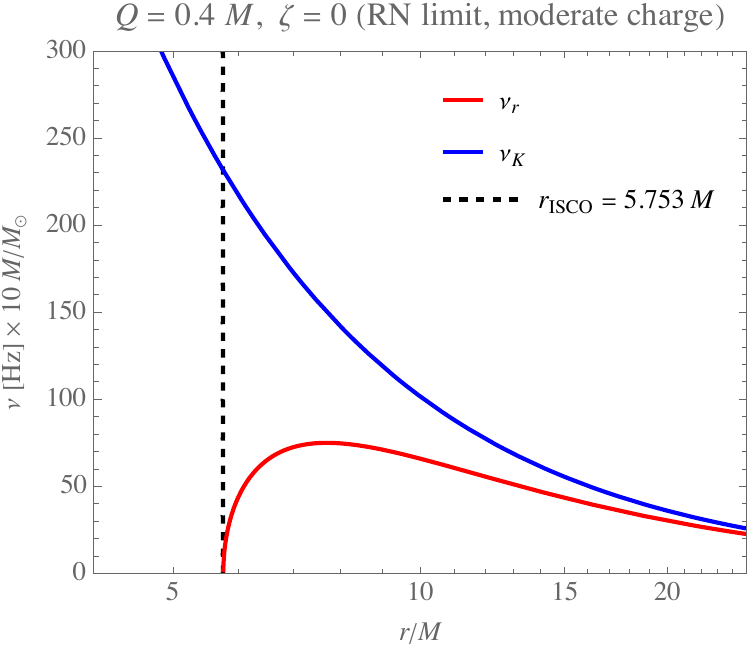}\\
    (a) \hspace{6cm} (b) \hspace{6cm} (c)\\
    \includegraphics[width=5.4cm]{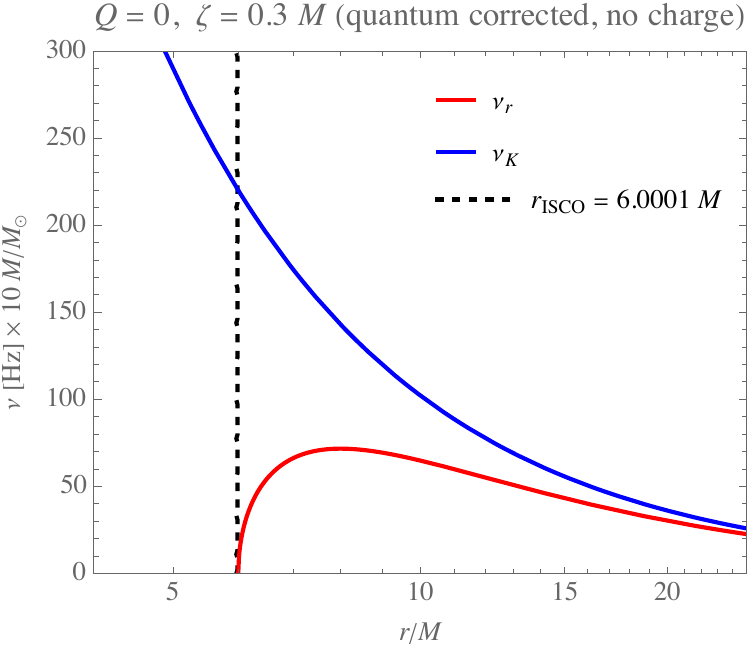}\quad
    \includegraphics[width=5.4cm]{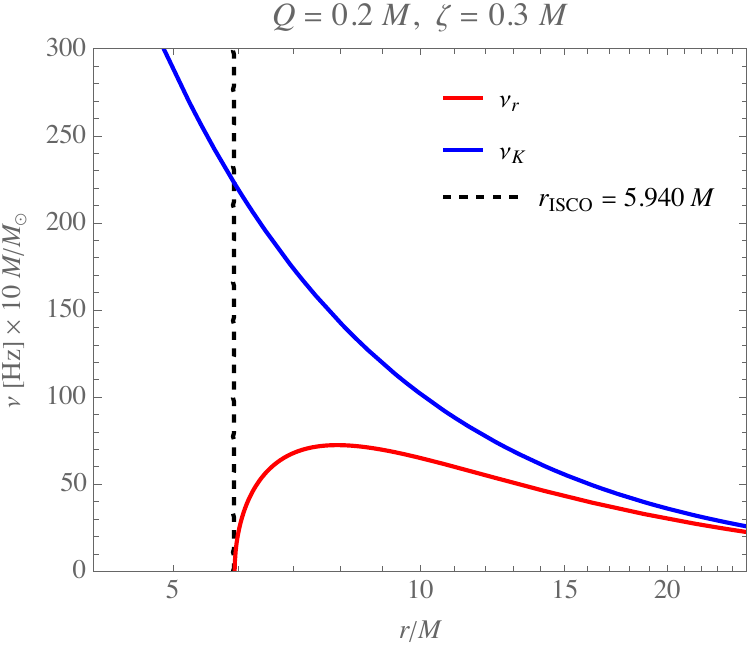}\quad
    \includegraphics[width=5.4cm]{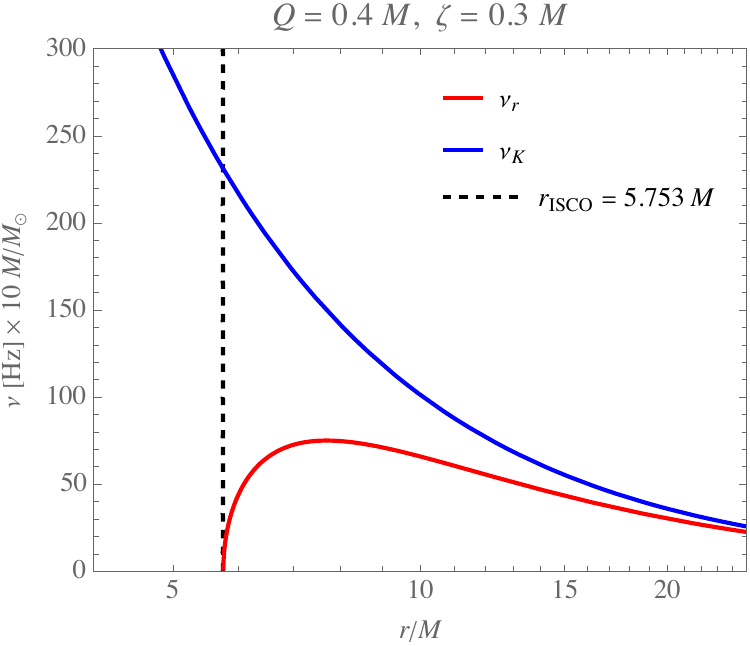}\\
     (d) \hspace{6cm} (e) \hspace{6cm} (f)\\
    \includegraphics[width=5.4cm]{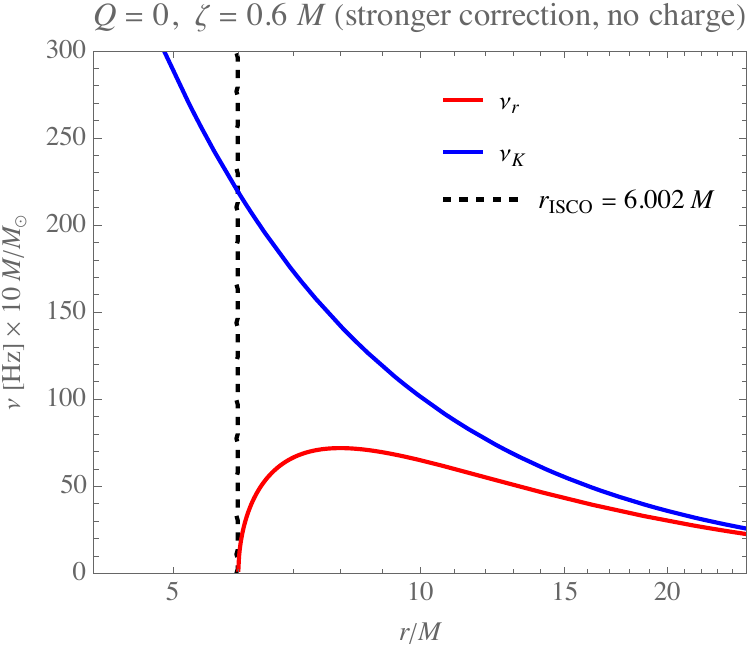}\quad
    \includegraphics[width=5.4cm]{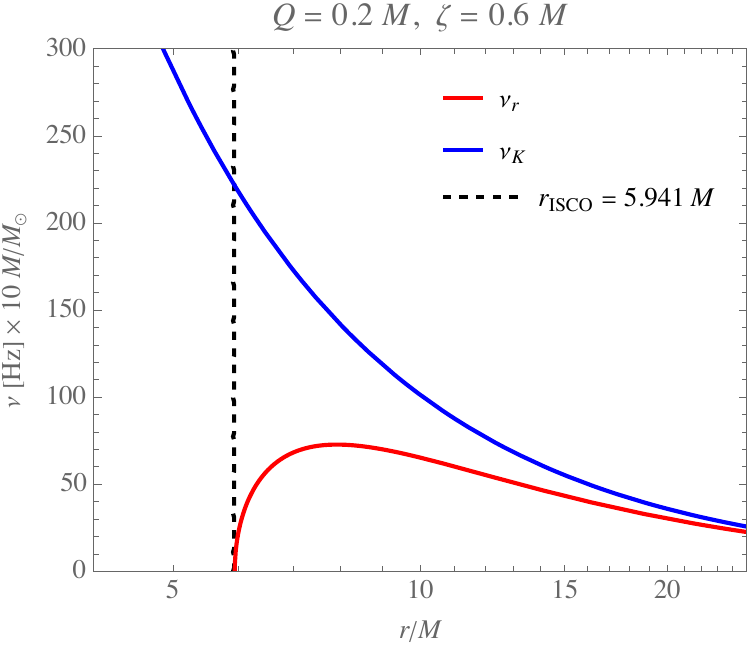}\quad
    \includegraphics[width=5.4cm]{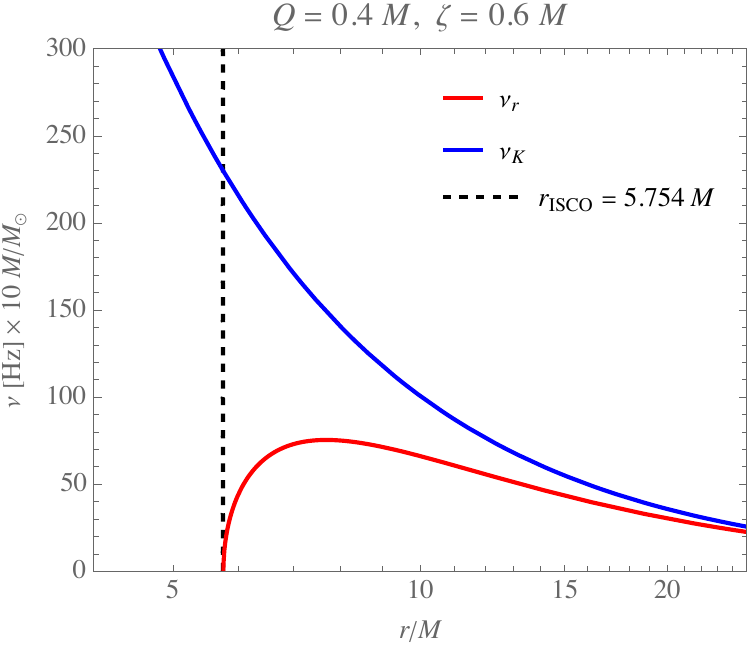}\\
     (g) \hspace{6cm} (h) \hspace{6cm} (i)\\
    \caption{Representative examples of the radial profiles of the observational frequencies $\nu_i$ for different values of the BH parameters.}
    \label{fig:nu_i}
\end{figure*}

In Fig.~\ref{fig:nu_i}, we plot the radial profiles of $\nu_r$ and $\nu_K$ for different values of the main BH parameters, namely $Q$ and $\zeta$.
As expected, at the ISCO we have $\nu_r = 0$, and therefore $r_{\mathrm{ISCO}}$ appears as the lower radial bound of the $\nu_r$ profile. On the other hand, the Schwarzschild and RN limits provide upper bounds for $r_{\mathrm{ISCO}}$ in the presence of both nonzero charge and quantum corrections. The latter typically induce only small increases in $r_{\mathrm{ISCO}}$, as can be seen, for example, in the case of BHs with zero charge and nonzero quantum corrections. Furthermore, one observes that $\nu_\theta = \nu_\phi = \nu_K$ remain nonzero at $r_{\mathrm{ISCO}}$ and increase rapidly as the radius approaches the event horizon. At larger radii, however, these frequencies approach the $\nu_r$ profile, and all of them eventually coincide in the asymptotic region.

\subsection{{QPO models}}\label{subsec:QPO_models}

The physical origin of electromagnetic emission from accretion disks around BHs is commonly associated with oscillatory motion of particles in the gravitational field. In particular, charged particles radiate at frequencies that coincide with their characteristic oscillation frequencies around the central object, which makes the dynamics of test particles in BH spacetimes a natural framework for interpreting QPOs in terms of (quasi-)harmonic oscillations in the radial and angular directions. A large number of QPOs have been observed in such systems, with their frequencies measured to high accuracy; however, a unique and universally accepted mechanism responsible for their formation has not yet been established. This problem remains under active discussion, especially in the context of testing gravity theories and probing the location of the inner edge of the accretion disk, which is closely related to the radius of the ISCO \cite{rayimbaev_quasiperiodic_2022,JR2022}. In this sense, QPO studies provide a powerful tool for constraining the characteristic radii of the innermost disk regions, estimating the mass of the central BH, and exploring the properties of gravity in the strong-field regime.

In this subsection, we investigate the relation between the upper and lower frequencies of twin-peaked QPOs around quantum-corrected RN BHs, and we compare our results with the corresponding frequency relations for neutral particles orbiting BHs \cite{stuchlik_models_2016}. To this end, we adopt several phenomenological models that can be confronted with astrophysical observations, following the frameworks discussed in Ref.~\cite{shahzadi_epicyclic_2021}, which provide practical prescriptions for interpreting QPO data in terms of fundamental frequencies. In particular, the proposed models include the relativistic precession (RP) model and its invariant version RP$\overline{1,2}$, the epicyclic resonance (ER) models, including the invariant ER$\overline{1,5}$, as well as the tidal disruption (TD) and warped disk (WD) models. In this subsection, we derive the expressions for the upper and lower QPO frequencies within the RP, ER and WD models. In the following, we briefly outline how these models are implemented in terms of the fundamental frequencies introduced in the previous subsection and present a graphical analysis of the resulting upper--lower frequency relations:
\begin{enumerate}

\item[(i)] In the RP model, the twin-peak QPO frequencies are associated with the orbital and radial epicyclic motions. Specifically, the upper and lower frequencies are identified as $\nu_U = \nu_K$ and $\nu_L = \nu_K - \nu_r$, respectively \cite{stella_correlations_1999}.

\item[(ii)] In the ER models, the accretion flow is assumed to be sufficiently thick for resonant oscillations to develop along geodesic trajectories of radiating particles. The observed QPOs are then interpreted as arising from different combinations of oscillation modes. Here, we consider three representative cases, denoted as ER2, ER3, and ER4, which differ in their mode couplings. In these models, the frequency pairs are given, respectively, by $(\nu_U, \nu_L) = (2\nu_K - \nu_r,\, \nu_r)$, $(\nu_U, \nu_L) = (2\nu_K + \nu_r,\, \nu_K)$, and $(\nu_U, \nu_L) = (2\nu_K + \nu_r,\, \nu_K - \nu_r)$ \cite{abramowicz_precise_2001}.

\item[(iii)] In the WD model, QPOs are assumed to originate from oscillatory motion of test particles in a thin accretion disk. Within this framework, the upper and lower frequencies are expressed as $\nu_U = 2\nu_K - \nu_r$ and $\nu_L = 2(\nu_K - \nu_r)$, respectively \cite{kato_resonant_2004,kato_frequency_2008}.

\end{enumerate}
In Fig.~\ref{fig:nuLnuU}, we illustrate the correlated behavior of the upper and lower frequencies of twin-peaked QPOs in the vicinity of quantum-corrected RN BHs, together with their Schwarzschild and RN limits. The frequencies are shown for radii ranging from the ISCO out to infinity, and for several representative values of the BH parameters $Q$ and $\zeta$. The diagrams display a set of inclined lines corresponding to specific frequency ratios, namely $\nu_U\!:\!\nu_L = 3\!:\!2$, $4\!:\!3$, $5\!:\!4$, and $1\!:\!1$. These lines indicate the allowed combinations of upper and lower frequencies compatible with each ratio \cite{kolos_possible_2017,kolos_quasi-periodic_2020,shahzadi_epicyclic_2021,shahzadi2023}. In particular, the $1\!:\!1$ line—often referred to as the ``graveyard'' of twin-peaked QPOs—marks the regime where the two frequencies coincide, so that the system effectively exhibits a single QPO peak.
\begin{figure*}[t]
%
\begin{subfigure}{0.3\textwidth}
\centering
  \includegraphics[width=5.3cm]{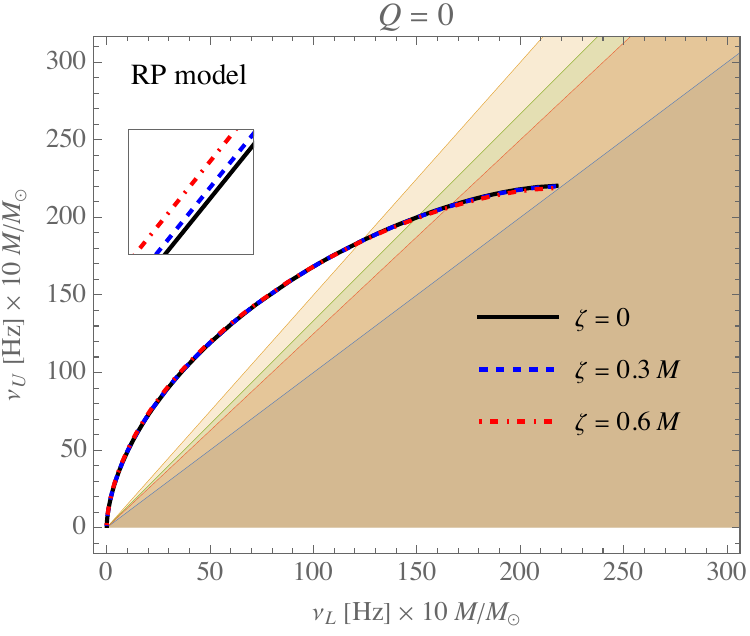}
\end{subfigure}
 \hfill
\begin{subfigure}{0.3\textwidth}
\centering
  \includegraphics[width=5.3cm]{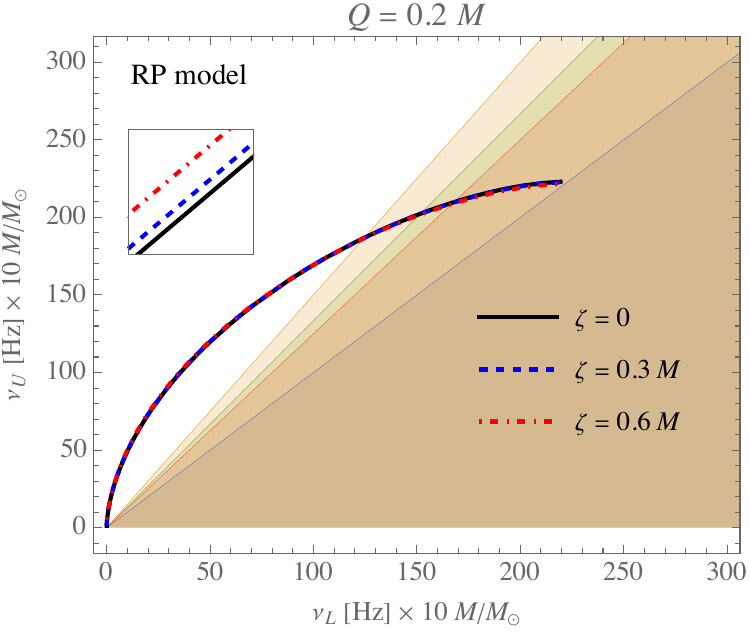}
\end{subfigure}
\hfill
\begin{subfigure}{0.3\textwidth}
\centering
  \includegraphics[width=5.3cm]{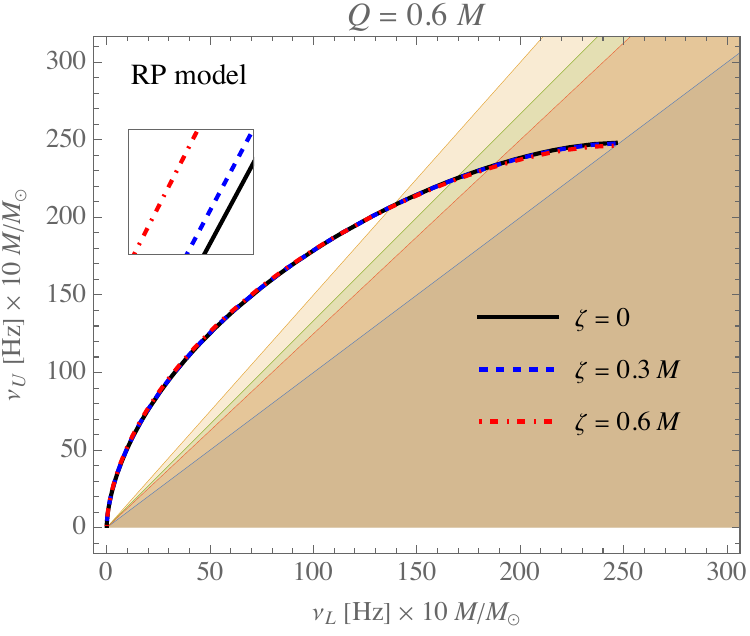}
\end{subfigure}
\hfill
\begin{subfigure}{0.3\textwidth}
\centering
  \includegraphics[width=5.3cm]{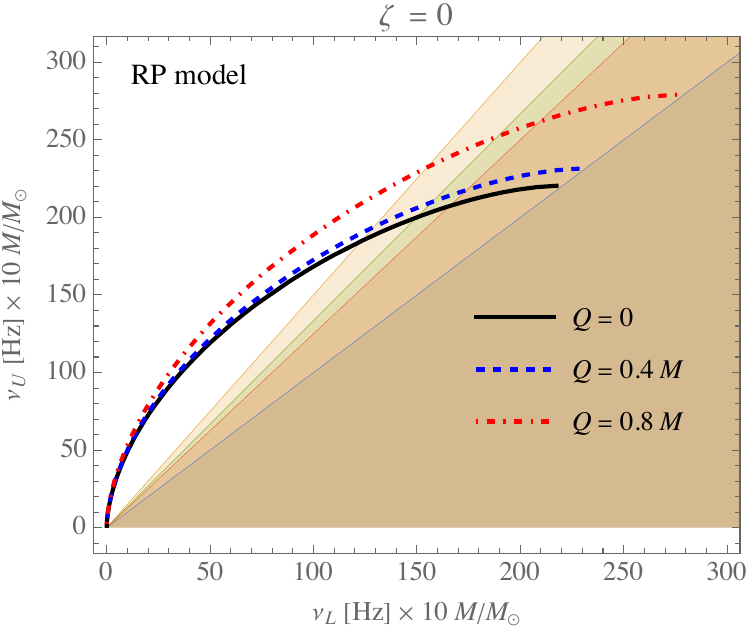}
\end{subfigure}
\hfill
\begin{subfigure}{0.3\textwidth}
\centering
  \includegraphics[width=5.3cm]{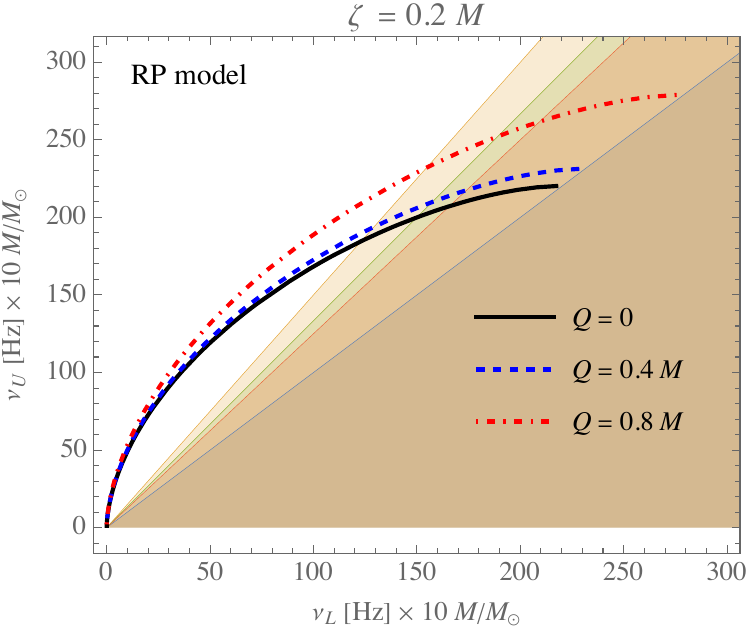}
\end{subfigure}
\hfill
\begin{subfigure}{0.3\textwidth}
\centering
  \includegraphics[width=5.3cm]{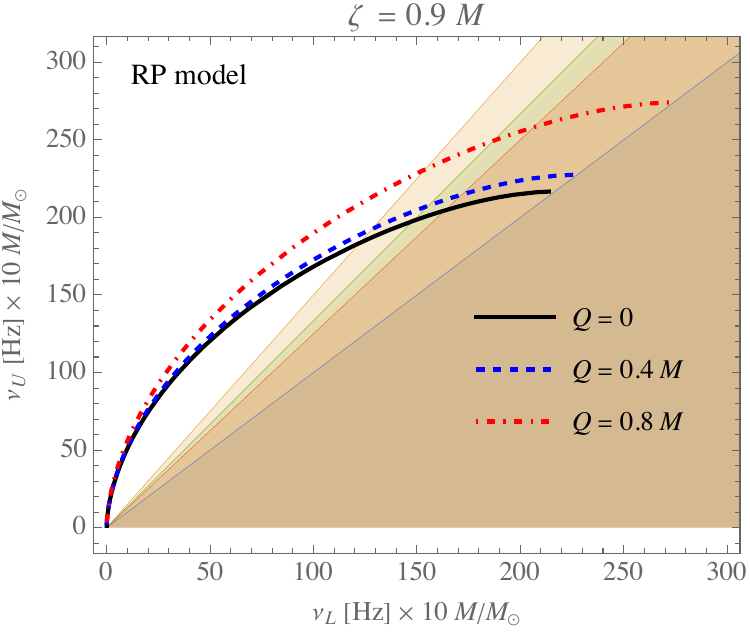}
\end{subfigure}
\hfill
\begin{subfigure}{0.3\textwidth}
\centering
 \includegraphics[width=5.3cm]{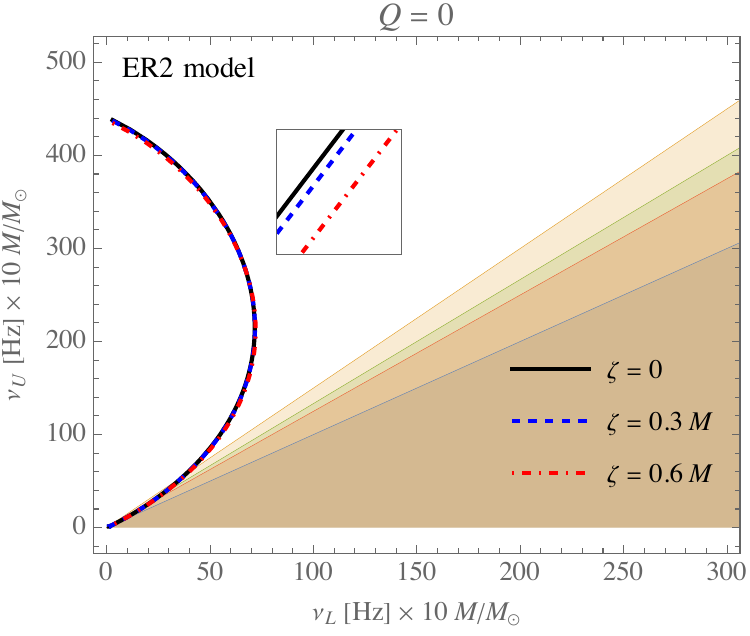}
\end{subfigure}
\hfill
\begin{subfigure}{0.3\textwidth}
\centering
 \includegraphics[width=5.3cm]{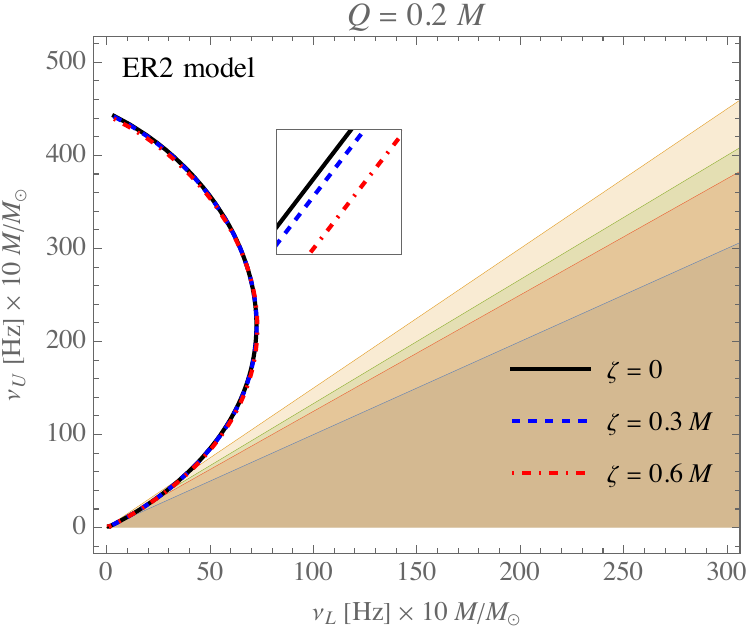}
\end{subfigure}
 \hfill
\begin{subfigure}{0.3\textwidth}
\centering
 \includegraphics[width=5.3cm]{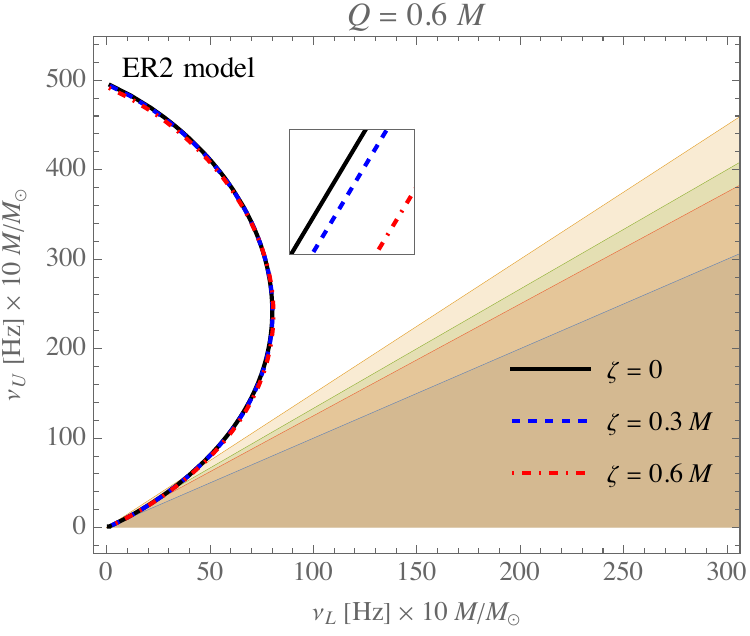}
\end{subfigure}
\hfill
\begin{subfigure}{0.3\textwidth}
\centering
\includegraphics[width=5.3cm]{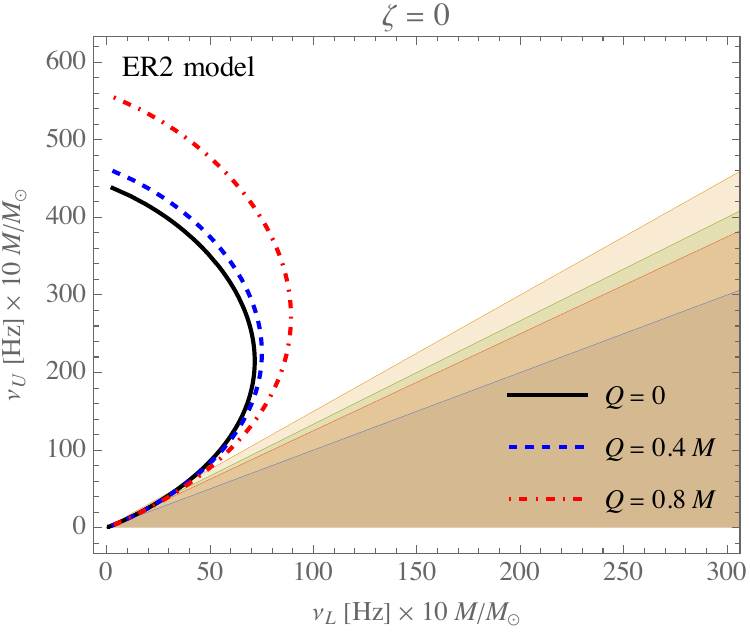}
\end{subfigure}
\hfill
\begin{subfigure}{0.3\textwidth}
\centering
\includegraphics[width=5.3cm]{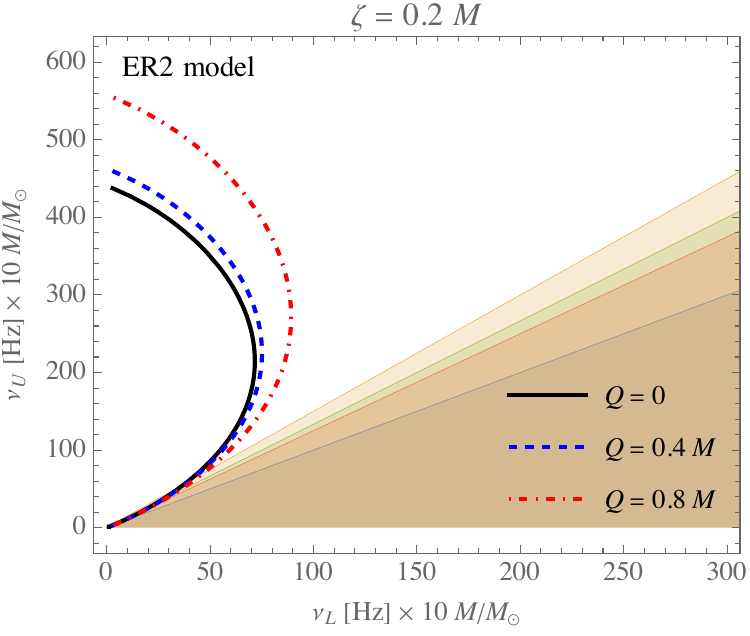}
\end{subfigure}
\hfill
\begin{subfigure}{0.3\textwidth}
\centering
\includegraphics[width=5.3cm]{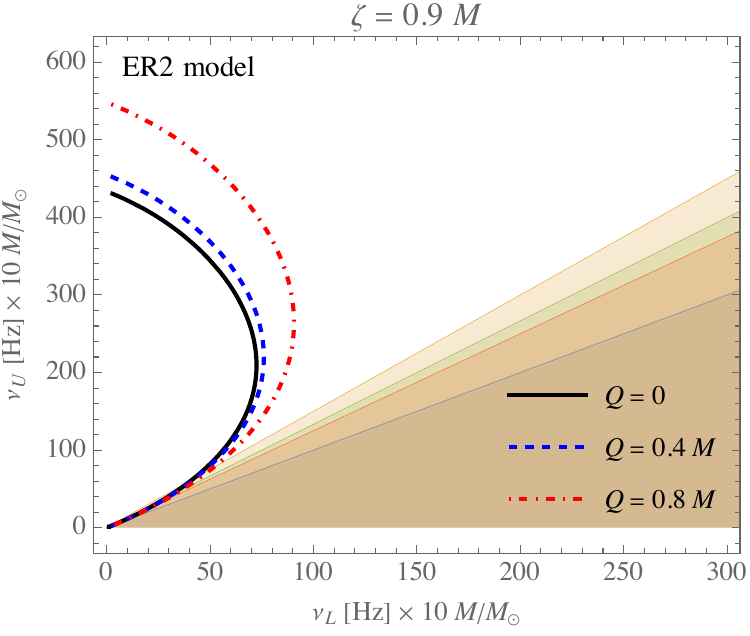}
\end{subfigure}
    \caption{Relations between the upper and lower frequencies of twin-peaked QPOs in the RP, ER2–ER4, and WD models around quantum-corrected RN BHs, shown for several values of $Q$ and $\zeta$. The straight lines indicate, from top to bottom, the frequency ratios $\nu_U\!:\!\nu_L = 3\!:\!2$, $4\!:\!3$, $5\!:\!4$, and $1\!:\!1$.
}
    \label{fig:nuLnuU}
\end{figure*}
\begin{figure*}[t]
  \ContinuedFloat
  \centering
  \begin{subfigure}{0.3\textwidth}
\centering
\includegraphics[width=5.3cm]{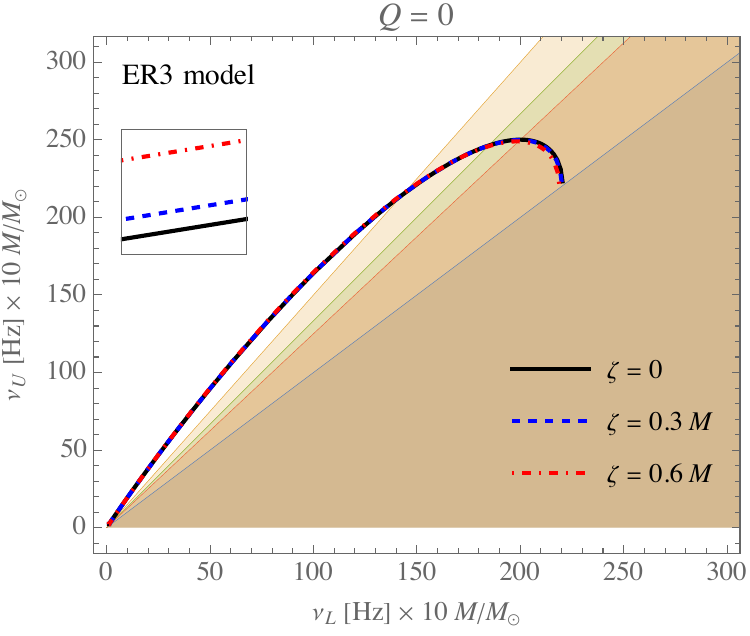}
\end{subfigure}
\hfill
  \begin{subfigure}{0.3\textwidth}
\centering
\includegraphics[width=5.3cm]{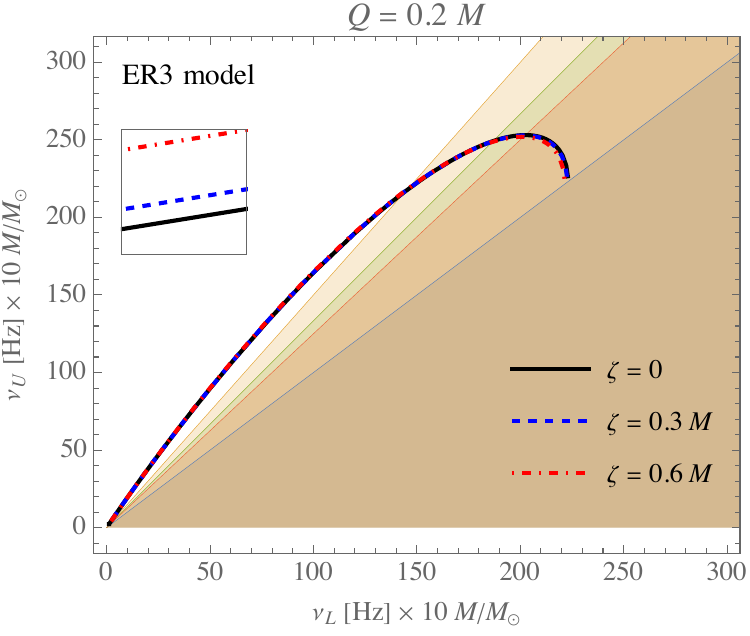}
\end{subfigure}
\hfill
  \begin{subfigure}{0.3\textwidth}
\centering
\includegraphics[width=5.3cm]{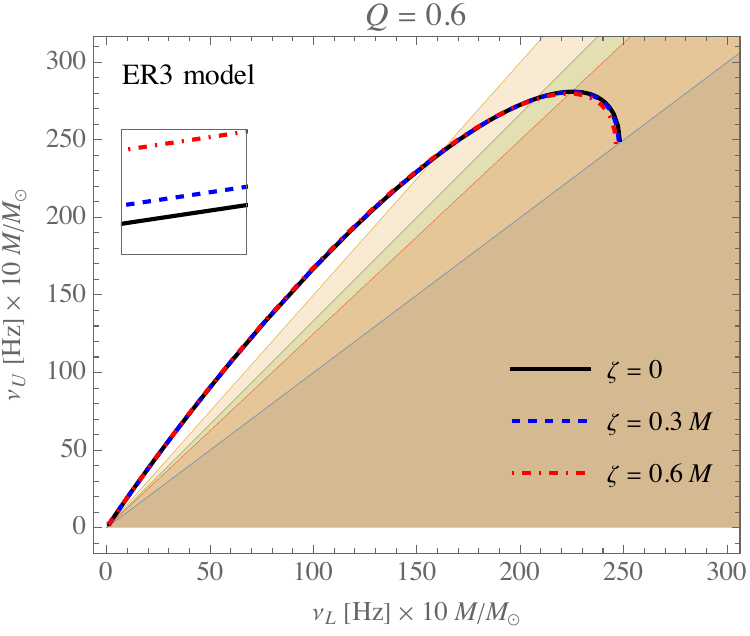}
\end{subfigure}
  \begin{subfigure}{0.3\textwidth}
\centering
\includegraphics[width=5.3cm]{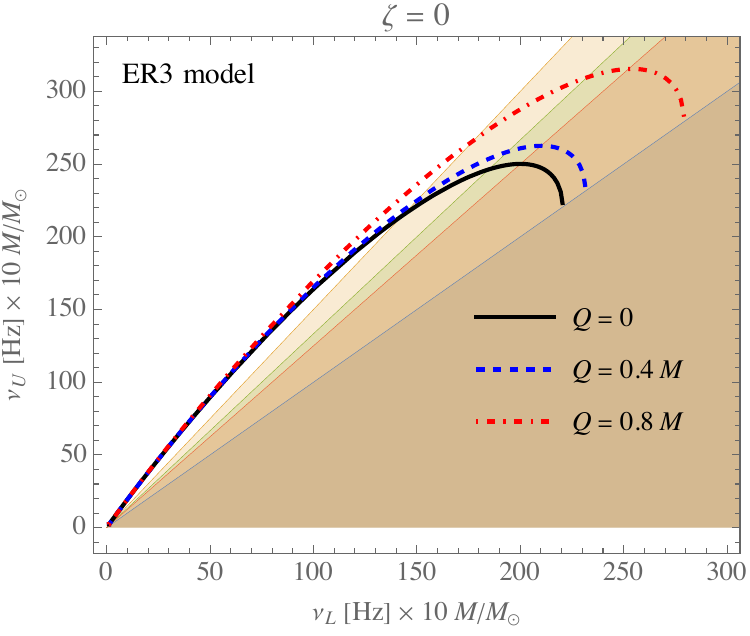}
\end{subfigure}
\hfill
  \begin{subfigure}{0.3\textwidth}
\centering
\includegraphics[width=5.3cm]{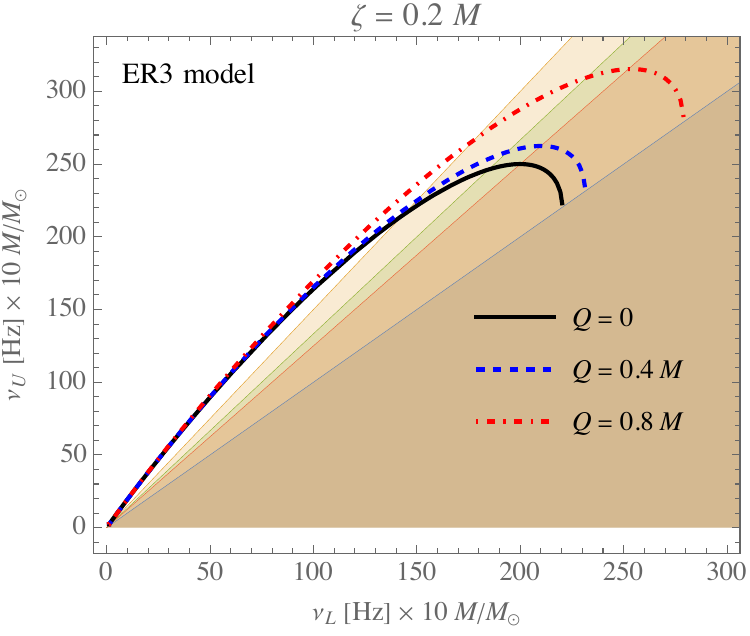}
\end{subfigure}
\hfill
  \begin{subfigure}{0.3\textwidth}
\centering
\includegraphics[width=5.3cm]{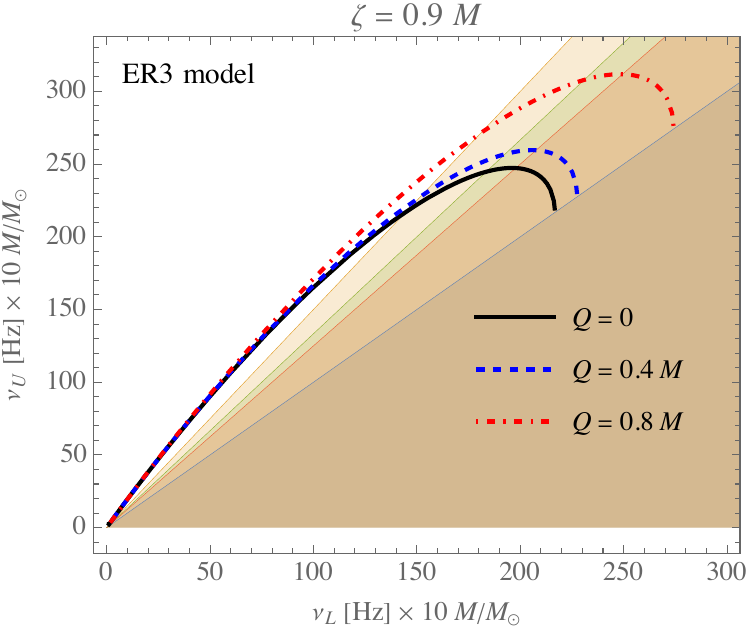}
\end{subfigure}
  \begin{subfigure}{0.3\textwidth}
\centering
\includegraphics[width=5.3cm]{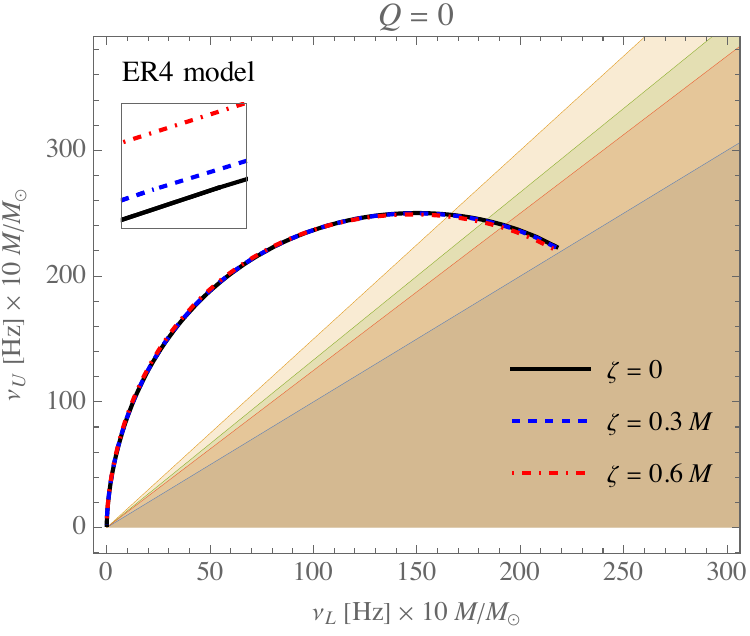}
\end{subfigure}
\hfill
  \begin{subfigure}{0.3\textwidth}
\centering
\includegraphics[width=5.3cm]{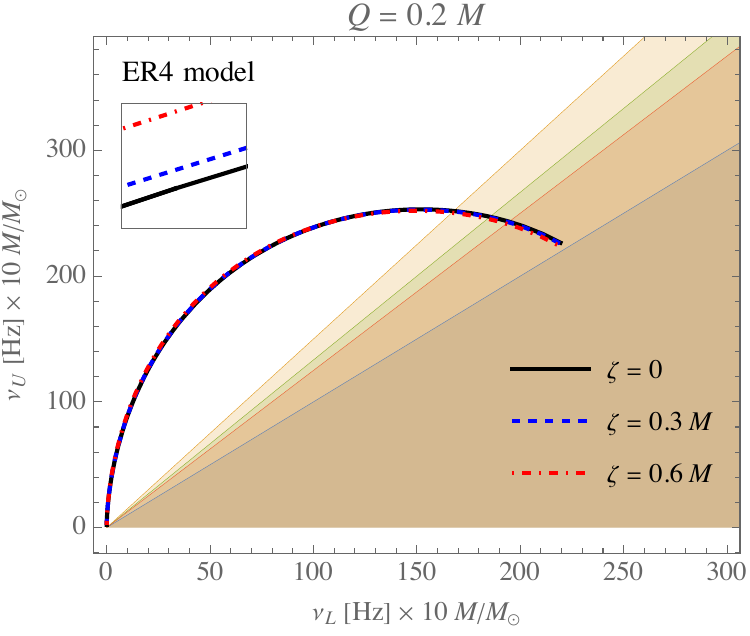}
\end{subfigure}
\hfill
  \begin{subfigure}{0.3\textwidth}
\centering
\includegraphics[width=5.3cm]{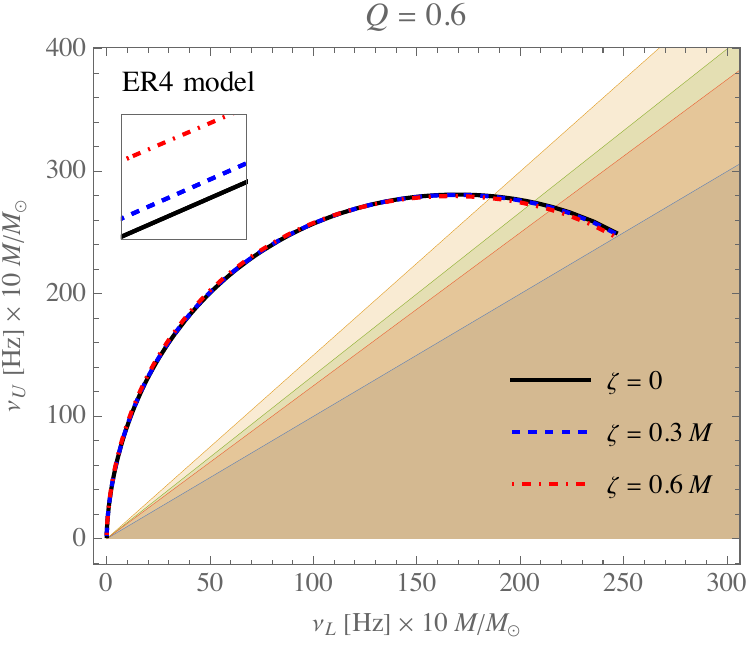}
\end{subfigure}
  \begin{subfigure}{0.3\textwidth}
\centering
\includegraphics[width=5.3cm]{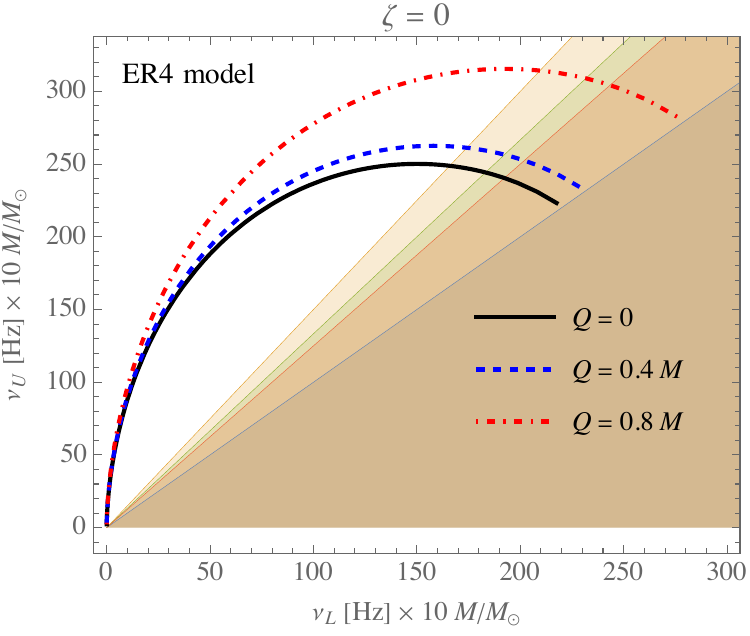}
\end{subfigure}
\hfill
  \begin{subfigure}{0.3\textwidth}
\centering
\includegraphics[width=5.3cm]{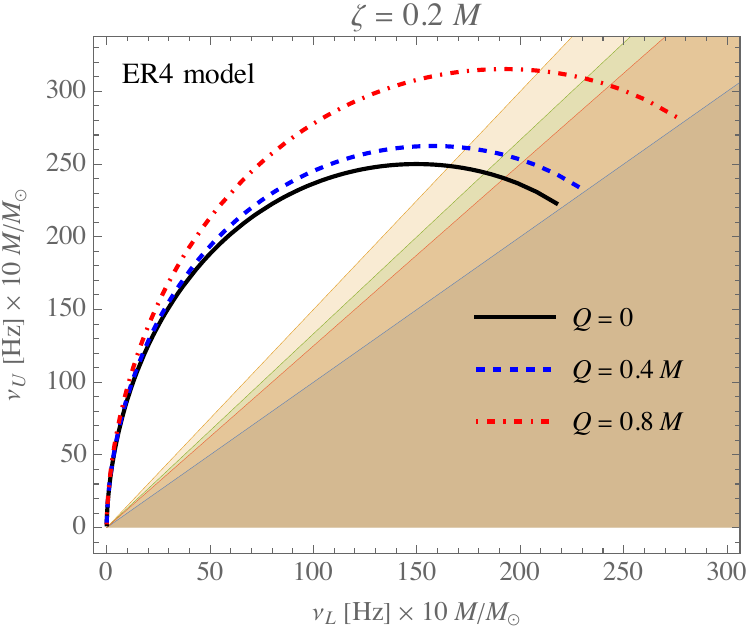}
\end{subfigure}
\hfill
  \begin{subfigure}{0.3\textwidth}
\centering
\includegraphics[width=5.3cm]{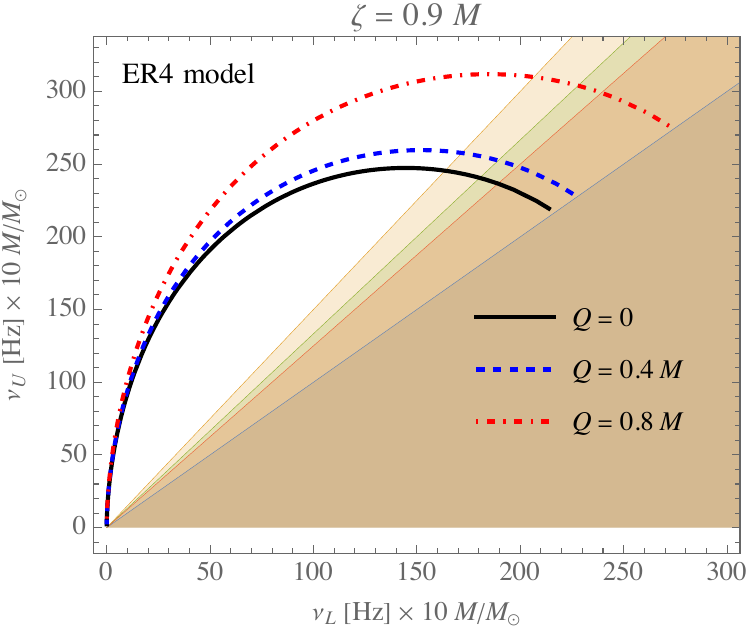}
\end{subfigure}
  \caption[]{(continued)}
\end{figure*}
\begin{figure*}[t]
  \ContinuedFloat
  \centering
  %
  \begin{subfigure}{0.3\textwidth}
\centering
\includegraphics[width=5.3cm]{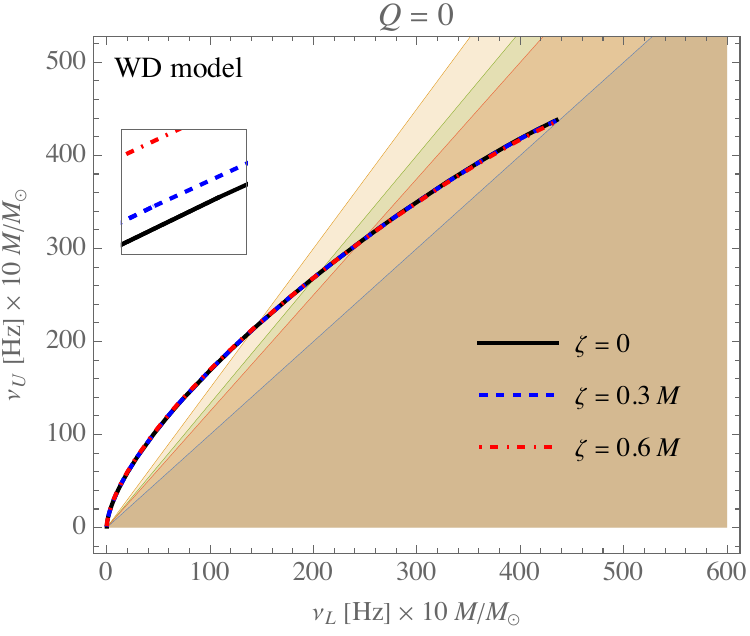}
\end{subfigure}
\hfill
  \begin{subfigure}{0.3\textwidth}
\centering
\includegraphics[width=5.3cm]{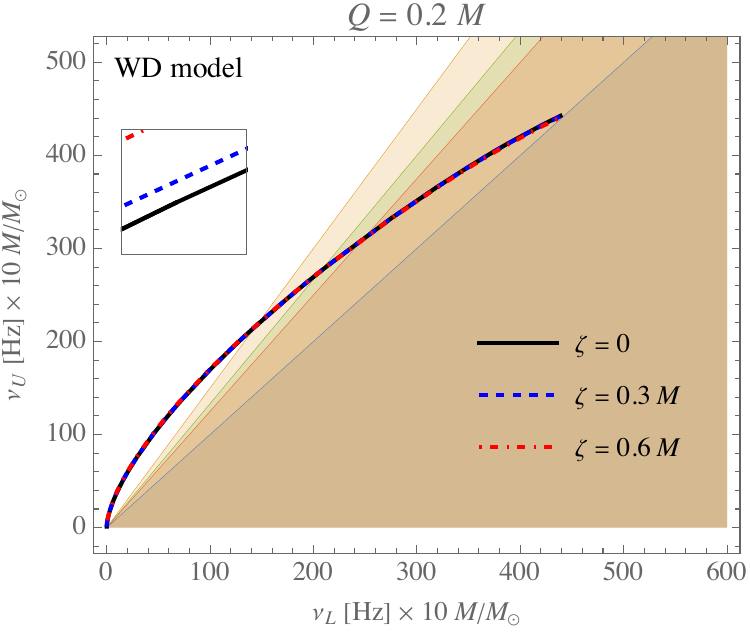}
\end{subfigure}
\hfill
  \begin{subfigure}{0.3\textwidth}
\centering
\includegraphics[width=5.3cm]{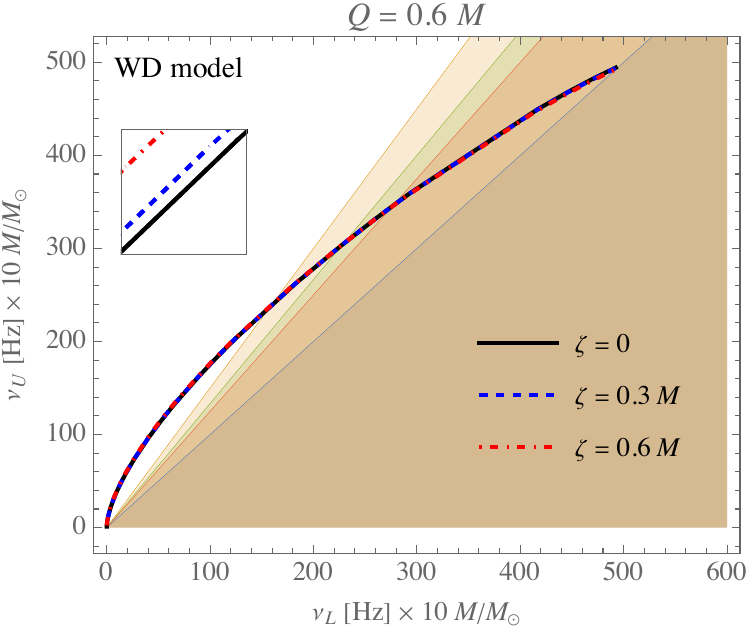}
\end{subfigure}
  \begin{subfigure}{0.3\textwidth}
\centering
\includegraphics[width=5.3cm]{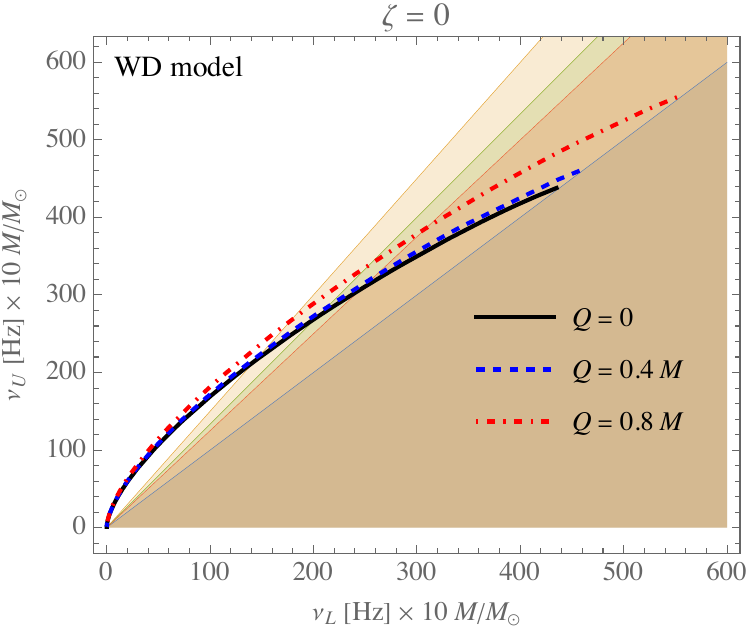}
\end{subfigure}
\hfill
  \begin{subfigure}{0.3\textwidth}
\centering
\includegraphics[width=5.3cm]{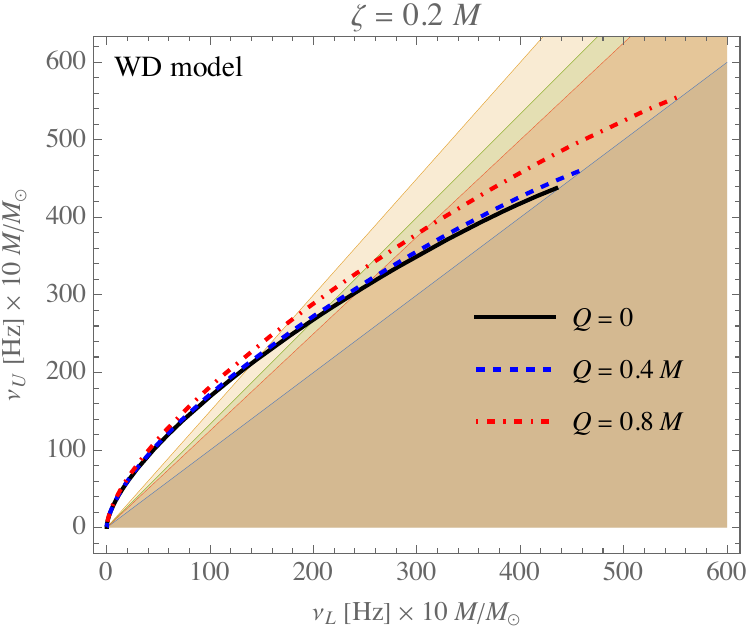}
\end{subfigure}
\hfill
  \begin{subfigure}{0.3\textwidth}
\centering
\includegraphics[width=5.3cm]{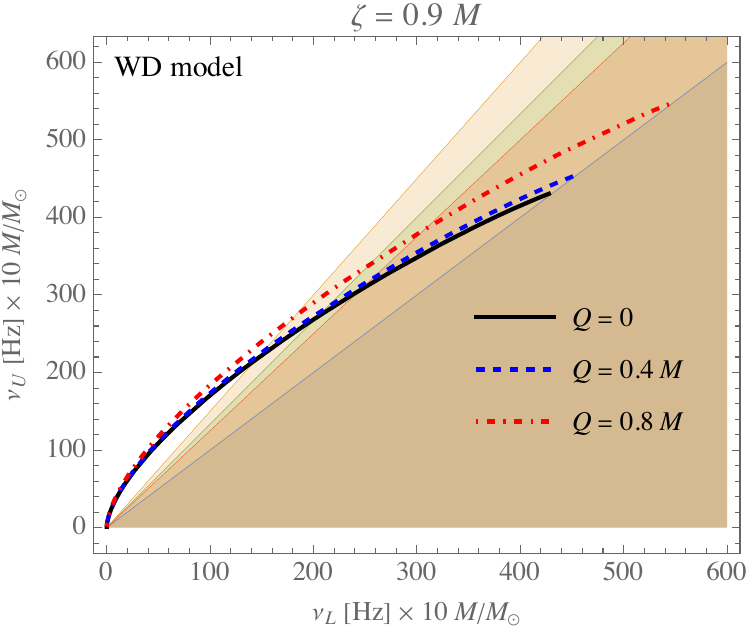}
\end{subfigure}
  \caption[]{(continued)}
\end{figure*}
The impact of the spacetime parameters is also imprinted in the orbital radii at which QPOs with the aforementioned twin-peak frequency ratios can be generated within the different models. This dependence is conveniently encoded through the resonance condition \cite{jumaniyozov_circular_2024}
\begin{equation}
p\,\nu_L(r,M,Q,\zeta) = q\,\nu_U(r,M,Q,\zeta),
\label{eq:pq}
\end{equation}
where $p$ and $q$ are integers specifying the corresponding frequency ratio. Solving this relation yields the QPO radii, $r_{\mathrm{QPO}}$, for the cases $(p,q) = (3,2), (4,3)$, and $(5,4)$. 

To quantify the influence of the parameters $Q$ and $\zeta$ on the separation between the QPO orbits and the ISCO, we introduce the dimensionless (normalized) radial separation
\begin{equation}
\delta_r = \frac{r_{\mathrm{QPO}}}{r_{\mathrm{ISCO}}} - 1,
\label{eq:deltar}
\end{equation}
which measures the fractional radial offset of the QPO radius with respect to the ISCO. This quantity can be evaluated for each of the QPO models considered above.

\begin{figure*}[ht!]
    \centering
    \includegraphics[width=5.3cm]{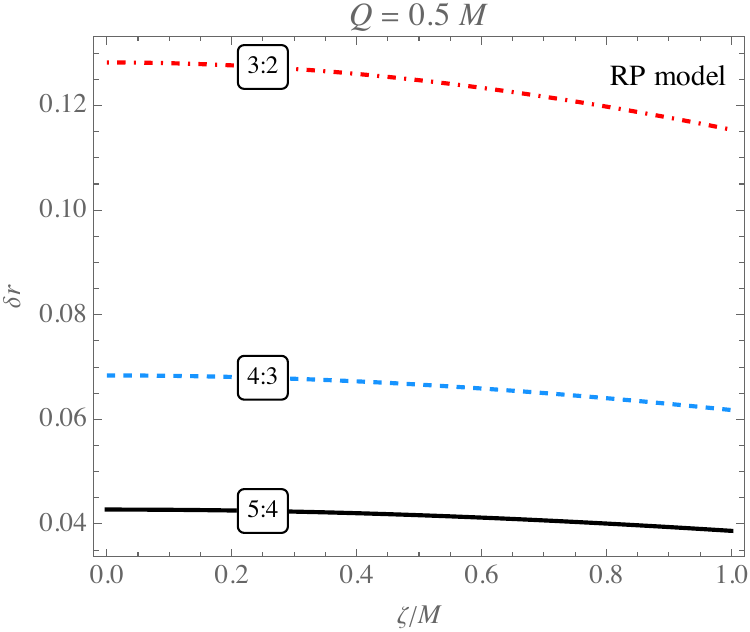}\,(a)
    \includegraphics[width=5.3cm]{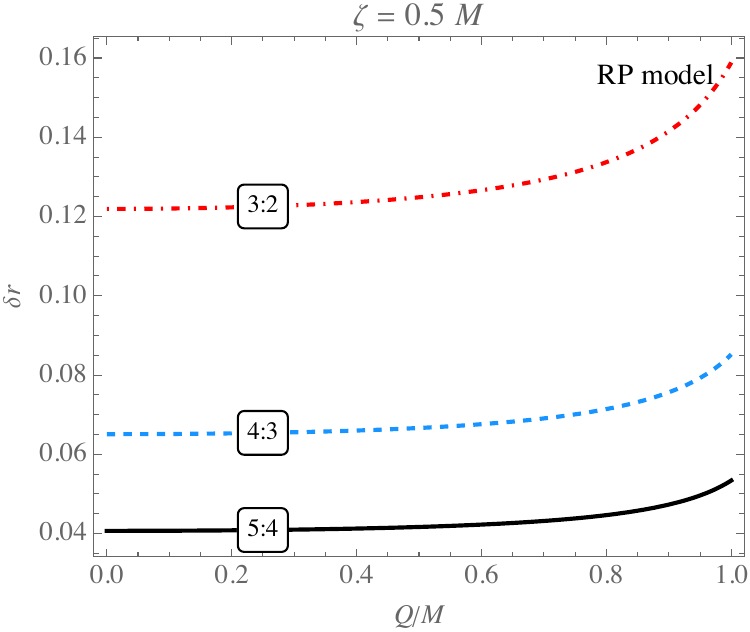}\,(b)
     \includegraphics[width=5.3cm]{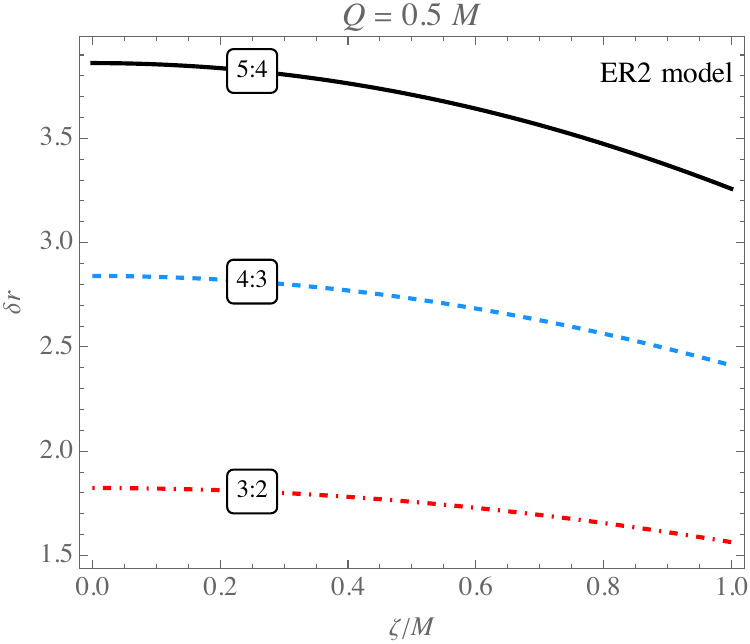}\,(c)
    \includegraphics[width=5.3cm]{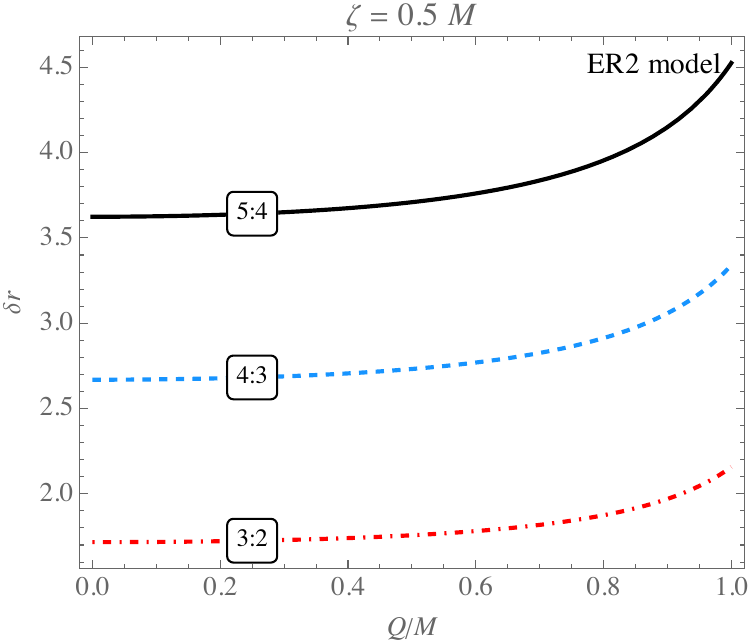}\,(d)
    \includegraphics[width=5.3cm]{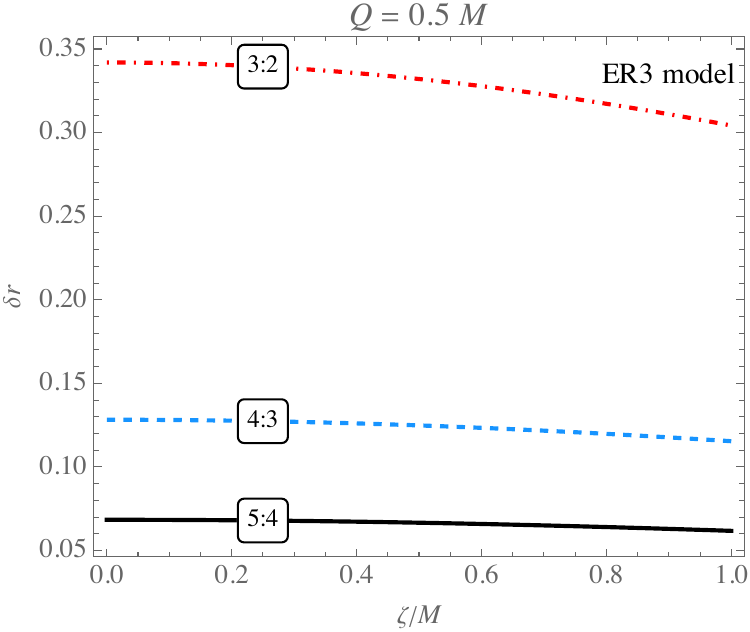}\,(e)
    \includegraphics[width=5.3cm]{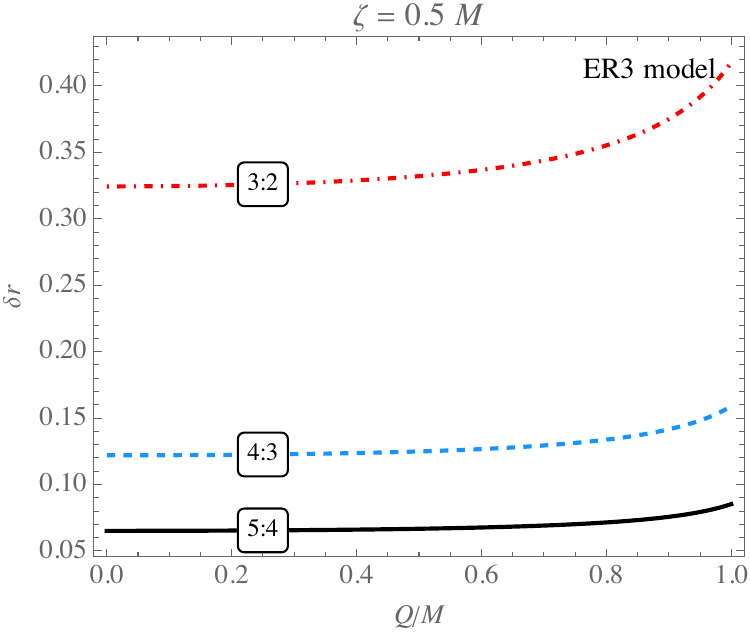}\,(f)
    \includegraphics[width=5.3cm]{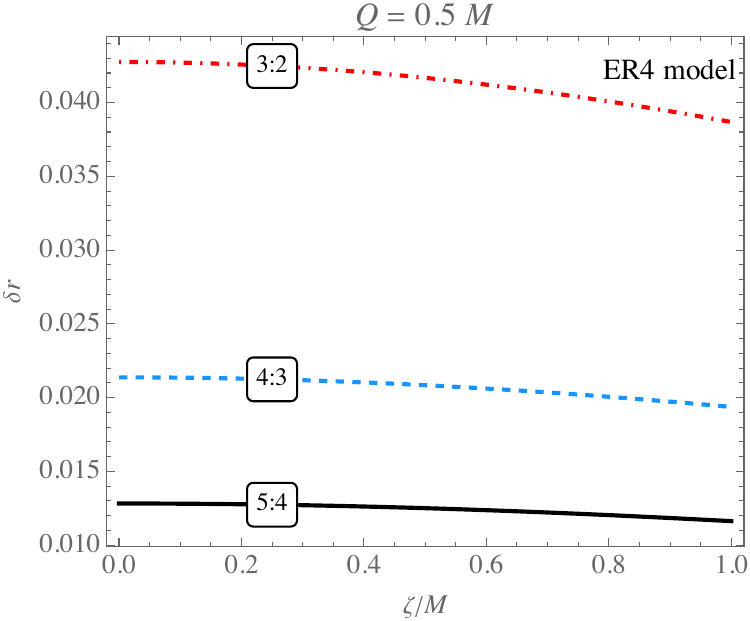}\,(g)
    \includegraphics[width=5.3cm]{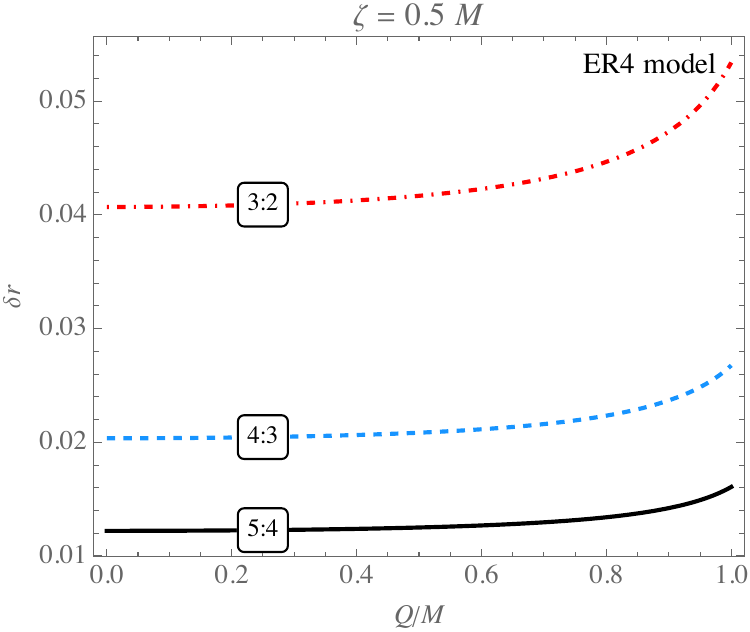}\,(h)
    \includegraphics[width=5.3cm]{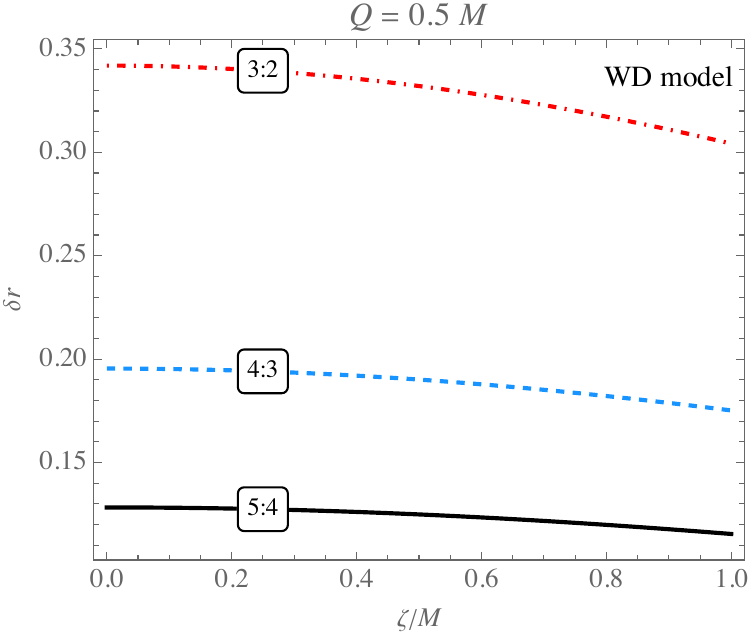}\,(i)
    \includegraphics[width=5.3cm]{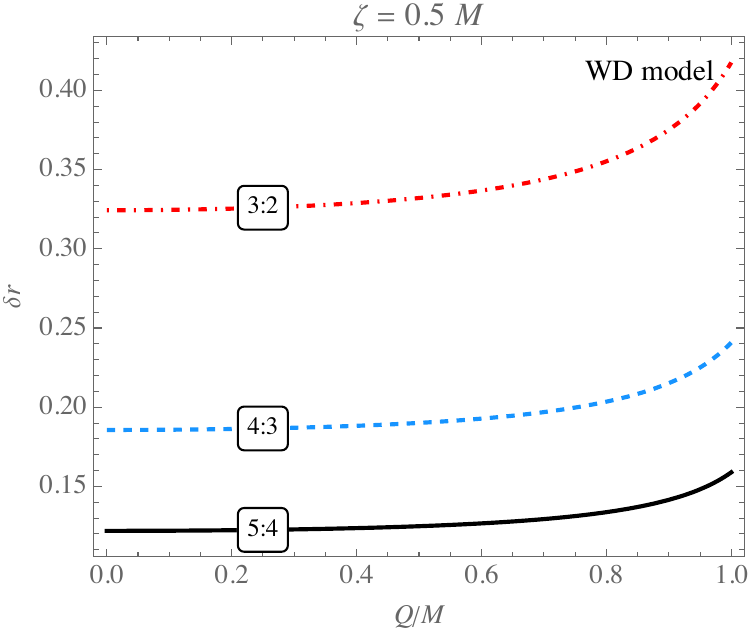}\,(j)
    \caption{The normalized radial separation between the QPO orbits and the ISCO as functions of $\zeta$ and $Q$, respectively, for moderate fixed values of $Q$ and $\zeta$, for the QPO models considered in this work.}
    \label{fig:deltar}
\end{figure*}

In Fig.~\ref{fig:deltar}, we display the dependence of $\delta_r$ on $Q$ and $\zeta$ for all the QPO models employed in this work.
As can be inferred from the diagrams, in all cases an increase in the quantum-correction parameter leads to a decrease in the normalized separation, while an increase in the total electric charge results in a larger value of this quantity. Furthermore, for the RP, ER3, ER4, and WD models, smaller frequency ratios $p\!:\!q$ correspond to smaller values of the normalized separation. In contrast, for the ER2 model, lower $p\!:\!q$ ratios are associated with larger normalized separations.

\subsection{BH parameter estimation from QPO data}\label{subsec:MCMC}
 
In this subsection, we aim to constrain the parameters associated with particle motion around the quantum-corrected RN BH by confronting the model with observational data from three classes of BH candidates: stellar-mass, intermediate-mass, and supermassive systems.

For the stellar-mass category, we consider the compact objects located at the centers of the microquasars GRO~J1655-40 and XTE~J1550-564. As a representative intermediate-mass BH candidate, we adopt M82~X-1, an ultra-luminous X-ray source in the M82 galaxy, whose mass is estimated to lie in the range of $\sim 10^{2}$--$10^{3}\,M_\odot$ \cite{fiorito_is_2004,torok_possible_2005,stuchlik_mass_2015}. 

In addition, we include the supermassive BH Sgr~A$^\ast$ at the center of the Milky Way, which is of particular interest in the context of microsecond QPOs. The QPO-frequency-related properties of these sources are summarized in Table~\ref{tab:1}.

\begin{table*}
    \centering
    \begin{tabular}{c|ccccc}
         source  & $\nu_U$ [Hz] & $\Delta\nu_U$ [Hz] & $\nu_L$ [Hz] & $\Delta\nu_L$ [Hz] & mass [$M_\odot$]\\
         \hline
        XTE J1550-564 \cite{orosz_improved_2011}& 276 & $\pm3$ & 184 & $\pm5$ & $9.1\pm0.61$ \\
        GRO J1655-40 \cite{strohmayer_discovery_2001} & 451 & $\pm5$ & 298  & $\pm4$ & $5.4\pm0.3$ \\
        M82 X-1 \cite{pasham_400-solar-mass_2014} & 5.07 & $\pm0.06$ & 3.32 & $\pm0.06$ & $415\pm63$ \\
        Sgr A* \cite{ghez_measuring_2008} & $1.445\times 10^{-3}$ &  $\pm0.16\times 10^{-3}$ & $0.886\times10^{-3}$ & $\pm0.04\times10^{-3}$ & $(4.1\pm0.6)\times 10^6$ \\
    \end{tabular}
    \caption{The twin-peak QPO frequencies for the source candidates.}
    \label{tab:1}
\end{table*}

We then employ the Python package \texttt{emcee} \cite{ascl,zhadyranova_exploring_2024,abdulkhamidov_parameter_2024} to carry out an MCMC analysis and to constrain the parameters describing the motion of test particles around the quantum-corrected RN BH.

Within a Bayesian framework, the posterior probability distribution is defined as \cite{liu_constraints_2023,mitra_charged_2024}
\begin{equation}
\mathcal{P}(\theta|\mathcal{M}) = \frac{P(\mathcal{D}|\theta,\mathcal{M})\,\pi(\theta|\mathcal{M})}{P(\mathcal{D}|\mathcal{M})},
\label{eq:posterior_0}
\end{equation}
where $\pi(\theta)$ and $P(\mathcal{D}|\theta,\mathcal{M})$, are, respectively, the prior and the likelihood. We choose the priors to be Gaussian (normal), within the suitably identified boundaries (listed in Table \ref{tab:2}). Hence, we set
\begin{equation}
\pi(\theta_i) \sim \exp\left[\frac{1}{2}\left(\frac{\theta_i-\theta_{0,i}}{\sigma_i}\right)^2\right],\qquad \theta_{\mathrm{low},i}<\theta_i<\theta_{\mathrm{high},i},
    \label{eq:pi_i}
\end{equation}
where $\theta_i=\{r/M,M/M_\odot,Q/M,\zeta/M\}$ for the spacetime of the quantum-corrected RN BH, and $\sigma_i$ are their respective variances. Furthermore, to perform our MCMC analysis, we use two distinctive data parts, which result in the likelihood functions 
\begin{equation}
\ln\mathfrak{L} = \ln\mathfrak{L}_U + \ln\mathfrak{L}_L,
    \label{eq:like_0}
\end{equation}
in which
\begin{subequations}
    \begin{align}
        & \ln\mathfrak{L}_U = -\frac{1}{2}\sum_j\left(\frac{\nu_{\phi,\mathrm{obs}}^j - \nu_{\phi,\mathrm{theo}}^j}{\sigma_{\phi,\mathrm{obs}}^j}\right)^2,\\
        & \ln\mathfrak{L}_L = -\frac{1}{2}\sum_j\left(\frac{\nu_{\mathrm{per},\mathrm{obs}}^j - \nu_{\mathrm{per},\mathrm{theo}}^j}{\sigma_{\mathrm{per},\mathrm{obs}}^j}\right)^2,
    \end{align}
\end{subequations}
are, respectively, the likelihoods of upper (orbital) astrometric frequency ($\nu_\phi$), and the lower (periapsis) frequency ($\nu_{\mathrm{per}}$), in which, $(\nu_{\phi,\mathrm{obs}}^j, \nu_{\mathrm{per},\mathrm{obs}}^j)$ are the direct observational results of the orbital (Keplerian) frequency (i.e. $\nu_K$), and the periapsis frequency (i.e. $\nu_\mathrm{per} = \nu_K - \nu_r$), and $(\nu_{\phi,\mathrm{theo}}^j, \nu_{\mathrm{per},\mathrm{theor}}^j)$ are their theoretical estimations.

\begin{table*}[ht!]
	\centering
	\begin{tabular}{cccccccccccc}
		\toprule
		 &\multicolumn{2}{c}{XTE J1550-564}& &\multicolumn{2}{c}{GRO J1655-40}& &\multicolumn{2}{c}{M82-X1}& &\multicolumn{2}{c}{Sgr A*}\\
		\cmidrule{2-3} \cmidrule{5-6} \cmidrule{8-9} \cmidrule{11-12}
		
		{} & {$\mu$} & {$\sigma$} &  {} & {$\mu$} & {$\sigma$}  & {} & {$\mu$} & {$\sigma$}&  {} & {$\mu$} & {$\sigma$}\\
		\midrule
		$r/M$ & 3.932 & 0.26 & & 5.206 & 0.194 & & 6.419 & 0.268 & & 5.865 & 0.847\\
		$M/M_\odot$ & 12.498 & 5.531 & & 4.607 & 0.477 & & 375.671 & 37.565 & & $1.481\times 10^6$ & $0.646\times 10^6$\\
        $Q/M$ & 1.092 & 0.072 & & $-0.894$ & 0.078 & & $-0.432$ & 0.094 & & 0.781 & 0.039\\
        $\zeta/M$ & 0.495 & 0.036 & & 0.503 & 0.024 & & 1.163 & 0.050 & & 0.618 & 0.051\\
		\bottomrule
	\end{tabular}
 \caption{The Gaussian priors (including the mean values $\mu$ and the variances $\sigma$) given for the quantum-corrected RN BH within the considered QPO data sources.}
 \label{tab:2}
\end{table*}

\begin{table*}
    \centering
    \begin{tabular}{c|cccc}
           & XTE J1550-564 & GRO J1655-40 & M82 X-1 & Sgr A*\\
         \hline
         $r/M$ & $3.97^{+0.25}_{-0.24}$ & $5.26^{+0.19}_{-0.19}$ & $6.47^{+0.26}_{-0.26}$ & $6.04^{+0.80}_{-0.78}$  \\
         $M/M_\odot$ & $12.50^{+0.13}_{-0.13}$ & $5.74^{+0.43}_{-0.42}$ & $387^{+34}_{-33}$  & $1.76^{+0.53}_{-0.46}\times 10^6$ \\
         $Q/M$ & $1.09^{+0.07}_{-0.07}$ & $-0.9^{+0.08}_{-0.08}$ & $-0.43^{+0.09}_{-0.09}$ & $0.78^{+0.04}_{-0.04}$ \\
        $\zeta/M$ & $0.50^{+0.04}_{-0.04}$ & $0.50^{+0.02}_{-0.02}$ & $1.16^{+0.05}_{-0.05}$ & $0.62^{+0.05}_{-0.05}$  \\
    \end{tabular}
    \caption{Best--fit values and constraints on the quantum-corrected RN BH parameters inferred from the QPO data listed in Table~\ref{tab:1}.}
    \label{tab:3}
\end{table*}

\begin{figure*}[ht!]
\begin{subfigure}{0.47\textwidth}
\centering
  \includegraphics[width=\linewidth]{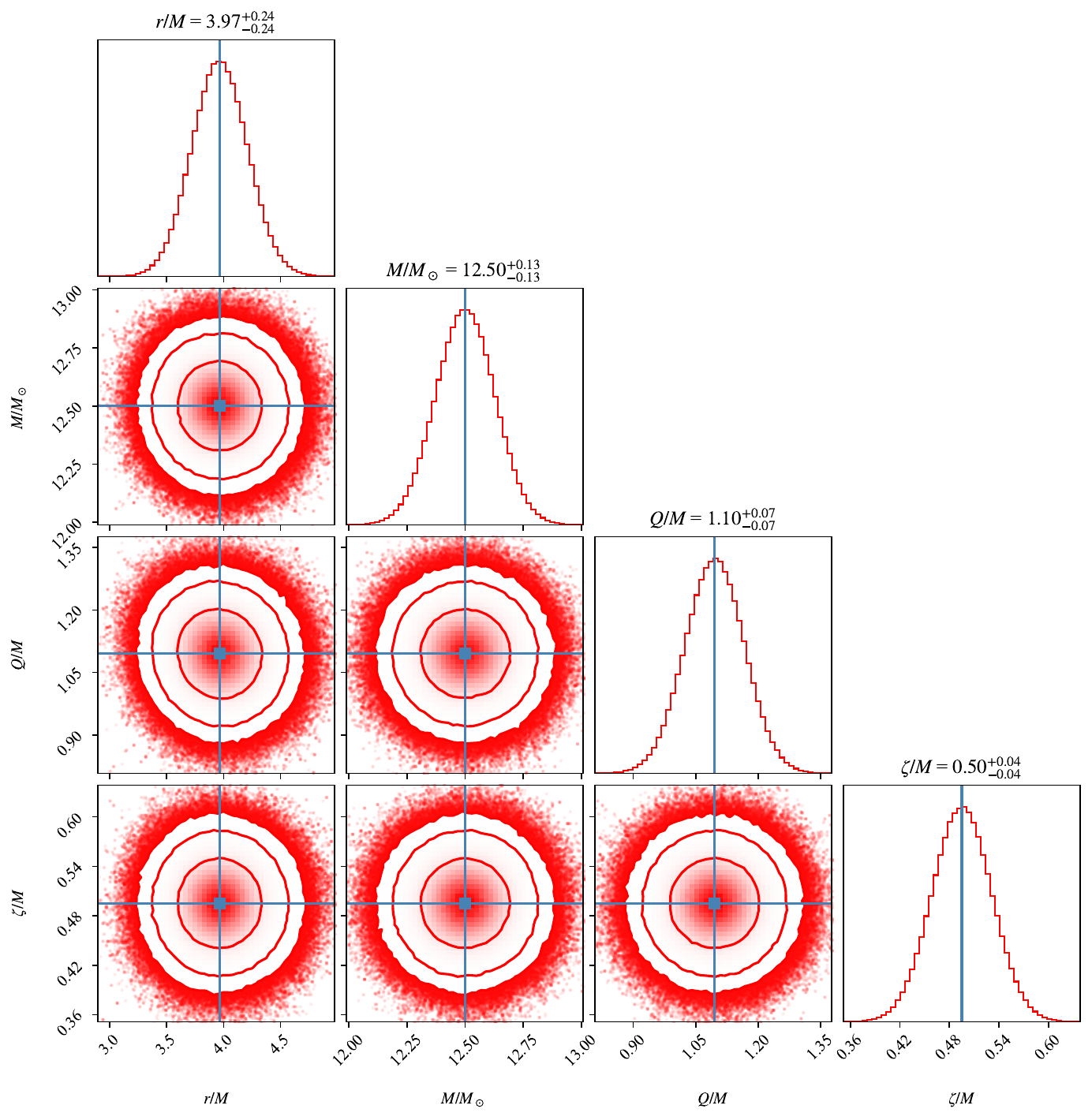}
  \caption{XTE J1550-564}
\end{subfigure}
 \qquad
\begin{subfigure}{0.47\textwidth}
\centering
  \includegraphics[width=\linewidth]{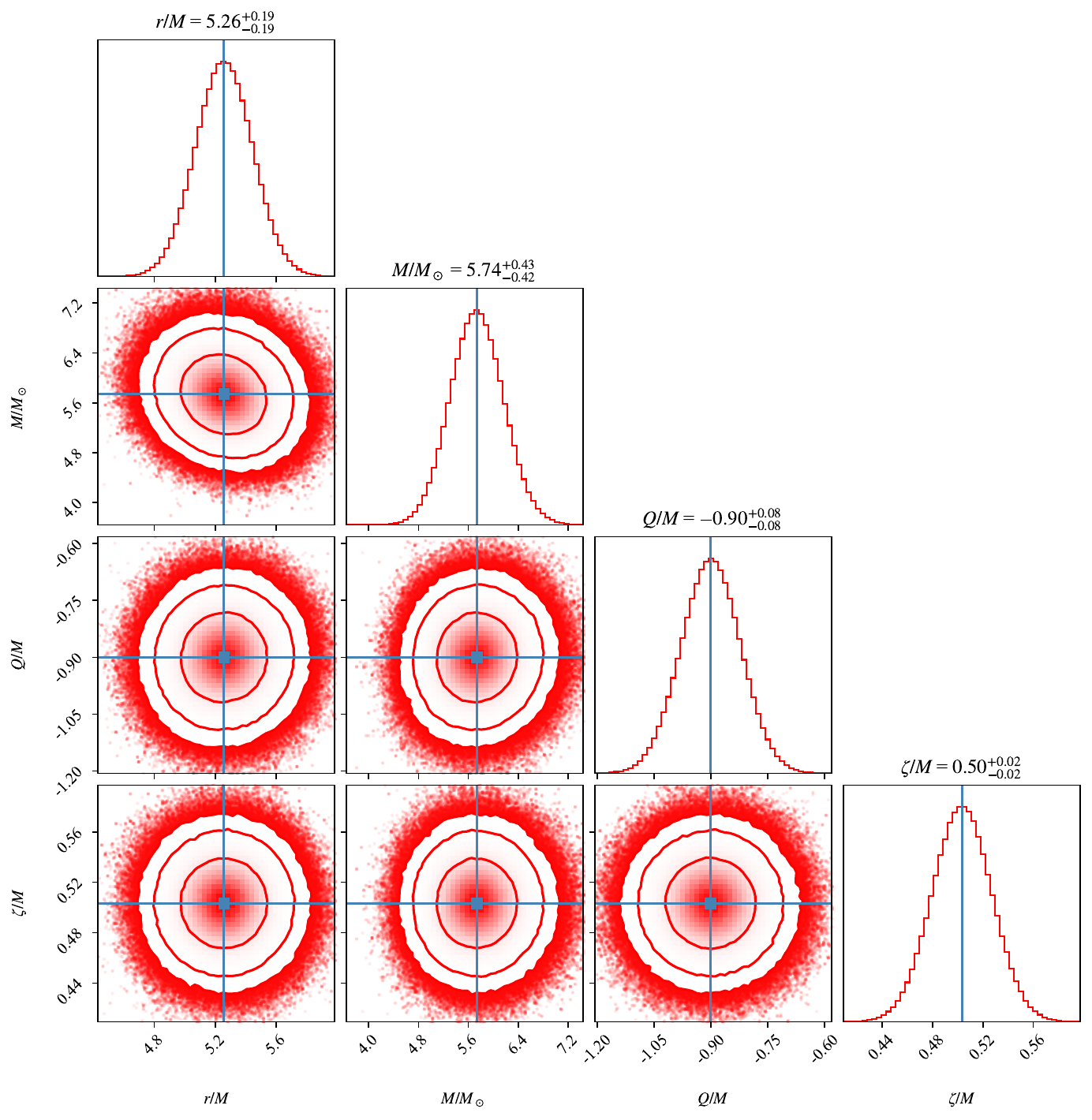}
  \caption{GRO J1655-40}
\end{subfigure}
\\
\begin{subfigure}{0.47\textwidth}
\centering
  \includegraphics[width=\linewidth]{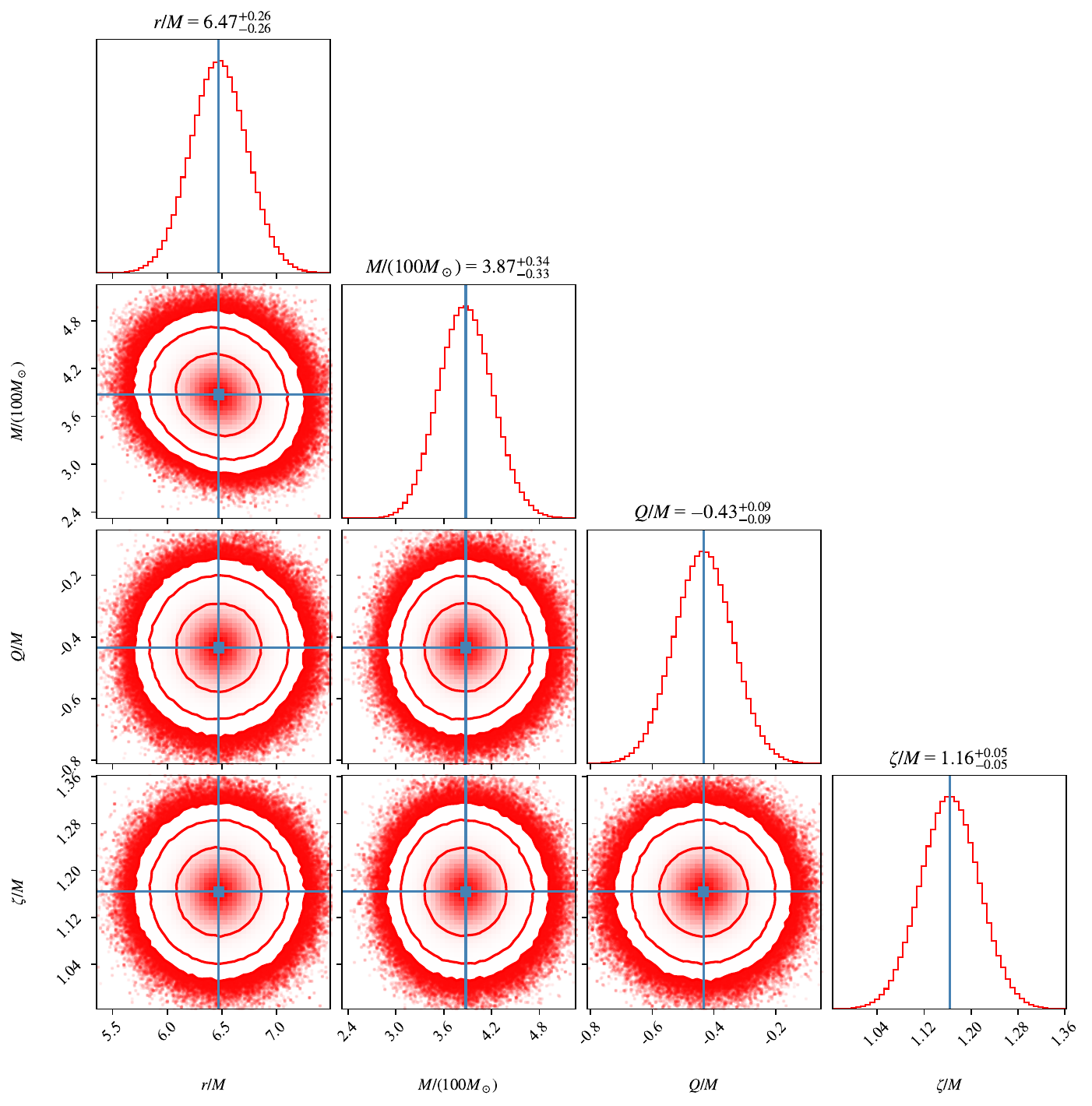}
  \caption{M82 X-1}
\end{subfigure}
\qquad
\begin{subfigure}{0.47\textwidth}
\centering
  \includegraphics[width=\linewidth]{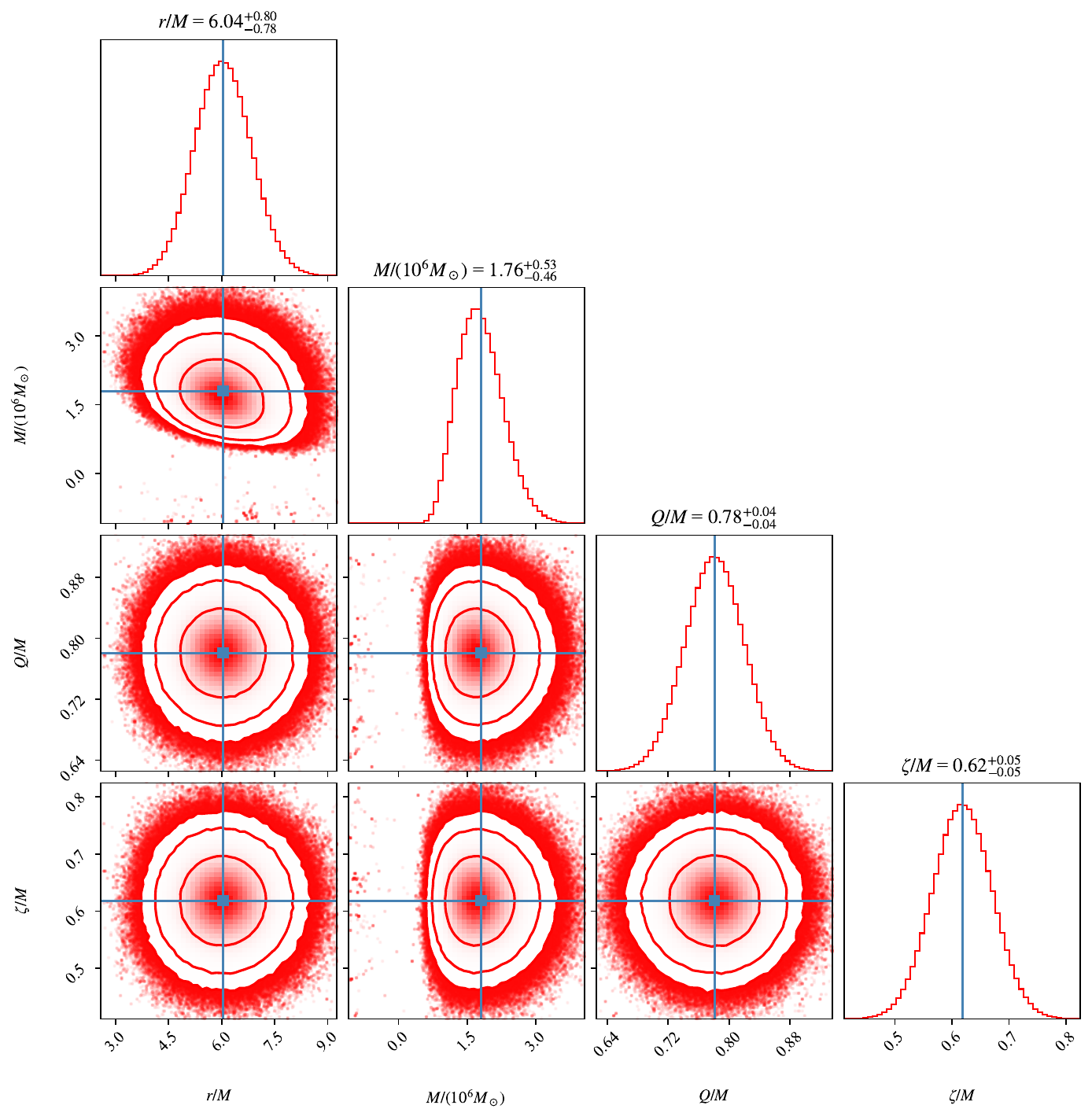}
  \caption{Sgr A*}
\end{subfigure}
    \caption{Constraints on the quantum-corrected RN BH parameters obtained from a four--dimensional MCMC analysis using the QPO data sets.}
    \label{fig:MCMC}
\end{figure*}

To constrain the parameter set $\{r/M,\, M/M_\odot,\, Q/M,\, \zeta/M\}$ of the quantum-corrected RN BH, we perform a MCMC analysis employing the Gaussian priors introduced above, which are based on the observational inputs for the QPO sources listed in Table~\ref{tab:1}. For each parameter, we draw approximately $10^5$ samples from the corresponding prior Gaussian distribution. This procedure enables an efficient exploration of the physically admissible region of the parameter space within the adopted bounds and allows us to determine the best--fit values of the model parameters, which are reported in Table~\ref{tab:3}.

In Fig.~\ref{fig:MCMC}, we present the outcome of the MCMC analysis in terms of the corresponding posterior probability distributions. The contour plots display the confidence regions at the $1\sigma$ ($68\%$), $2\sigma$ ($95\%$), and $3\sigma$ ($99\%$) levels for the full parameter set. The shaded areas indicate these confidence intervals and illustrate the correlations among the parameters.

According to the diagrams, one can infer that in all cases, the MCMC chains converge to well-defined regions in the parameter space $\{r/M,\, M/M_\odot,\, Q/M,\, \zeta/M\}$, indicating that the quantum-corrected RN model is capable of providing meaningful constraints from the observational data. The stellar-mass systems XTE~J1550-564 and GRO~J1655-40 exhibit relatively tight posteriors, especially for $M/M_\odot$ and $Q/M$, reflecting the higher precision of their QPO measurements. In contrast, the supermassive source Sgr~A* shows broader credible regions, particularly in $\zeta/M$, signaling a larger degree of degeneracy among the model parameters. The intermediate-mass candidate M82-X1 lies between these two regimes, with uncertainties larger than those of the stellar-mass sources but smaller than those of Sgr~A*. Despite these differences, all four sources display consistent, well-behaved posterior structures, supporting the robustness of the inferred constraints within the quantum-corrected RN framework.

\section{Scalar Field Perturbations}\label{sec:5}

In this section, we investigate the dynamics of a massless scalar field around a quantum-corrected charged black hole, highlighting how quantum corrections and other black hole parameters modify the perturbation potential. Using this potential, we analyze the greybody factors, which encode the modification of Hawking radiation due to the black hole's potential barrier, determining the fraction of particles that can escape to infinity. Greybody factors are therefore crucial for understanding the black hole's energy emission spectrum and its observational signatures. A large body of work has been devoted to studying scalar perturbations of black holes, both within the framework of general relativity and in various modified gravity theories (see \cite{Chandrasekhar1983,Kokkotas1999,Schutz1985,Iyer1987,Konoplya2011,Berti2009,FA3,FA5,FA7,FA9}).

\subsection{The Massless Klein-Gordon Equation}

The massless scalar field dynamics are governed by the Klein-Gordon equation \cite{Chandrasekhar1983}
\begin{equation}
\frac{1}{\sqrt{-g}}\,\partial_{\mu}\left(\sqrt{-g}\,g^{\mu\nu}\,\partial_{\nu}\Psi\right)=0\quad\quad (\mu,\nu=0,\cdots,3), \label{ff1}    
\end{equation}
where $\Psi$ represents the scalar field wave function, $g_{\mu\nu}$ denotes the covariant metric tensor, $g=\det(g_{\mu\nu})$ is the metric determinant, $g^{\mu\nu}$ corresponds to the contravariant metric components, and $\partial_{\mu}$ represents coordinate partial derivatives.

For our KRG-QF spacetime geometry given in Eq.~(\ref{aa1}), the metric components are:
\begin{align*}
g_{\mu\nu}=\left(-f,\,f^{-1},\,r^2,\,r^2\,\sin^2 \theta\right),
g^{\mu\nu}=\left(-1/f,\,f,\,1/r^{2},\,1/(r^{2}\sin^{2} \theta)\right),
\end{align*}
\begin{equation}
    g=\det (g_{\mu\nu})=-r^4\,\sin^2 \theta.\label{ff2}
\end{equation}

We employ the standard separable ansatz for the scalar field wave function:
\begin{equation}
\Psi(t, r,\theta, \phi)=\exp(-i\,\omega\,t)\,Y^{m}_{\ell} (\theta,\phi)\,\frac{\psi(r)}{r},\label{ff3}
\end{equation}
where $\omega$ represents the (possibly complex) temporal frequency characterizing QNM behavior, $\psi(r)$ denotes the radial wave function, and $Y^{m}_{\ell}(\theta,\phi)$ are the spherical harmonics satisfying the eigenvalue equation:
\begin{equation}
\left[ \frac{1}{\sin\theta} \frac{\partial}{\partial\theta} \left( \sin\theta \frac{\partial}{\partial\theta} \right) + \frac{1}{\sin^2\theta} \frac{\partial^2}{\partial\phi^2} \right] Y_{\ell m}(\theta, \phi) = -\ell(\ell+1) Y_{\ell m}(\theta, \phi).\label{ff4}
\end{equation}

Substituting the separable ansatz into the Klein-Gordon equation yields the radial wave equation:
\begin{equation}
f\, \psi''(r) + f'\, \psi'(r) + \left( \frac{\omega^2}{f} - \frac{f'}{r} - \frac{\ell\,(\ell + 1)}{r^2} \right)\, \psi(r) = 0,\label{ff5}
\end{equation}
where the prime denotes differentiation with respect to the radial coordinate $r$.

To eliminate the first-order derivative term and obtain a Schr\"{o}dinger-like wave equation, we implement the tortoise coordinate transformation:
\begin{eqnarray}
r_*=\int\,\frac{dr}{f}\quad,\quad \partial_{r_{*}}=f\,\partial_r\quad,\quad \partial^2_{r_{*}}=f^2\,\partial^2_r+f\,f'\,\partial_r.\label{ff6}
\end{eqnarray}

This coordinate transformation converts the radial equation into the standard Schr\"{o}dinger form:
\begin{equation}
\frac{\partial^2 \psi(r_*)}{\partial r^2_{*}}+\left(\omega^2-V_\text{s}\right)\,\psi(r_*)=0,\label{ff7}
\end{equation}
where the effective scalar perturbation potential is given by:

\begin{align}
V_\text{s}(r)&=\left(\frac{\ell\,(\ell+1)}{r^2}+\frac{f'}{r}\right)\,f\nonumber\\
&=\frac{\left(1-\frac{2 M}{r}+\frac{Q^2}{r^2}\right)\left[1+\frac{\zeta ^2}{r^2}\left(1-\frac{2 M}{r}+\frac{Q^2}{r^2}\right)\right]}{r^2}\times\nonumber\\
&\Bigg[\ell(\ell+1)+\left( \frac{2 M}{r} - \frac{2 Q^2}{r^2} \right) \left\{1 + \frac{2 \zeta^2}{r^2} \left( 1 - \frac{2 M}{r} + \frac{Q^2}{r^2} \right) \right\}\nonumber\\ 
&- \frac{2 \zeta^2}{r^2} \left( 1 - \frac{2 M}{r} + \frac{Q^2}{r^2} \right)^2\Bigg].\label{ff8}
\end{align}

\begin{figure}[ht!]
    \centering
  \includegraphics[width=1.1\linewidth]{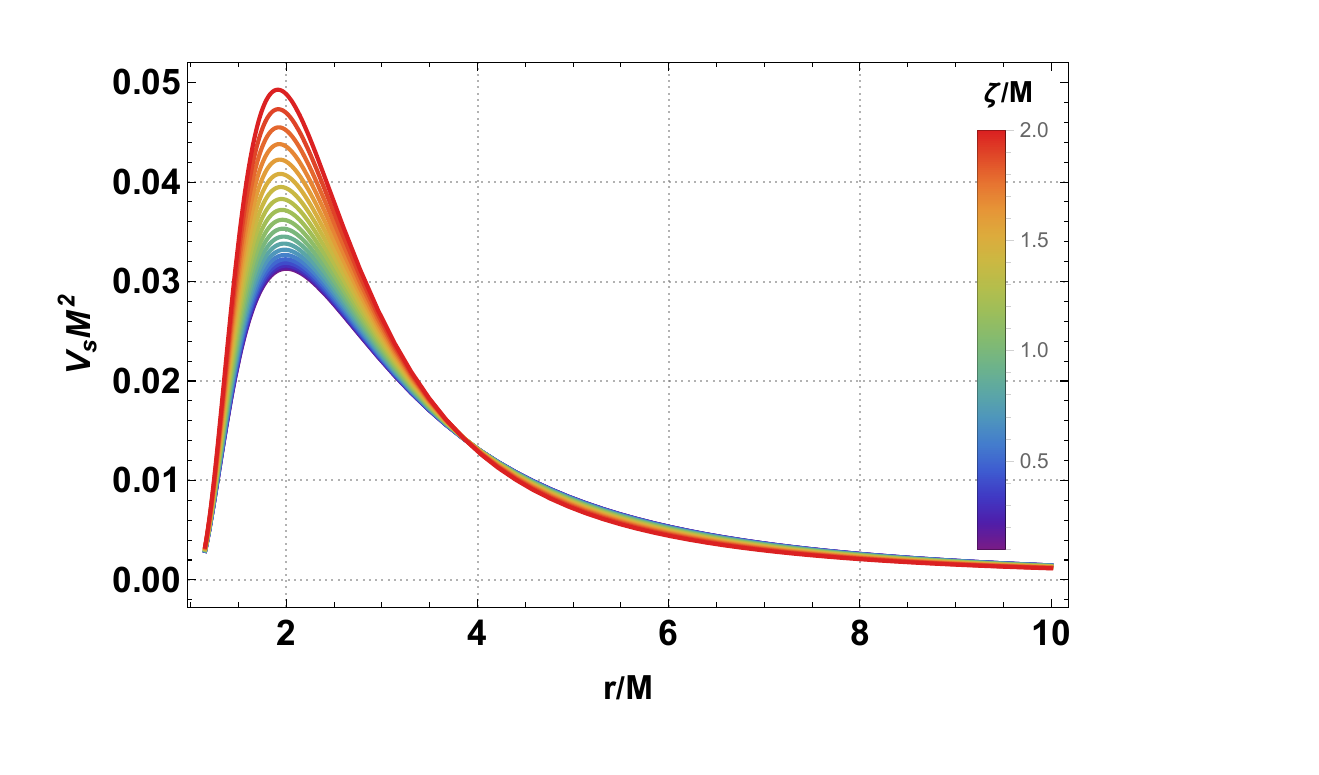}\\
  (i) $\ell=0$ \\
  \includegraphics[width=1.1\linewidth]{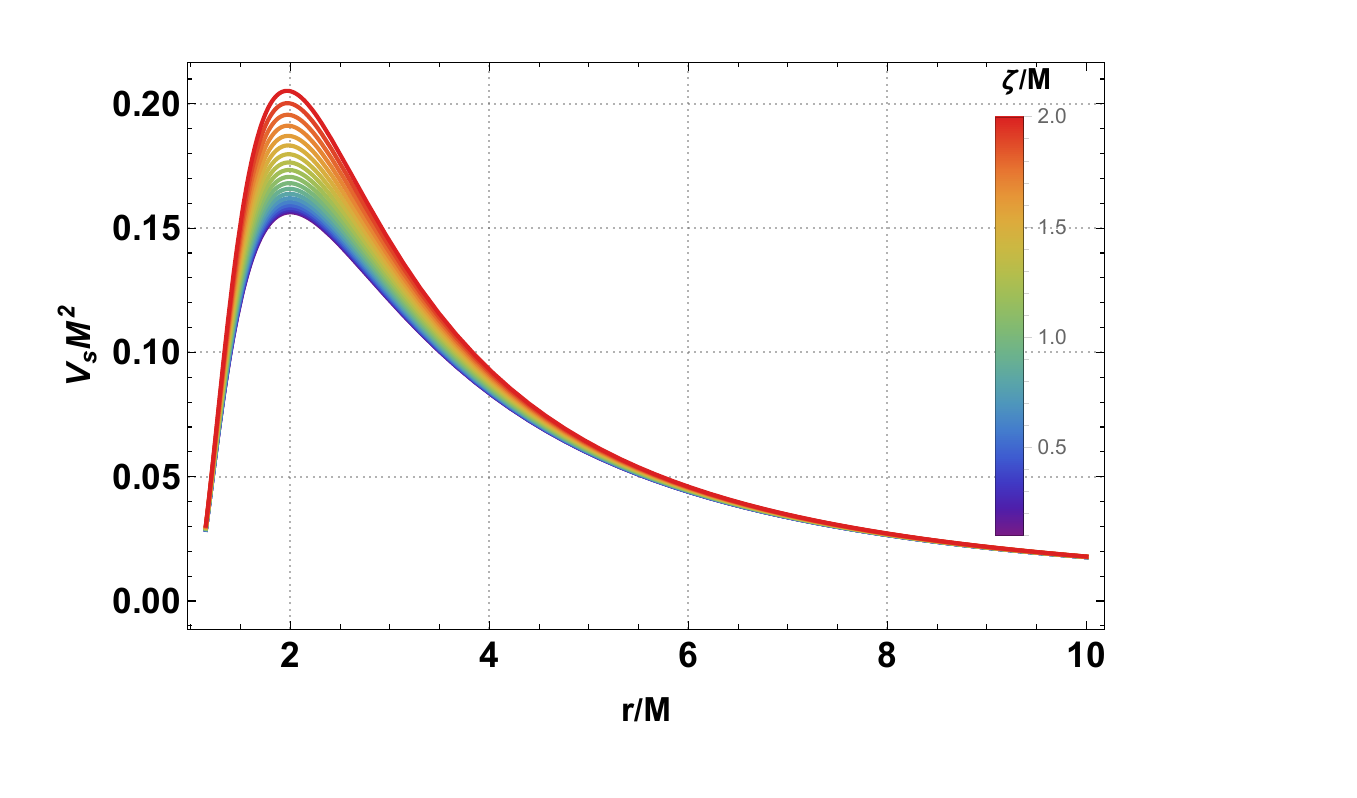}\\
   (ii) $\ell=1$
    \caption{Behaviors of the scalar perturbation potential of quantum-corrected RN BH for various values of the quantum correction parameter $\zeta$. The rest of the arguments are set as: $Q/M=1$.}
    \label{fig:scalar-potential}
\end{figure}

\begin{itemize}
    \item When $\zeta=0$, corresponding to the absence of quantum correction, the scalar perturbation potential simplifies as,
\begin{eqnarray}
V_\text{s}(r)=\frac{1}{r^2}\left(\ell(\ell+1)+\frac{2 M}{r} - \frac{2 Q^2}{r^2}\right)\left(1-\frac{2 M}{r}+\frac{Q^2}{r^2}\right)\label{ff9}
\end{eqnarray}
which is similar to that for RN BH case.

\item When $Q=0$, corresponding to the absence of electric charge, the selected space-time reduces to the quantum-corrected Schwarzschild black hole metric. In that limit, the perturbation potential simplifies as,
\begin{align}
V_\text{s}(r) &= 
\frac{\left(1-\frac{2 M}{r}\right)\left[1+\frac{\zeta^2}{r^2}\left(1-\frac{2 M}{r}\right)\right]}{r^2}\times\nonumber\\ 
&\Bigg[\ell(\ell+1)+
\frac{2 M}{r} \left\{1 + \frac{2 \zeta^2}{r^2} \left( 1 - \frac{2 M}{r} \right) \right\}
- \frac{2 \zeta^2}{r^2} \left( 1 - \frac{2 M}{r} \right)^2
\Bigg].
\label{ff10}
\end{align}
\end{itemize}

From the scalar perturbation analysis, we observe that both the electric charge and the quantum correction modifies the perturbation potential in comparison to the standard BH case.

\begin{figure}[ht!]
    \centering
    \includegraphics[width=1\linewidth]{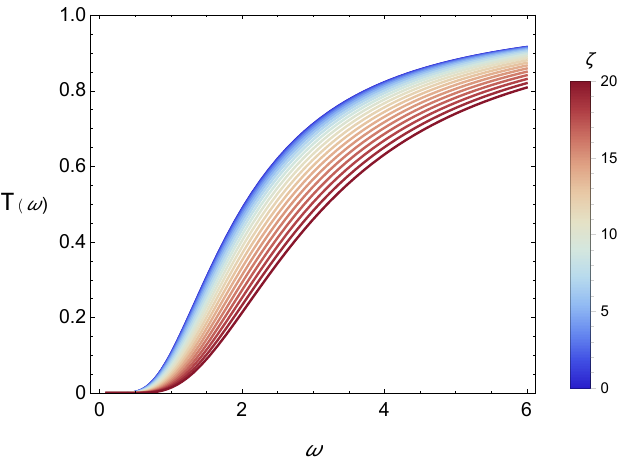}
    \caption{Greybody factor of quantum-corrected RN BH for various values of the quantum parameter $\zeta$. The rest of the arguments are set  as: $M=1, Q=0.6$, and $L=2 $.}
    \label{fig:greybody}
\end{figure}

\subsection{Greybody Factor}

We will use the general method of semi-analytical bounds for the greybody factors \cite{gf1,gf2,gf3,FA7,FA9} to compute the greybody factor of the  quantum-corrected RN BH spacetime as follows, 
\begin{equation}
T\left( \omega \right) \geq \sec h^{2}\left( \int_{-\infty
}^{+\infty }\wp dr_{\ast }\right) ,  \label{is8}
\end{equation}

in which $r_{\ast }$ is the tortoise coordinate and 
\begin{equation}
\wp =\frac{1}{2h}\sqrt{\left( \frac{dh\left( r_{\ast }\right) }{dr_{\ast }}%
\right) ^{2}+(\omega ^{2}-V_{s}-h^{2}\left( r_{\ast }\right) )^{2}},
\label{is9}
\end{equation}

where $h(r\ast )$ is a positive function satisfying $h\left( -\infty \right)
=h\left( -\infty \right) =\omega $. For more details, one can see [39, 40]. We
select $h=\omega $. Thus, Eq. \eqref{is9} becomes
\begin{equation}
T \left( \omega \right) \geq \sec h^{2}\left( \int_{r_{h}}^{+\infty }%
\frac{V_{s}}{2\omega }dr_{\ast }\right) .  \label{gb1}
\end{equation}
We manage to obtain analytical expression for the greybody factor:
\begin{widetext}
\begin{equation}
T(\omega) = 
\mathrm{sech}^{2} \left[\frac{1}{2\omega} 
\left\{
- \frac{\ell(\ell+1)}{r_h} 
- \frac{M}{r_h^2} 
+ \frac{2(\zeta^2 + Q^2) r_h - 9 \zeta^2 M}{3 r_h^4} 
+ \frac{8 \zeta^2 (2 M^2 + Q^2)}{r_h^5} 
+ \frac{2 \zeta^2 Q^2 (9 Q^2 - 35 M r_h)}{21 r_h^7}\right\}\right].
\label{gf3}
\end{equation}
\end{widetext}

Figure \ref{fig:greybody} displays the greybody factor $T(\omega)$ for the quantum-corrected RN BH for different values of the parameter $\zeta$. It is evident that increasing $\zeta$ leads to a systematic suppression of the transmission probability over the entire frequency range.
This behavior can be understood from the corresponding effective potential analysis. As $\zeta$ increases, the height of the effective potential barrier surrounding the BH is enhanced, which strengthens the backscattering of outgoing modes. Consequently, a smaller fraction of scalar waves can tunnel through the potential barrier and reach asymptotic infinity, resulting in a reduced greybody factor.
Overall, the parameter $\zeta$ plays a crucial role in shaping the greybody spectrum of the quantum-corrected RN BH. Larger values of $\zeta$ effectively increase the filtering strength of the spacetime, leading to a suppression of Hawking radiation and potentially influencing the BH evaporation rate.

\section{Energy Emission Rate: Energy Flux }\label{sec:6}

Quantum effects in curved spacetime imply that BHs are not perfectly black: a thermal flux is emitted from the near-horizon region with temperature fixed by the surface gravity. In semiclassical language, this phenomenon may be interpreted through tunneling/particle production near the event horizon, which leads to a gradual decrease of the black-hole mass as energy is carried away to infinity \cite{Javed2019}. From the viewpoint of a distant observer, the radiated power depends not only on the Hawking temperature, but also on the probability that the emitted quanta escape the gravitational potential barrier. This effect is encoded in the frequency-dependent absorption cross section (greybody factor). 

In the geometric-optics (high-frequency) regime, the absorption cross section oscillates around a constant limiting value $\sigma_{\rm lim}$. Motivated by the fact that the capture of high-energy quanta is governed by null geodesics, the shadow cast by the black hole causes the high energy
cross section of absorption by the black hole. The limiting constant
value of cross section area, which is related to the radius of the photon
sphere is given as \cite{Misner1973,Mashhoon1973,Wei2013}
\begin{equation}
\sigma_{\rm lim}\approx \pi R^2_{\rm sh},
\end{equation}
where \(R_{\rm sh}\) denotes the radius of the black hole shadow.

Within this approximation, the spectral energy emission rate of black hole is given by the following equation \cite{Wei2013}
\begin{equation}
\frac{d^{2}\mathcal{E}}{d\omega\,dt}=\frac{2\pi ^{2}\sigma _{\rm lim}}{e^{\omega/T}-1}\,\omega ^{3},\label{eee1}
\end{equation}
where $\omega$ is the emitted frequency and $T$ is the Hawking temperature.

The Hawking temperature of the black hole system is given by
\begin{equation}
T=\frac{f'(r_h)}{4\pi}=\frac{1}{2\pi r_h}\left(\frac{M}{r_h}-\frac{Q^2}{r_h^2}\right),\quad r_h=M+\sqrt{M^2-Q^2}.\label{eee4}
\end{equation}
As the selected space-time is asymptotically flat, the shadow radius of the black hole is given by
\begin{align}
    R_{\rm sh}&=\frac{r_s}{\sqrt{f(r_s)}}=\beta_c\nonumber\\
    &=\frac{r_s}{\sqrt{\left(1-\frac{2 M}{r_s}+\frac{Q^2}{r^2_s}\right)+\frac{\zeta ^2}{r^2_s}\left(1-\frac{2 M}{r_s}+\frac{Q^2}{r^2_s}\right)^2}},\label{shadow}
\end{align}
where $r_s$ is the photon sphere radius satisfying the following polynomial relation ($2 \,f-r\,f'=0$):
\begin{align}
&r^3_s - 3Mr^2_s + 2Q^2 r_s\nonumber\\
&+ \zeta^2\left(
r_s - 5M + \frac{8M^2+4Q^2}{r_s}
- \frac{12MQ^2}{r^2_s}
+ \frac{4Q^4}{r^3_s}
\right)=0.
\end{align}

Thereby, substituting $T$ from (\ref{eee4}) into the (\ref{eee1}), we find the following expression:
\begin{align}
\frac{d^{2}\mathcal{E}}{d\omega\,dt}=\frac{2 \pi^3 \omega^3\,r_s^2\,\left[\exp\left\{2\pi r_h\, \omega\,\left(\frac{M}{r_h}-\frac{Q^2}{r_h^2}\right)^{-1} \right\}-1\right]^{-1}}{\left(1-\frac{2 M}{r_s}+\frac{Q^2}{r^2_s}\right)+\frac{\zeta ^2}{r^2_s}\left(1-\frac{2 M}{r_s}+\frac{Q^2}{r^2_s}\right)^2}.\label{eee5}
\end{align}
From the above analysis, we observe that the high-energy energy emission rate of the black hole is influenced by the quantum-correction parameter $\zeta$, the black hole mass $M$, and its electric charge $Q$.

\begin{figure}[ht!]
    \centering
    \includegraphics[width=1.1\linewidth]{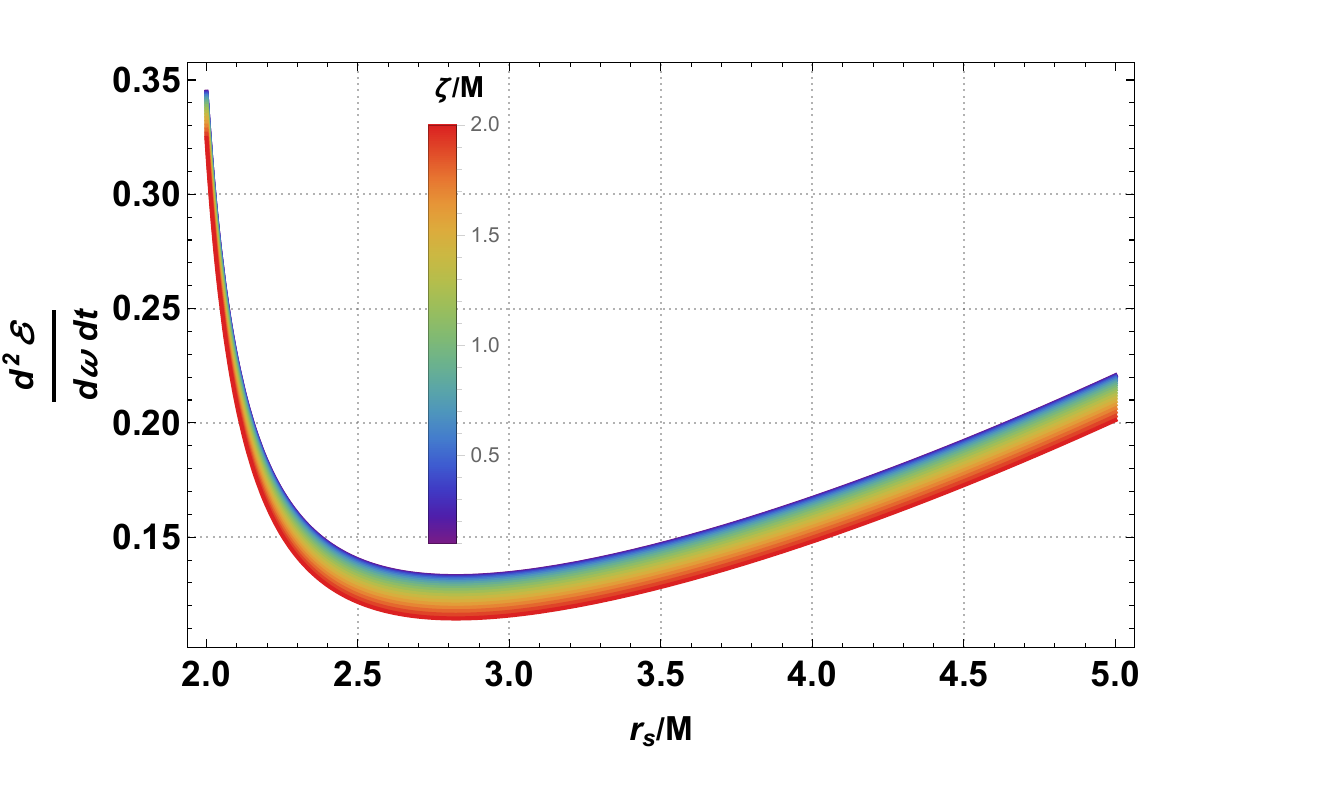}\\
    (i) $\omega=0.1$ \\
    \includegraphics[width=1.1\linewidth]{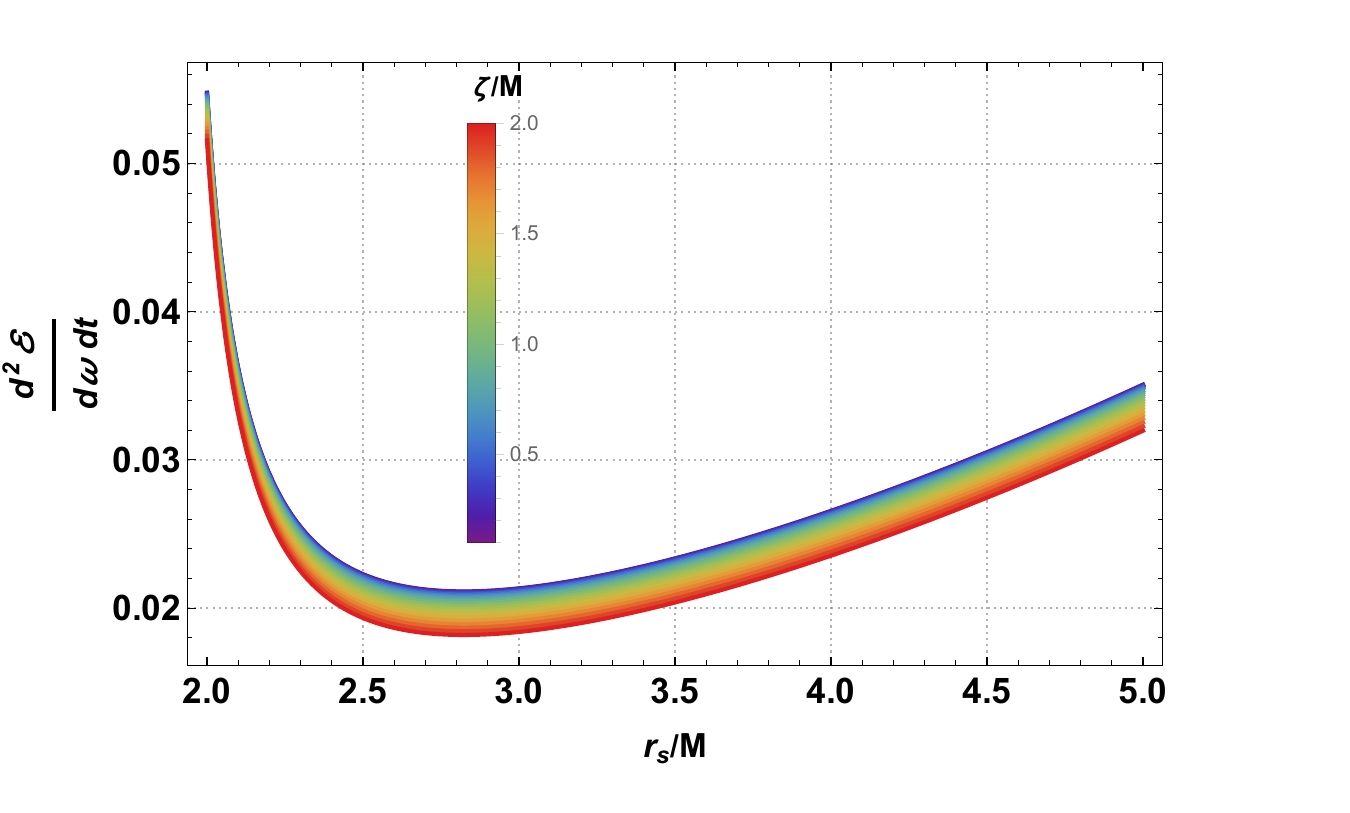}\\
    (ii) $\omega=0.3$
    \caption{Behaviors of the energy emission rate given in Eq.~(\ref{eee5}) as a function of photon sphere radius $r_s$ by varying quantum-correction parameter $\zeta$. Here $Q/M=0.5$}
    \label{fig:energy-flux}
\end{figure}

Below, we discuss some special cases.
\begin{itemize}
    \item When $\zeta=0$, corresponding to the absence of quantum-correction in the black hole, the selected space-time simplifies to the Reissner-Nordstrom metric. In this limit, the energy flux simplifies as
\begin{align}
\frac{d^{2}\mathcal{E}}{d\omega\,dt}=\frac{2 \pi^3 \omega^3\,r_s^2\,\left[\exp\left\{2\pi r_h\, \omega\,\left(\frac{M}{r_h}-\frac{Q^2}{r_h^2}\right)^{-1} \right\}-1\right]^{-1}}{\left(1-\frac{2 M}{r_s}+\frac{Q^2}{r^2_s}\right)},\label{eee6}
\end{align}
where $r_h$ is given earlier and $r_s$ is now as follows:
\begin{equation}
    r_s=\frac{3M+\sqrt{9 M^2-8 Q^2}}{2},\label{eee7}
\end{equation}
the photon sphere radius for RN-black hole.

\item When $Q=0$, corresponding to the absence of the electric charge of the black hole, the selected space-time simplifies to the quantum-corrected black hole \cite{Zhang2025}. In that limiting case, the energy flux reduces to (setting now $r_h=2M$):
\begin{align}
\frac{d^{2}\mathcal{E}}{d\omega\,dt}&=\frac{2 \pi^3 \omega^3\,r_s^2\,\left[e^{8\pi M \omega}-1\right]^{-1}}{\left(1-\frac{2 M}{r_s}\right)+\frac{\zeta ^2}{r^2_s}\left(1-\frac{2 M}{r_s}\right)^2},\label{eee8}
\end{align}
where $r_s$ here satisfies the following quartic equation:
\begin{align}
r^4_s - 3Mr^3_s + \zeta^2 r_s^2-5 M \zeta^2 r_s+8 M^2 \zeta^2=0.
\end{align}

We observe that the energy flux is modified by the quantum-correction parameter $\zeta$, which consequently reduces the emission compared to the standard Schwarzschild black hole case.
\end{itemize}

\begin{figure}[ht!]
    \centering
    \includegraphics[width=1.1\linewidth]{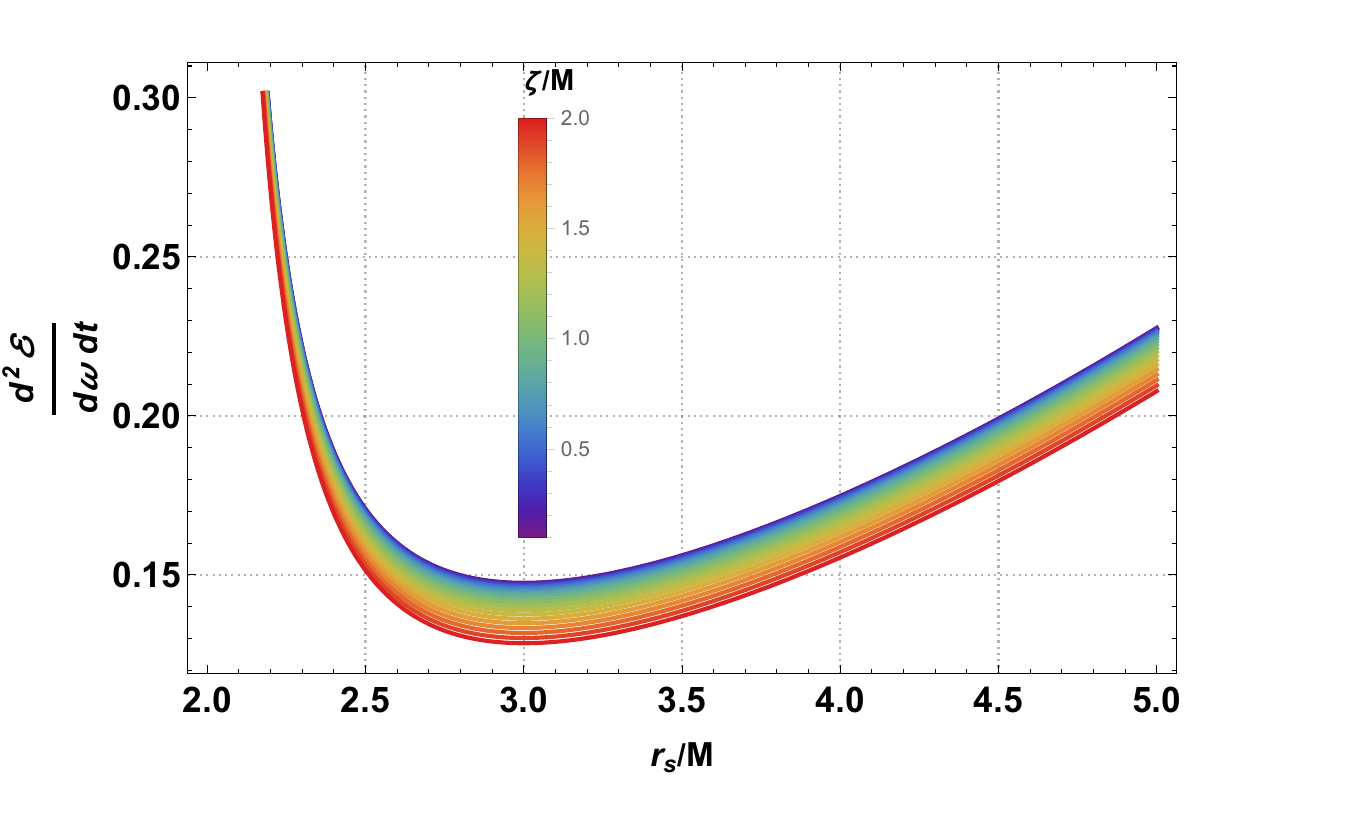}\\
    (i) $\omega=0.1$ \\
    \includegraphics[width=1.1\linewidth]{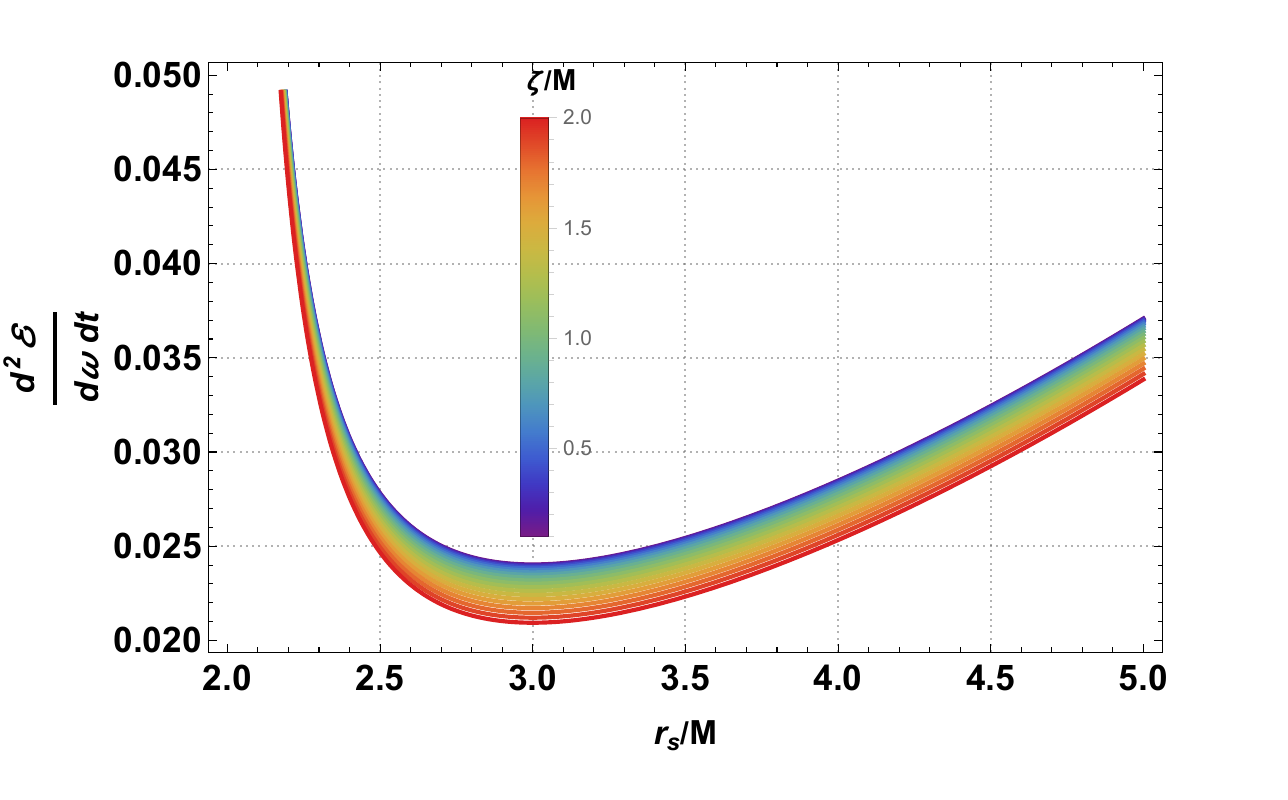}\\
    (ii) $\omega=0.3$
    \caption{Behaviors of the energy emission rate given in Eq.~(\ref{eee8}) as a function of photon sphere radius $r_s$ by varying quantum-correction parameter $\zeta$.}
    \label{fig:energy-flux2}
\end{figure}

\section{Thermal fluctuations}\label{sec:7}

We now investigate the effect of thermal fluctuations on the BH entropy. Within the Euclidean quantum gravity framework, the partition function is obtained by performing a Wick rotation of the time coordinate and expanding the gravitational action about the saddle-point (classical) geometry \cite{Pourhassan2015}. In this approach, thermal (or quantum) fluctuations around equilibrium lead to corrections to the leading semiclassical entropy. In the saddle-point approximation, the leading contribution reproduces the Bekenstein-Hawking entropy, $S=A/4=\pi r_h^2$, while Gaussian fluctuations around equilibrium generate subleading corrections. For a canonical ensemble, incorporating thermal fluctuations leads to a corrected entropy of the form \cite{Sadeghi2016,Pourhassan2023}:
\begin{equation}
S_c=S-\frac{\lambda}{2}\, \ln \left(S\,T^2\right),\label{tf1}
\end{equation}
where S is the original entropy of the BH without thermal fluctuations, and the correction parameter $\lambda$ added by hand to see the effect of logarithmic correction in the analytical expressions. These logarithmic corrections are typically negligible for large BHs, where the temperature is small and the heat capacity is sufficiently large to suppress fluctuations. However, for small BHs-characterized by higher temperatures and reduced thermodynamic stability-the canonical ensemble becomes increasingly sensitive to fluctuations, and the correction terms can become significant \cite{Mustafa2025}.

Using the standard Bekenstein-Hawking entropy, $S=\pi r_h^2$, the corrected entropy becomes
\begin{equation}
    S_c=\pi r_h^2-\frac{\lambda}{2}\,\ln\!\left[\frac{1}{4\pi}\left(\frac{M}{r_h}-\frac{Q^2}{r_h^2}\right)^2\right]\label{tf2}
\end{equation}
which is similar to that for the RN black hole.

\begin{figure}[ht!]
    \centering
    \includegraphics[width=1\linewidth]{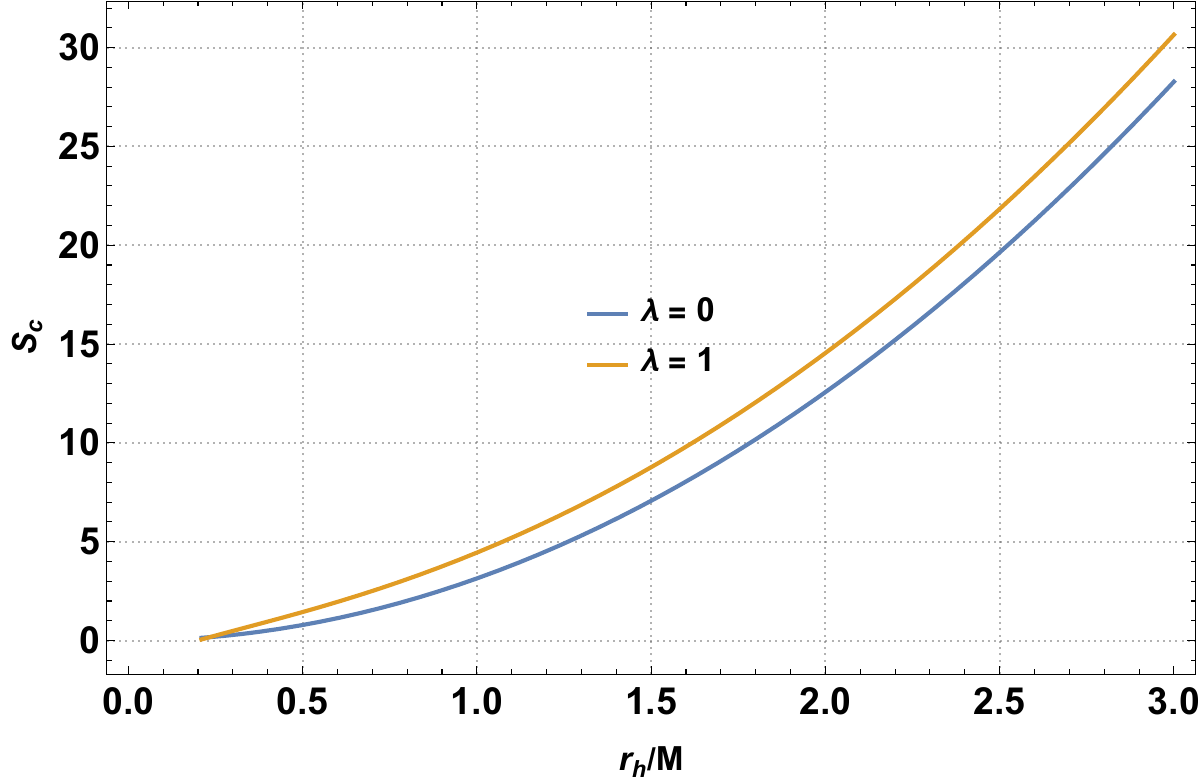}
    \caption{Behavior of the corrected entropy expression as a function of horizon $r_h$. Here $Q/M=0.2$.}
    \label{fig:entropy}
\end{figure}

\section{Summary and conclusions}\label{sec:8}

In this paper, we have performed a detailed phenomenological analysis of a covariant quantum-corrected RN BH, focusing on the role of the quantum correction parameter $\zeta$ together with the electric charge $Q$ and the mass $M$. The study has been carried out in a systematic way, starting from the particle dynamics and QPO observables, and then moving to the scalar perturbations, greybody factor, energy emission, and thermodynamic fluctuations.

We first analyzed the dynamics of neutral test particles in this spacetime by using the Hamiltonian formalism and deriving the effective potential. From the conditions for the stable circular motion, we obtained the Keplerian, radial, and vertical epicyclic frequencies as measured by a distant observer. These fundamental frequencies were then employed to construct several QPO models, including the relativistic precession and epicyclic resonance scenarios. In all cases, we observed that the quantum correction modifies the effective force acting on the particles and shifts the location of the characteristic orbits.

By introducing the normalized radial separation
$\delta r = r_{\mathrm{QPO}}/r_{\mathrm{ISCO}} - 1$,
we quantified the influence of the spacetime parameters on the QPO radii. We found that increasing the quantum parameter $\zeta$ generally reduces the separation between the QPO orbit and the ISCO, while a larger electric charge tends to increase this separation. This behavior is common for most of the QPO models that we considered, although the detailed dependence on the frequency ratio $p:q$ is different for the ER2 model.

In order to connect the theoretical model with the observations, we confronted the predicted QPO frequencies with the available data from the stellar-mass sources GRO~J1655-40 and XTE~J1550-564, the intermediate--mass BH candidate M82~X-1, and the supermassive BH Sgr~A*. By using a Bayesian framework together with an MCMC analysis, we obtained the constraints on the parameters $(r/M, M, Q/M, \zeta/M)$. The posterior distributions show well-behaved structures, which indicate that the model provides consistent fits to the observational data. In particular, the quantum correction parameter is constrained to finite nonzero values, suggesting that the QPO observations can in principle probe the quantum-modified BH geometries.

We then studied the scalar perturbations of the spacetime by separating the massless Klein-Gordon equation and transforming the radial part into a Schr\"{o}dinger-like equation. The corresponding effective potential explicitly shows how the parameters $Q$ and $\zeta$ modify the propagation of scalar waves. In the limit $\zeta \to 0$, the standard RN result is recovered. The positivity and the regular behavior of the potential outside the event horizon indicate the stability of the spacetime under scalar perturbations.

Next, we calculated the greybody factor and the spectral energy emission rate in the geometric-optics regime. In this approximation, the absorption cross section approaches a constant value proportional to the horizon area. We showed that the quantum correction affects the Hawking temperature through the horizon radius and therefore modifies the radiation spectrum as well as the absorption probability.

Finally, we investigated the effect of thermal fluctuations within the canonical ensemble. These fluctuations lead to logarithmic corrections to the Bekenstein-Hawking entropy, $S_c = S - \frac{\lambda}{2}\ln(S T^2)$, which become important for small BHs with high temperature and low heat capacity. In the large-horizon limit, the correction term becomes negligible and the standard area law is recovered.

To summarize, our analysis shows that the quantum correction parameter $\zeta$ leaves clear signatures in both the dynamical and thermodynamical sectors. From the observational point of view, the QPO data provide a powerful tool to constrain the parameter space of the model. From the theoretical point of view, the spacetime remains stable under the scalar perturbations and exhibits a consistent thermodynamic behavior with the quantum-induced logarithmic corrections.

The present work can be extended in several directions. It would be interesting to study the QNMs and compare them with the future GW observations, and also to consider rotating quantum-corrected BHs together with more realistic accretion disk models. Such investigations may provide further insight into the possible observational signatures of the quantum effects in the strong gravitational fields.

\scriptsize

\begin{acknowledgments}
F.A. gratefully acknowledges the Inter University Center for Astronomy and Astrophysics (IUCAA), Pune, India, for the conferment of a visiting associateship. The work of M.F. has been supported by Universidad Central de Chile through the project No. PDUCEN20240008.
\end{acknowledgments}



\bibliographystyle{apsrev4-2}
\bibliography{references}

\end{document}